\documentclass[12pt]{article} %***
\usepackage[sectionbib]{natbib}
\usepackage{array,epsfig,fancyheadings,rotating}
\usepackage[]{hyperref}  %<----modified by Ivan
\usepackage{sectsty, secdot}

\usepackage{chngcntr}

\counterwithin*{equation}{section}

\sectionfont{\fontsize{12}{14pt plus.8pt minus .6pt}\selectfont}
\renewcommand{\theequation}{\thesection\arabic{equation}}
\subsectionfont{\fontsize{12}{14pt plus.8pt minus .6pt}\selectfont}
\usepackage{amsmath}
\usepackage{amssymb}
\usepackage{amsfonts}
\usepackage{multirow}
\usepackage{amsthm}
\usepackage{verbatim}

\newtheorem{Theorem}{Theorem}

\theoremstyle{definition}

\newtheorem{Assumption}{Assumption}

\DeclareMathOperator{\tr}{tr}

\begin{document}
	
	%%%%%%%%%%%%%%%%%%%%%%%%%%%%%%%%%%%%%%%%%%%%%%%%%%%%%%%%%%%%%%%%%%%%%%%%%%%%%%%%%%%%%%%%%%%%%%%%%%%%%%%%%%%%%%%%%%%%%%%%%%%%
	%%%%%%%%%%%%%%%%%%%%%%%%%%%%%%%%%%%%%%%%%%%%%%%%%%%%%%%%%%%%%%%%%%%%%%%%%%%%%%%%%%%%%%%%%%%%%%%%%%%%%%%%%%%%%%%%%%%%%%%%%%%%
	
	\renewcommand{\baselinestretch}{1.5}
	
	\markright{ \hbox{\footnotesize\rm Statistica Sinica
			%{\footnotesize\bf 24} (201?), 000-000
		}\hfill\\[-13pt]
		\hbox{\footnotesize\rm
			%\href{http://dx.doi.org/10.5705/ss.20??.???}{doi:http://dx.doi.org/10.5705/ss.20??.???}
		}\hfill }
	
	\markboth{\hfill{\footnotesize\rm FIRSTNAME1 LASTNAME1 AND FIRSTNAME2 LASTNAME2} \hfill}
	{\hfill {\footnotesize\rm VI for Dynamic Latent Space Models} \hfill}
	
	\renewcommand{\thefootnote}{}
	$\ $\par
	
	%%%%%%%%%%%%%%%%%%%%%%%%%%%%%%%%%%%%%%%%%%%%%%%%%%%%%%%%%%%%%%%%%%%%%%%%%%%%%%%%%%%%%%%%%%%%%%%%%%%%%%%%%%%%%%%%%%%%%%%%%%%%
	
	\fontsize{12}{14pt plus.8pt minus .6pt}\selectfont \vspace{0.8pc}
	\centerline{\large\bf Variational Inference for Latent Space Models for Dynamic Networks}
	%\vspace{2pt} \centerline{\large\bf IF A SECOND LINE IS NEEDED}
	\vspace{.4cm} \centerline{Yan Liu and Yuguo Chen} \vspace{.4cm} \centerline{\it
		University of Illinois at Urbana-Champaign} \vspace{.55cm} \fontsize{9}{11.5pt plus.8pt minus
		.6pt}\selectfont
	
	%%%%%%%%%%%%%%%%%%%%%%%%%%%%%%%%%%%%%%%%%%%%%%%%%%%%%%%%%%%%%%%%%%%%%%%%%%%%%%%%%%%%%%%%%%%%%%%%%%%%%%%%%%%%%%%%%%%%%%%%%%%%
	
	\begin{quotation}
		\noindent {\it Abstract:}
		%{\bf Contents of the Abstract.}\\
		Latent space models are popular for analyzing dynamic network data. We propose a variational approach to estimate the model parameters as well as the latent positions of the nodes in the network. The variational approach is much faster than Markov chain Monte Carlo algorithms, and is able to handle large networks. Theoretical properties of the variational Bayes risk of the proposed procedure are provided. We apply the variational method and latent space model to simulated data as well as real data to demonstrate its performance.

		\vspace{9pt}
		\par
	\end{quotation}\par

	\def\thefigure{\arabic{figure}}
	\def\thetable{\arabic{table}}
	
	\renewcommand{\theequation}{\thesection.\arabic{equation}}

	\fontsize{12}{14pt plus.8pt minus .6pt}\selectfont
	
	\setcounter{section}{0} %***
	\setcounter{equation}{0} %-1
	
	\lhead[\footnotesize\thepage\fancyplain{}\leftmark]{}\rhead[]{\fancyplain{}\rightmark\footnotesize\thepage}%Put this line in Page 2
	
	\section{Introduction}
	Network data analysis has become an increasingly important research topic in various scientific disciplines in recent years. Most existing work on network data focuses on static networks, which means the inference is based on a static list of nodes and edges in an observed network at a given point in time (see \cite{goldenberg2010survey} for a survey). However, the network structures of real-world systems are often time-varying, or dynamic, in nature, with the set of nodes or the set of edges or both evolving over time. In this paper, we focus on a time series of observed networks, with the same set of nodes and a sequence of sets of edges observed at multiple time points. Analyzing such dynamic networks is crucial to understanding the dynamic aspect of networks, such as how social relations and structures, gene-protein interactions, and co-authorship patterns evolve over time.
	
	Many models for dynamic networks have been proposed in the literature, and some of them are extensions of existing static network models,
	including the dynamic versions of the stochastic blockmodel (SBM) \citep{yang2011detecting,xu2014dynamic,xu2014adaptive,xu2015stochastic,matias2017statistical}, the degree-corrected stochastic blockmodel \citep{wilson2016modeling}, the mixed membership stochastic blockmodel (MMSB) \citep{fu2009dynamic,xing2010state,ho2011evolving}, the exponential random graph model \citep{guo2007recovering,ahmed2009recovering,hanneke2010discrete,krivitsky2014separable,lee2019model}, the latent factor model \citep{ward2013gravity,durante2014bayesian,durante2014nonparametric,hoff2015multilinear}, the latent feature relational model \citep{foulds2011dynamic,heaukulani2013dynamic}, and the latent space model \citep{sarkar2006dynamic,sewell2015latent,friel2016interlocking,sewell2017latent}. See \cite{kim2018review} for a review of dynamic network models with latent variables.

	The latent space model embeds dynamic networks into a low-dimensional Euclidean space, and has the advantage of meaningful visualization and interpretation. The latent space model has also been used for multilayer networks \citep{gollini2016joint, durante2017nonparametric}. Although sometimes dynamic networks are treated as multilayer networks \citep{han2015consistent}, the temporal aspect of dynamic networks is not considered in multilayer networks.

	The latent space model is a popular approach for modeling dynamic networks, but the estimation of the model parameters and latent positions is often computationally intensive. \cite{sewell2015latent} used Markov chain Monte Carlo (MCMC) with a case-control approximate likelihood to reduce computational cost, but it still has difficulties in handling large networks. As an alternative, the variational inference approach \citep{jordan1999introduction, wainwright2008graphical} has recently become increasingly popular in the statistics community, see \cite{blei2017variational} for a comprehensive review. \cite{daudin2008mixture} proposed a variational approach to estimate the parameters of SBM, and this idea was later generalized by \cite{mariadassou2010uncovering} to deal with valued graphs and possible covariates.
\cite{yang2011detecting} and \cite{matias2017statistical} designed variational expectation-maximization (EM) algorithms for the dynamic version of SBM. \cite{airoldi2008mixed} used a variational approach to fit the MMSB, and \cite{xing2010state} and \cite{ho2011evolving} proposed variational EM algorithms for approximate inference for the dynamic version of the MMSB. \cite{salter2013variational} proposed a variational Bayes algorithm for a static latent space model with community structure.	\cite{sewell2017latent} proposed a variational Bayes algorithm for projection models in dynamic networks.

	Despite the empirical success of variational algorithms in estimating posterior distributions of the parameters of various network models, theoretical studies of such algorithms are limited.	Some results for variational approach have been obtained under the SBM \citep{celisse2012consistency, bickel2013asymptotic, zhang2017theoretical}, but	there is no theoretical results on variational algorithms for latent space models. Recently, \cite{yang2017alpha} proposed a more general class of variational Bayes algorithms, with the standard variational approximation as a special case. They also developed variational inequalities and convergence results on the Bayes risk of the proposed variational approximation. Their results indicate that the parameter estimates given by the variational algorithm converge to the true parameter values under certain conditions. This work provides a framework to analyze the theoretical properties of variational Bayes algorithms for latent space models.
	
	In this paper, we consider a class of latent space models for dynamic networks, and propose a variational algorithm to perform posterior inference for large scale networks. Furthermore, we show that the parameter estimation based on the variational posterior is consistent. The simulation studies demonstrate that the proposed variational algorithm is much faster than the MCMC algorithm while still gives satisfactory results, so it is more suitable for large scale dynamic networks. We also apply our method to analyze a friendship network from ``Teenage Friends and Lifestyle Study" and a Wiki-talk communication network.
	
	The rest of the paper is organized as follows. Section \ref{sec:model} considers a class of latent space models for dynamic networks. Section \ref{sec:VB} proposes a variational algorithm for posterior inference. Section \ref{sec:verify} gives finite sample upper bounds to variational Bayes risk of the proposed procedure and shows consistency of parameter estimation. Sections \ref{sim} and \ref{real_data} illustrate the performance of the proposed method with simulation studies and real-world examples, respectively. Section \ref{sec:discussion} concludes the paper with a discussion.

	\setcounter{equation}{0} %-1
	
	\section{Dynamic Latent Space Model}
     \label{sec:model}
	Latent space models for network data are a popular class of models first proposed by \cite{hoff2002latent} for static networks, and later generalized by \cite{sewell2015latent} and \cite{sewell2016latent} to dynamic networks. This class of models embed the nodes of a network into an unobserved latent space, which can provide visualization and insight into the evolution of the network.

	Suppose there are $n$ nodes and $T$ time steps in a dynamic network. Let $\boldsymbol{Y}_1,\ldots,\boldsymbol{Y}_T$ be the observed $n\times n$ adjacency matrices at the $T$ time steps, where $Y_{ijt}=1$ if there is an edge from node $i$ to node $j$ at time $t$, and 0 otherwise. The latent space we consider will be the $d$-dimensional Euclidean space $\mathbb{R}^d$ throughout the paper. Let $\boldsymbol{X}_{it} \in \mathbb{R}^d$ be the latent position of the $i$-th node at time $t$, $1\leq i \leq n$, $1\leq t \leq T$. The dynamic latent space model represents the time series of networks by a hidden Markov model. The latent positions of the nodes (which are the hidden states) evolve independently of each other, and the latent position of each node is modeled by a Markov process with the initial distribution
	$
	\boldsymbol{X}_{i1} \sim \mathcal{N}(\boldsymbol{0},\sigma^2 \mathbb{I}_{d})
	(i=1,\dots,n),
	$
	and transition distribution
	$
	\boldsymbol{X}_{it}|\boldsymbol{X}_{i(t-1)} \sim \mathcal{N}(\boldsymbol{X}_{i(t-1)},\tau^2 \mathbb{I}_{d})
	(t=2,\dots,T,~ i=1,\dots,n),
	$
	where $\mathbb{I}_{d}$ is the $d\times d$ identity matrix.
	
	The observed networks $\boldsymbol{Y}_1,\ldots,\boldsymbol{Y}_T$ are conditionally independent given the latent positions across the time span. In addition, for the network $\boldsymbol{Y}_t$ at time $t$, we assume that all edges $Y_{ijt}$ are formed independently conditioning on the latent positions at time $t$. A general expression for the probability that there is an edge between nodes $i$ and $j$ can be written as
	$
	p(Y_{ijt}=1|\boldsymbol{X}_{it},\boldsymbol{X}_{jt},\boldsymbol{\beta})
	= h(||\boldsymbol{X}_{it}-\boldsymbol{X}_{jt}||,\boldsymbol{\beta}), \quad
	1\leq i \neq j \leq n,\ \  1\leq t \leq T,
	$
	where $||\boldsymbol{X}_{it}-\boldsymbol{X}_{jt}||$ is the Euclidean distance between nodes $i$ and $j$ at time $t$, $\boldsymbol{\beta}\in\mathbb{R}^p$ denote the model parameters that do not change with $t$, and
	$h: \mathbb{R}^{p+1}\rightarrow [0,1]$ is a function that is strictly decreasing in its first argument. This model assumes that a smaller distance between two nodes in the latent space yields a larger link probability between them.

Different forms of the function $h$ have been suggested in the literature. The distance model in \cite{hoff2002latent} assumes that $\boldsymbol{\beta}$ is a one-dimensional intercept term, and
	$\text{logit} [h(||\boldsymbol{X}_{it}-\boldsymbol{X}_{jt}||,\beta)]= \beta - ||\boldsymbol{X}_{it}-\boldsymbol{X}_{jt}||$. However, the variation inference based on this formulation involves several additional approximation steps \citep{salter2013variational}.	\cite{gollini2016joint} suggested using $\text{logit} [h(||\boldsymbol{X}_{it}-\boldsymbol{X}_{jt}||,\beta)]= \beta - ||\boldsymbol{X}_{it}-\boldsymbol{X}_{jt}||^2$, which can reduce the number of approximation steps in variational inference. \cite{sewell2015latent} proposed a more complicated $h$ function to distinguish the effect of the sender and the receiver in edge formation.
	
	In this paper, we adopt the $h$ function suggested by \cite{gollini2016joint} to derive the variational inference. More precisely, for any $1\leq i \neq j \leq n$ and $1\leq t\leq T$,
	$
	\text{logit} [p(Y_{ijt}=1|\boldsymbol{X}_{it},\boldsymbol{X}_{jt},\beta)]
	= \beta - ||\boldsymbol{X}_{it}-\boldsymbol{X}_{jt}||^2.
	$
	Let $\boldsymbol{\mathcal{X}}=\left\{\boldsymbol{X}_{it}: 1 \leq i \leq n,\ 1 \leq t \leq T \right\}$ and $\boldsymbol{\mathcal{Y}}=\left\{\boldsymbol{Y}_{t}: 1\leq t \leq T\right\}$. Then the model can be written as
	\begin{align} \label{likelihood}
	p(\boldsymbol{\mathcal{Y}}|\boldsymbol{\mathcal{X}},\beta)
	= \prod_{t=1}^{T}\prod_{1\leq i\neq j \leq n}	\frac{\exp\{Y_{ijt}\left(\beta-||\boldsymbol{X}_{it}-\boldsymbol{X}_{jt}||^2\right)\}}{1+\exp\{\beta-||\boldsymbol{X}_{it}-\boldsymbol{X}_{jt}||^2\}}.
	\end{align}

	\section{A Variational Algorithm for Posterior Inference}
    \label{sec:VB}
	
	Variational inference (VI) or variational Bayes (VB) \citep{jordan1999introduction,wainwright2008graphical} is a powerful tool for approximating intractable complex distributions. The basic idea of VB is to approximate the posterior distribution by the closest member in a certain family of distributions (which is usually called the variational family). The closest member, which is referred to as the variational distribution, is then used for posterior inference. Thus the posterior inference problem is turned into an optimization problem of finding the member in the variational family that minimizes a divergence measure between the approximate posterior and the true posterior.
	
	The most popular approach for variational inference is the mean-field method, which approximates the target distribution by a fully factorized distribution. In this section, we further restrict each component of the factorized distribution to be in a family of tractable distributions indexed by variational parameters. These variational parameters are chosen to minimize the Kullback-Leibler (KL) divergence between the approximate posterior and the true posterior.
	
	Now we derive a variational algorithm for posterior inference of the dynamic latent space model described in Section \ref{sec:model}. We are interested in $p(\boldsymbol{\mathcal{X}},\beta|\boldsymbol{Y}_1,\ldots,\boldsymbol{Y}_T)$,
	the posterior distribution of the intercept $\beta$ as well as the latent positions of the nodes $\boldsymbol{\mathcal{X}}$.
	We assign a normal prior $\mathcal{N}(\xi,\psi^2)$ for the intercept $\beta$, and view $\xi$, $\psi^2$, $\sigma^2$ and $\tau^2$ as hyperparameters. Then
	$
	p(\boldsymbol{\mathcal{X}},\beta|\boldsymbol{Y}_1,\ldots,\boldsymbol{Y}_T)
	\propto p(\beta)p(\boldsymbol{\mathcal{X}}) \prod_{t=1}^{T}p(\boldsymbol{Y}_t|\boldsymbol{\mathcal{X}},\beta).
	$
We approximate the posterior by the following family of distributions
	%\vspace{-0.3cm}
	$$
	q(\boldsymbol{\mathcal{X}},\beta) = q(\beta=\cdot|\tilde{\xi},\tilde{\psi}^2)\prod_{t=1}^{T}\prod_{i=1}^{n}q(\boldsymbol{X}_{it}=\cdot|\tilde{\boldsymbol{\mu}}_{it},\tilde{\Sigma}),
	$$
where $q(\beta=\cdot|\tilde{\xi},\tilde{\psi}^2)$ is a normal distribution with mean $\tilde{\xi}$ and variance $\tilde{\psi}^2$, and $q(\boldsymbol{X}_{it}=\cdot|\tilde{\boldsymbol{\mu}}_{it},\tilde{\Sigma})$ is a $d$-dimensional normal distribution with mean vector $\tilde{\boldsymbol{\mu}}_{it}$ and covariance matrix $\tilde{\Sigma}$. Note that we can also allow the covariance matrix $\tilde{\Sigma}$ to vary with $i$ and $t$, and the derivation of the variational algorithm is similar.

	The main hindrance in deriving the analytical form of the KL divergence between $q(\boldsymbol{\mathcal{X}},\beta)$ and $p(\boldsymbol{\mathcal{X}},\beta|\boldsymbol{\mathcal{Y}})$ is	the expectation of the log-likelihood $\mathbb{E}_{q}[\log p(\boldsymbol{\mathcal{Y}}|\boldsymbol{\mathcal{X}},\beta)]$, which does not have an analytical form. Therefore, instead of working with the original KL divergence, we approximate it by a lower bound derived below:
	%\vspace{-0.3cm}
	\begin{align*}
	\mathbb{E}_{q}[\log p(\boldsymbol{\mathcal{Y}}|\boldsymbol{\mathcal{X}},\beta)]
	&= \sum_{t=1}^{T}\sum_{i\neq j}\mathbb{E}_{q}
	\left[
	Y_{ijt}(\beta - ||\boldsymbol{X}_{it}-\boldsymbol{X}_{jt}||^2)
	-\log\left(1+e^{\beta - ||\boldsymbol{X}_{it}-\boldsymbol{X}_{jt}||^2}\right)
	\right]\\
	&\geq \sum_{t=1}^{T}\sum_{i\neq j}
	\left[
	Y_{ijt}(\tilde{\xi} - \mathbb{E}_{q}[||\boldsymbol{X}_{it}-\boldsymbol{X}_{jt}||^2])
	\right]
	-\log\left(
	1+\mathbb{E}_{q}\left[e^{\beta - ||\boldsymbol{X}_{it}-\boldsymbol{X}_{jt}||^2}\right]
	\right)\\
	&= \sum_{t=1}^{T}\sum_{i\neq j}
	\biggr\{ Y_{ijt}
	\left(
	\tilde{\xi}-2\tr(\tilde{\Sigma})-||\tilde{\boldsymbol{\mu}}_{it}-\tilde{\boldsymbol{\mu}}_{jt}||^2
	\right)\\
	&\quad -\log\biggr(
	1+\frac{\exp\{\tilde{\xi}+\frac{1}{2}\tilde{\psi}^2\}}{\det(\mathbb{I}+4\tilde{\Sigma})^{1/2}}
	\cdot
	\exp\{
	-(\tilde{\boldsymbol{\mu}}_{it}-\tilde{\boldsymbol{\mu}}_{jt})^T
	(\mathbb{I}+4\tilde{\Sigma})^{-1}
	(\tilde{\boldsymbol{\mu}}_{it}-\tilde{\boldsymbol{\mu}}_{jt})
	\}
	\biggr)
	\biggr\},
	\end{align*}
where the inequality is due to Jensen's inequality. Then the KL divergence $D\left[q(\boldsymbol{\mathcal{X}},\beta)||p(\boldsymbol{\mathcal{X}},\beta|\boldsymbol{\mathcal{Y}})\right]$ between the approximate posterior and the true posterior can be approximated by an upper bound as follows:
	\begin{align}
	\nonumber
	&\;\;\;\;\;\; D\left[q(\boldsymbol{\mathcal{X}},\beta)||p(\boldsymbol{\mathcal{X}},\beta|\boldsymbol{\mathcal{Y}})\right]\\ \nonumber
	&:=\mathbb{E}_{q}[\log q(\boldsymbol{\mathcal{X}})]-\mathbb{E}_{q}[\log p(\boldsymbol{\mathcal{X}})]
	+\mathbb{E}_{q}[\log q(\beta)]-\mathbb{E}_{q}[\log p(\beta)]
	-\mathbb{E}_{q}[\log p(\boldsymbol{\mathcal{Y}}|\boldsymbol{\mathcal{X}},\beta)] + constant\\ \nonumber
	&\leq -\frac{nT}{2}\log(\det(\tilde{\Sigma}))
	+ \left(\frac{n}{2\sigma^2}+\frac{n(T-1)}{\tau^2}\right)\tr(\tilde{\Sigma})\\ \nonumber
	&\quad + \frac{1}{2\sigma^2}\sum_{i=1}^{n}
	\left\lVert \tilde{\boldsymbol{\mu}}_{i1} \right\rVert^2
	+ \frac{1}{2\tau^2}\sum_{t=2}^{T}\sum_{i=1}^{n} \left\lVert \tilde{\boldsymbol{\mu}}_{it}-\tilde{\boldsymbol{\mu}}_{i(t-1)}\right\rVert^2
	+ \frac{1}{2}\left(\frac{\tilde{\psi}^2}{\psi^2}-\log\frac{\tilde{\psi}^2}{\psi^2} + \frac{(\tilde{\xi}-\xi)^2}{\psi^2} \right)\\ \nonumber
	&\quad - \sum_{t=1}^{T}\sum_{i\neq j}\biggr\{ Y_{ijt}
	\left(
	\tilde{\xi}-2\tr(\tilde{\Sigma})-||\tilde{\boldsymbol{\mu}}_{it}-\tilde{\boldsymbol{\mu}}_{jt}||^2
	\right) \\
	&\quad -\log\left(
	1+\frac{\exp\{\tilde{\xi}+\frac{1}{2}\tilde{\psi}^2\}}{\det(\mathbb{I}+4\tilde{\Sigma})^{1/2}}
	\cdot
	\exp\{
	-(\tilde{\boldsymbol{\mu}}_{it}-\tilde{\boldsymbol{\mu}}_{jt})^T
	(\mathbb{I}+4\tilde{\Sigma})^{-1}
	(\tilde{\boldsymbol{\mu}}_{it}-\tilde{\boldsymbol{\mu}}_{jt})
	\}
	\right)
	\biggr\} + constant.
	\label{KL_approx}
	\end{align}
Two different constant terms in the above derivation are not written out explicitly and will be omitted later because they do not play a role in the optimization procedure. Compared with the multiple approximation steps in the derivation of variational algorithm in \cite{salter2013variational}, only one approximation step based on Jensen's inequality is used in our derivation. It is also worth mentioning that this Jensen's bound is tighter than the lower bound given by first order approximation.
	
	Our goal is to minimize the approximated KL divergence (\ref{KL_approx}) over variational parameters $\{\tilde{\boldsymbol{\mu}}_{it}\}$, $\tilde{\Sigma}$, $\tilde{\xi}$ and $\tilde{\psi}^2$.
	For convenience, we define
	$$
	f(\tilde{\boldsymbol{\mu}},\tilde{\Sigma},\tilde{\xi},\tilde{\psi}^2)
	:= \sum_{t=1}^{T}\sum_{i\neq j}
	\log\biggr( 1+\frac{\exp\{\tilde{\xi}+\frac{1}{2}\tilde{\psi}^2\}}{\det(\mathbb{I}+4\tilde{\Sigma})^{1/2}}
	\cdot
	\exp\{
	-(\tilde{\boldsymbol{\mu}}_{it}-\tilde{\boldsymbol{\mu}}_{jt})^T
	(\mathbb{I}+4\tilde{\Sigma})^{-1}
	(\tilde{\boldsymbol{\mu}}_{it}-\tilde{\boldsymbol{\mu}}_{jt})
	\}
	\biggr)
	\biggr\}.
	$$
In order to avoid numerical root searching in the optimization over $\{\tilde{\boldsymbol{\mu}}_{it}\}$, we use the Taylor expansion  to approximate the gradient function
	$
	f'(x)=f'(x_0)+f''(x_0)(x-x_0)+o(x-x_0).
	$
	
	Assume we have the estimates of the variational parameters after $s$ iterations. Then the update equation for each variational parameter in the $(s+1)$-th iteration is as follows.
	\begin{itemize}
		\item Update of $\tilde{\Sigma}$:
		$$
		\tilde{\Sigma}^{(s+1)} \leftarrow
		\frac{nT}{2}
		\left[
		\left(
		\frac{n}{2\sigma^2}+\frac{n(T-1)}{\tau^2}+\sum_{t=1}^{T}\sum_{i\neq j}2Y_{ijt}
		\right)\mathbb{I}_{d}
		+ J(\tilde{\Sigma}^{(s)})
		\right]^{-1},
		$$
where $J(\tilde{\Sigma}^{(s)})$ is the Jacobian matrix of $f$ evaluated at $\tilde{\Sigma}^{(s)}$, and has the following expression
		$$
		J(\tilde{\Sigma}^{(s)})
		= \sum_{t=1}^{T}\sum_{i\neq j} \frac{1}{A_{ijt}}
		\left(
		-8(\mathbb{I}_{d}+4\tilde{\Sigma}^{(s)})^{-1} (\tilde{\boldsymbol{\mu}}_{it}^{(s)}-\tilde{\boldsymbol{\mu}}_{jt}^{(s)}) (\tilde{\boldsymbol{\mu}}_{it}^{(s)}-\tilde{\boldsymbol{\mu}}_{jt}^{(s)})^T (\mathbb{I}_{d}+4\tilde{\Sigma}^{(s)})^{-1}
		\right),
		$$
where
		$
		A_{ijt} = 1 + \frac{\exp\{-\tilde{\xi}-\frac{1}{2}\tilde{\psi}^2\}}{\det(\mathbb{I}+4\tilde{\Sigma}^{(s)})^{-1/2}}
		\cdot
		\exp\{
		(\tilde{\boldsymbol{\mu}}_{it}-\tilde{\boldsymbol{\mu}}_{jt})^T
		(\mathbb{I}+4\tilde{\Sigma}^{(s)})^{-1}
		(\tilde{\boldsymbol{\mu}}_{it}-\tilde{\boldsymbol{\mu}}_{jt})
		\}.
		$
		
		\item Update of $\tilde{\boldsymbol{\mu}}_{it}$:
		\begin{align*}
		\tilde{\boldsymbol{\mu}}_{it}^{(s+1)} \leftarrow
		&\left(
		H(\tilde{\boldsymbol{\mu}}_{it}^{(s)})
		+\left(\frac{2}{\tau^2} + \sum_{j\neq i}2(Y_{ijt}+Y_{jit})\right)\mathbb{I}_{d}
		\right)^{-1}\\
		&\cdot \left[
		\sum_{j\neq i}2(Y_{ijt}+Y_{jit})\tilde{\boldsymbol{\mu}}_{jt}^{(s)}
		+\frac{1}{\tau^2}(\tilde{\boldsymbol{\mu}}_{i(t-1)}^{(s)}+\tilde{\boldsymbol{\mu}}_{i(t+1)}^{(s)})
		+ H(\tilde{\boldsymbol{\mu}}_{it}^{(s)})\tilde{\boldsymbol{\mu}}_{it}^{(s)}
		-G(\tilde{\boldsymbol{\mu}}_{it}^{(s)})
		\right],
		\end{align*}
		where $G(\tilde{\boldsymbol{\mu}}_{it}^{(s)})$ is the gradient of $f$ evaluated at $\tilde{\boldsymbol{\mu}}_{it}^{(s)}$, and $H(\tilde{\boldsymbol{\mu}}_{it}^{(s)})$ is the Hessian matrix of $f$ evaluated at $\tilde{\boldsymbol{\mu}}_{it}^{(s)}$, i.e.,
		$$
		G(\tilde{\boldsymbol{\mu}}_{it}^{(s)})
		= \sum_{i\neq j} \frac{-2}{A_{ijt}}
		(\mathbb{I}_{d}+4\tilde{\Sigma}^{(s)})^{-1}(\tilde{\boldsymbol{\mu}}_{it}^{(s)}-\tilde{\boldsymbol{\mu}}_{jt}^{(s)}),
		$$
		\begin{align*}
		H(\tilde{\boldsymbol{\mu}}_{it}^{(s)})
		=& \sum_{i\neq j}
		\frac{4}{1+A_{ijt}+\frac{1}{A_{ijt}-1}}(\mathbb{I}_{d}+4\tilde{\Sigma}^{(s)})^{-1} (\tilde{\boldsymbol{\mu}}_{it}^{(s)}-\tilde{\boldsymbol{\mu}}_{jt}^{(s)}) (\tilde{\boldsymbol{\mu}}_{it}^{(s)}-\tilde{\boldsymbol{\mu}}_{jt}^{(s)})^T (\mathbb{I}_{d}+4\tilde{\Sigma}^{(s)})^{-1}\\
		&-\frac{2}{1+A} (\mathbb{I}_{d}+4\tilde{\Sigma}^{(s)})^{-1}.
		\end{align*}

		\item Update of $\tilde{\xi}$:
		$
		\tilde{\xi}^{(s+1)} \leftarrow
		\left(1+\psi^2 f''(\tilde{\xi}^{(s)})\right)^{-1}
		\left[
		\xi+\psi^2
		\left(
		\sum_{t=1}^{T}\sum_{i\neq j}Y_{ijt} + f''(\tilde{\xi}^{(s)})\tilde{\xi}^{(s)}-f'(\tilde{\xi}^{(s)})
		\right)
		\right].
		$
		
		\item Update of $\tilde{\psi}^2$:
		$
		\tilde{\psi}^{2~(s+1)} \leftarrow
		\left(\frac{1}{\psi^2}+2f'(\tilde{\psi}^{2~(s)})\right)^{-1}.
		$
		
	\end{itemize}
	
	We declare that convergence of the algorithm is achieved if the relative change in the log-likelihood at two consecutive steps is smaller than some threshold value. After convergence, the variational parameter $\tilde{\boldsymbol{\mu}}_{it}$ will be the estimated posterior mean of the latent position of node $i$ at time $t$, and $\tilde{\xi}$ will be the estimated posterior mean of the intercept.

	%%%%%%%%%%%%%%%%%%%%%%%%%%%%%%%%%%%%%%%%%%%%%%%%%%%%%%%%%%%%%%%%%%%%%%%%%%%
	\section{Theoretical Properties}\label{sec:verify}
	
	In this section, we provide theoretical properties of variational algorithms for latent space models, which have not been studied in the literature.	In particular, we show that under certain regularity conditions, the point estimates of model parameters resulted from the VB procedure converge to the true parameters as the number of nodes goes to infinity. We will achieve this result by showing that the variational Bayes risk goes to 0 as the number of nodes goes to infinity.
	
	We follow the framework of \cite{yang2017alpha} in which the authors studied a slightly modified VB procedure, which is called $\alpha$-VB, with the standard VB algorithm as a special case. The authors showed that the point estimates given by this VB procedure converge to the true parameters. The main idea of the proof is to establish a finite-sample upper bound of the variational Bayes risk using the VB objective function, and then give the convergence rate of this upper bound. Note that the theoretical results in \cite{yang2017alpha} are mainly for independent and identically distributed data, while our results are tailored to dependent random variables in the network setting.

	We first introduce some notations used in this section. Let $D(p||q)=\int p \log (\frac{p}{q}) d\mu$ denote the Kullback-Leibler divergence between probability density functions $p$ and $q$ with respect to a measure $\mu$. For $\alpha\in (0,1)$, let
	$
	D_{\alpha}(p||q)
	:= \frac{1}{\alpha} \log \int p^{\alpha} q^{1-\alpha} d\mu
	$
denote the $\alpha$-divergence between probability density functions $p$ and $q$. In this section, we use $\theta:=(\beta,\sigma^2,\tau^2)$ to denote all the parameters in the model, and  $\pi:=(\sigma^2,\tau^2)$ to denote the parameters that are related to the latent variables. We use $q_{\theta}$ and $q_{\boldsymbol{\mathcal{X}}}$ to denote the variational distribution of the model parameters $\theta$ and the variational distribution of the latent variables $\boldsymbol{\mathcal{X}}$, respectively. Also, for the rest of this section, we assume $\theta$ has a true value $\theta^*$.
	
	Now we follow the framework of \cite{yang2017alpha} and derive a further decomposition of the VB objective function. If we adopt the standard VB algorithm, we minimize the KL divergence between the variational posterior and the true posterior, i.e., we minimize the following objective function over the members of the variational family:
	\begin{align}
	&\quad D(q(\theta,\boldsymbol{\mathcal{X}})||p(\theta,\boldsymbol{\mathcal{X}}|\boldsymbol{\mathcal{Y}})) \nonumber \\
	&=\mathbb{E}_{q}[\log q(\boldsymbol{\mathcal{X}})] + \mathbb{E}_{q}[\log q(\theta)]
	- \mathbb{E}_{q}[\log p(\boldsymbol{\mathcal{Y}}|\boldsymbol{\mathcal{X}},\theta)]
	- \mathbb{E}_{q}[\log p(\boldsymbol{\mathcal{X}}|\theta)] - \mathbb{E}_{q}[\log p(\theta)] + constant \nonumber \\
	&= D(q(\theta)||p(\theta))
	-\int\int \log\left[\frac{p(\boldsymbol{\mathcal{Y}}|\boldsymbol{\mathcal{X}},\theta)p(\boldsymbol{\mathcal{X}}|\theta)}{q_{\boldsymbol{\mathcal{X}}}(\boldsymbol{\mathcal{X}})}\right] q_{\theta}(d\theta) q_{\boldsymbol{\mathcal{X}}}(d\boldsymbol{\mathcal{X}}) + constant.
	\label{decom1}
	\end{align}
	
	Let
	$
	l_n(\theta):=\log p(\boldsymbol{\mathcal{Y}}|\theta)
	=\log \left(
	\int p(\boldsymbol{\mathcal{Y}}|\theta,\boldsymbol{\mathcal{X}})p(\boldsymbol{\mathcal{X}}|\theta) d\boldsymbol{\mathcal{X}}
	\right),
	$ and
	$
	\hat{l}_n(\theta)
	:=\int \log \left( \frac{p(\boldsymbol{\mathcal{Y}}|\theta,\boldsymbol{\mathcal{X}})p(\boldsymbol{\mathcal{X}}|\theta)}{q_{\boldsymbol{\mathcal{X}}}(\boldsymbol{\mathcal{X}})}
	\right)
	q_{\boldsymbol{\mathcal{X}}}(d\boldsymbol{\mathcal{X}}).
	$
By Jensen's inequality, $l_n(\theta) \geq \hat{l}_n(\theta)$. Then the KL divergence (\ref{decom1}) can be decomposed into three parts as follows
	\begin{align}	D(q(\theta,\boldsymbol{\mathcal{X}})||p(\theta,\boldsymbol{\mathcal{X}}|\boldsymbol{\mathcal{Y}}))
	&= D(q(\theta)||p(\theta))
	+ \int \left(l_n(\theta)-\hat{l}_n(\theta)\right) q_{\theta}(d\theta)
	-\int l_n(\theta) q_{\theta}(d\theta) + constant,
	\label{decom3}
	\end{align}
where the first term is the discrepancy between the variational distribution and the prior of the model parameters, the second term is an average Jensen gap (will be denoted by $\Delta_{J}(q_{\theta},q_{\boldsymbol{\mathcal{X}}})$) due to the variational approximation (which is the only term that involves the variational distribution $q(\boldsymbol{\mathcal{X}})$), and the third term is the integrated log-likelihood. 	The constant term will be omitted later.
	
	Define the following objective function
	\begin{align*}
	\Psi_n(q_{\theta},q_{\boldsymbol{\mathcal{X}}})
	&:= -\int_{\Theta}(l_n(\theta)-l_n(\theta^*)) q(d\theta)
	+\Delta_{J}(q_{\theta},q_{\boldsymbol{\mathcal{X}}})
	+ D(q(\theta)||p(\theta)),
	\end{align*}
where the subscript $n$ indicates the dependence of the objective function on the number of nodes $n$. Notice that minimizing $\Psi_n$ over $(q_{\theta},q_{\boldsymbol{\mathcal{X}}})$ is equivalent to minimizing the KL divergence (\ref{decom3}) over $(q_{\theta},q_{\boldsymbol{\mathcal{X}}})$, since $l_n(\theta^*)$ does not depend on the variational distribution $q(\boldsymbol{\mathcal{X}})$.
	
	\cite{yang2017alpha} proposed a slightly different procedure, called the $\alpha$-variational Bayes ($\alpha$-VB), in which a stronger penalty on the discrepancy between the variational distribution and the prior is introduced into the objective function
	\begin{align}
	\Psi_{n,\alpha}(q_{\theta},q_{\boldsymbol{\mathcal{X}}})
	:&= -\int_{\Theta}(l_n(\theta)-l_n(\theta^*)) q(d\theta)
	+\Delta_{J}(q_{\theta},q_{\boldsymbol{\mathcal{X}}})
	+\frac{1}{\alpha}D(q(\theta)||p(\theta)), \label{alphaVBobj}
	\end{align}
where $\alpha\in(0,1]$ is a tuning parameter, $q_{\theta}$ and $q_{\boldsymbol{\mathcal{X}}}$ are variational distributions which are restricted in some variational family $\Gamma_{\theta}$ and $\Gamma_{\boldsymbol{\mathcal{X}}}$, respectively. Note that the $\alpha$-VB objective function $\Psi_{n,\alpha}$ reduces to $\Psi_{n}$ when $\alpha=1$. The $\alpha$-VB solution is defined by
	$
	(\hat{q}_{\theta,\alpha},\hat{q}_{\boldsymbol{\mathcal{X}},\alpha})
	:= \arg\min_{q_{\theta} \in \Gamma_{\theta}, q_{\boldsymbol{\mathcal{X}}} \in \Gamma_{\boldsymbol{\mathcal{X}}}} \Psi_{n,\alpha}(q_{\theta},q_{\boldsymbol{\mathcal{X}}}).
	$ 
It is worth noting that all the presented theoretical results are on this global optimum $(\hat{q}_{\theta,\alpha},\hat{q}_{\boldsymbol{\mathcal{X}},\alpha})$, which may not be achieved in practice.
	
	We use the average $\alpha$-divergence $\frac{1}{n(n-1)T} D_{\alpha}^{(n)}(\theta,\theta^*):= \frac{1}{n(n-1)T} D_{\alpha}\left[ p_{\theta}^{(n)}||p_{\theta^*}^{(n)} \right]$ as the loss function,
	where $p_{\theta}^{(n)}$ denotes the distribution of $\boldsymbol{\mathcal{Y}}$ given the model parameter $\theta$. This loss function measures the discrepancy between the distribution of $\boldsymbol{\mathcal{Y}}$ with model parameter $\theta$ and the distribution of $\boldsymbol{\mathcal{Y}}$ with the true model parameter $\theta^*$.
	
	The following theorem gives a finite-sample upper bound of the variational Bayes risk for the case of $\alpha<1$.
	
	\begin{Theorem}\label{theorem1}
		With a certain choice of variational family $\Gamma_{\boldsymbol{\mathcal{X}}}$ and the variational distribution $q_{\theta}(\theta)$ restricted to a certain KL-neighborhood (which will be defined later), for any $\zeta\in(0,1)$, $D>1$, and $(\epsilon_{\beta},\epsilon_{\pi})\in (0,1)^2$, it holds with probability at least $\left(1-\frac{2}{(D-1)^2 n (\epsilon_{\beta}^2+\epsilon_{\pi}^2)}\right)$ that
		\begin{align}\label{thm1}
		\int \frac{D_{\alpha}^{(n)}(\theta,\theta^*)}{n(n-1)T}  \hat{q}_{\theta,\alpha}(d\theta)
		\leq & \frac{D\alpha}{1-\alpha}(\epsilon_{\pi}^2 + \epsilon_{\beta}^2)
		- \frac{\log P_{\pi}(\mathcal{B}_{n}(\pi^*,\epsilon_{\pi}))}{n(n-1)T(1-\alpha)}
		- \frac{\log P_{\beta}(\mathcal{B}_{n}(\beta^*,\epsilon_{\beta}))}{n(n-1)T(1-\alpha)},
		\end{align}
		where $P_{\pi}$ and $P_{\beta}$ are probability measures corresponding to prior densities $p_{\pi}$ and $p_{\beta}$, respectively.
		Here $\mathcal{B}_{n}(\pi^*,\epsilon_{\pi})$ and $\mathcal{B}_{n}(\beta^*,\epsilon_{\beta})$ are KL-neighborhoods for model parameters defined in the following way:
		$$
		\mathcal{B}_{n}(\pi^*,\epsilon_{\pi})
		:= \left\{ \pi:		D\left(p(\boldsymbol{\mathcal{X}}_{1}|\pi^*)||p(\boldsymbol{\mathcal{X}}_{1}|\pi)\right)\leq \epsilon_{\pi}^2, \quad
		V\left(p(\boldsymbol{\mathcal{X}}_{1}|\pi^*)||p(\boldsymbol{\mathcal{X}}_{1}|\pi)\right)\leq \epsilon_{\pi}^2
		\right\},
		$$
		\begin{align*}
		\mathcal{B}_{n}(\beta^*,\epsilon_{\beta})
		:= \biggr\{ \beta: &\sup_{\boldsymbol{X}_{11},\boldsymbol{X}_{21}}D(p(Y_{121}|\beta^*,\boldsymbol{X}_{11},\boldsymbol{X}_{21})||p(Y_{121}|\beta,\boldsymbol{X}_{11},\boldsymbol{X}_{21}))\leq \epsilon_{\beta}^2,\\ &\sup_{\boldsymbol{X}_{11},\boldsymbol{X}_{21}}V(p(Y_{121}|\beta^*,\boldsymbol{X}_{11},\boldsymbol{X}_{21})||p(Y_{121}|\beta,\boldsymbol{X}_{11},\boldsymbol{X}_{21}))\leq \epsilon_{\beta}^2 \biggr\},
		\end{align*}
where
		$\mathcal{X}_1 = \{\boldsymbol{X}_{1t}: 1 \leq t \leq T\}$, and
		$V(p||q) := \int p \log^2(\frac{p}{q}) d\mu$ denotes the discrepancy measure between two probability density functions $p$ and $q$.
	\end{Theorem}
	
	The left-hand side of inequality (\ref{thm1}) is the variational Bayes risk. The upper bound here is obtained based on a certain choice of variational family $\Gamma_{\boldsymbol{\mathcal{X}}}$ and a theorem in \cite{yang2017alpha}, which connects the variational Bayes risk to the $\alpha$-VB objective function (\ref{alphaVBobj}). The proofs of all theorems are in the supplementary material.
	
	The following theorem gives the convergence rate of the variational Bayes risk.
	\begin{Theorem}\label{theorem2}
		Assume that the prior densities $p_{\beta}$ and $p_{\pi}$ are thick and continuous at $\beta^*$ and $\pi^*$, respectively (here ``thick" means $p_{\beta}(\beta^*)>0$ and $p_{\pi}(\pi^*)>0$). Then there exists a constant $C>0$, such that as $n\rightarrow\infty$, it holds with probability tending to 1 that
		$$
		\int \frac{1}{n(n-1)T} D_{\alpha}^{(n)}(\theta,\theta^*) \hat{q}_{\theta,\alpha}(d\theta)
		\lesssim \frac{C}{n}.
		$$
	\end{Theorem}
	
	Theorem \ref{theorem2} implies that the point estimate of the model parameter based on optimizing the $\alpha$-VB objective function ($\alpha<1$) converges to the true parameter value as $n\rightarrow\infty$, i.e.,  $\alpha$-VB provides consistent parameter estimation.

For the usual VB ($\alpha=1$), stronger conditions are required to obtain the variational risk bound.	Note that for this part, we let the loss function be the squared Hellinger distance
	$
	h^2(p||q):=\int (\sqrt{p}-\sqrt{q})^2 d\mu.
	$
Also, we define
	$
	h^2(\theta||\theta^*)
	:= h^2 (p(\boldsymbol{\mathcal{Y}}|\theta)||p(\boldsymbol{\mathcal{Y}}|\theta^*))
	$
to be the squared Hellinger distance between the distributions of the observed networks generated by parameters $\theta$ and $\theta^*$. We restrict the model parameters in the compact set	$[-M,M] \times [m,M]^2$, where $M>m>0$ are some constants.	We first state the assumptions.
	
	\begin{Assumption}\label{thickness}
	The prior densities $p_{\beta}$ and $p_{\pi}$ satisfy $\inf_{\beta}p_{\beta}(\beta)>0$ and $\inf_{\pi}p_{\pi}(\pi)>0$.
	\end{Assumption}

	\begin{Assumption}\label{P}
	There exists a constant $C>0$ such that the following inequalities hold for any $(\beta,\pi)$ and $(\beta',\pi')$ in the parameter space:
	\begin{align*}
	&D\left(p(Y_{121}=\cdot|\beta,\boldsymbol{X}_{11},\boldsymbol{X}_{21})||p(Y_{121}=\cdot|\beta',\boldsymbol{X}_{11},\boldsymbol{X}_{21})\right)
	\leq C\lvert \beta-\beta' \rvert^2, \\
	&V\left(p(Y_{121}=\cdot|\beta,\boldsymbol{X}_{11},\boldsymbol{X}_{21})||p(Y_{121}=\cdot|\beta',\boldsymbol{X}_{11},\boldsymbol{X}_{21})\right)
	\leq C\lvert \beta-\beta' \rvert^2, \\
	&D\left(p(\boldsymbol{\mathcal{X}}_{1}=\cdot|\pi)||p(\boldsymbol{\mathcal{X}}_{1}=\cdot|\pi')\right)
	\leq C\lVert \pi-\pi' \rVert^2,
	\quad
	V\left(p(\boldsymbol{\mathcal{X}}_{1}=\cdot|\pi)||p(\boldsymbol{\mathcal{X}}_{1}=\cdot|\pi')\right)
	\leq C\lVert \pi-\pi' \rVert^2.
	\end{align*}
	\end{Assumption}

	Assumption \ref{thickness} is the prior thickness condition. Assumption \ref{P} is a regularity condition for justifying the prior concentration condition, which ensures that the prior mass of the KL neighborhood around the true parameter values is not too small. The priors used in our implementation can be shown to satisfy these assumptions by simple calculation.
	
	The following theorem gives a high probability variational Bayes risk bound as well as the asymptotic variational posterior concentration result for the usual VB.
	
	\begin{Theorem}\label{theorem3}
		Under Assumptions \ref{thickness} and \ref{P}, for any $D>1$, it holds with probability at least $\left(1-\frac{1}{(D-1)^2 n\epsilon_n^2}\right)$ that for any $\epsilon\in\left[\epsilon_n, e^{cn(n-1)\epsilon_{n}^2}\right]$,
		$$
		\hat{Q}_{\theta}\left(
		h^2(\theta||\theta^*) \leq \epsilon^2
		\right)
		\rightarrow 1 \text{ as }n\rightarrow\infty,
		$$
		where $\hat{Q}_{\theta}$ is the probability measure corresponding to the variational distribution $\hat{q}_{\theta}$ given by the VB procedure.
	\end{Theorem}
	
	With this result, we can obtain the convergence rate for a truncated version of variational Bayes risk. See the proof of Theorem \ref{theorem3} in the supplementary material for details.

	%%%%%%%%%%%%%%%%%%%%%%%%%%%%%%%%%%%%%%%%%%%%%%%%%%%%%%%%%%%%%%%%%%%%%%%%%%%
	\section{Simulation Results}\label{sim}
	Computation is performed on a Linux machine with 2.20 GHz processors. In all simulation and real data analysis, we set the dimension of the latent space $d=2$ for better visualization. More details of the implementation and additional simulation studies can be found in the supplementary material.

The performance was evaluated by the AUC (area under the ROC curve) values of in-sample predictions. To calculate this criterion, we plugged the estimated posterior means of the model parameters and latent positions into  (\ref{likelihood}) and calculated the estimated link probabilities. Then we compared the estimated link probabilities with the observed data $\mathcal{Y}$. A value of 1 implies perfect model fitting, while a value of 0.5 implies random predictions. Note that while the AUC criterion does not directly measure the discrepancy between the approximate posterior and the true posterior, it measures the goodness-of-fit of the model to the observed data.

	\setcounter{equation}{0} %-1

	We carried out simulation studies for networks with $n=100$ and $n=1000$ nodes under various settings (see the supplementary material for the simulation with $n=5000$ nodes). We simulated 20 dynamic networks with the number of time steps $T=10$ for each case.	The average AUC values are reported in Figure \ref{sim100} for $n=100$ and Figure \ref{sim1000} for $n=1000$. The variational method performed well in all cases. While the variance of the transition distribution does not seem to affect the performance very much, the variational method performed better on dense networks than sparse networks.

	\begin{figure}[h]
%\vspace{-0.3cm}
		\centering
		\begin{minipage}{\linewidth}
			\begin{minipage}{0.5\textwidth}
			%\vspace{-0.5cm}	\includegraphics[width=0.8\linewidth]{aucplot100_small.pdf}
			\end{minipage}
			\begin{minipage}{0.5\textwidth}
			%\vspace{-0.5cm}	\includegraphics[width=0.8\linewidth]{aucplot100_large.pdf}
			\end{minipage}
		\end{minipage}
%\vspace{-1cm}
		\caption{The average AUC values for VB on simulated networks with $n=100$ nodes and (left) small transition and (right) large transition.}\label{sim100}
	\end{figure}
	
	\begin{figure}[h]
		\centering
		\begin{minipage}{\linewidth}
			\begin{minipage}{0.5\textwidth}
			%\vspace{-.8cm}	\includegraphics[width=0.8\linewidth]{aucplot1000_small.pdf}
			\end{minipage}
			\begin{minipage}{0.5\textwidth}
			%\vspace{-1cm}	\includegraphics[width=0.8\linewidth]{aucplot1000_large.pdf}
			\end{minipage}
		\end{minipage}
		\caption{The average AUC values for VB on simulated networks with $n=1000$ nodes and (left) small transition and (right) large transition.}\label{sim1000}
	\end{figure}

\begin{figure}[h]
		\centering
		%\vspace{-1.1cm}
		\includegraphics[width=.45\textwidth]{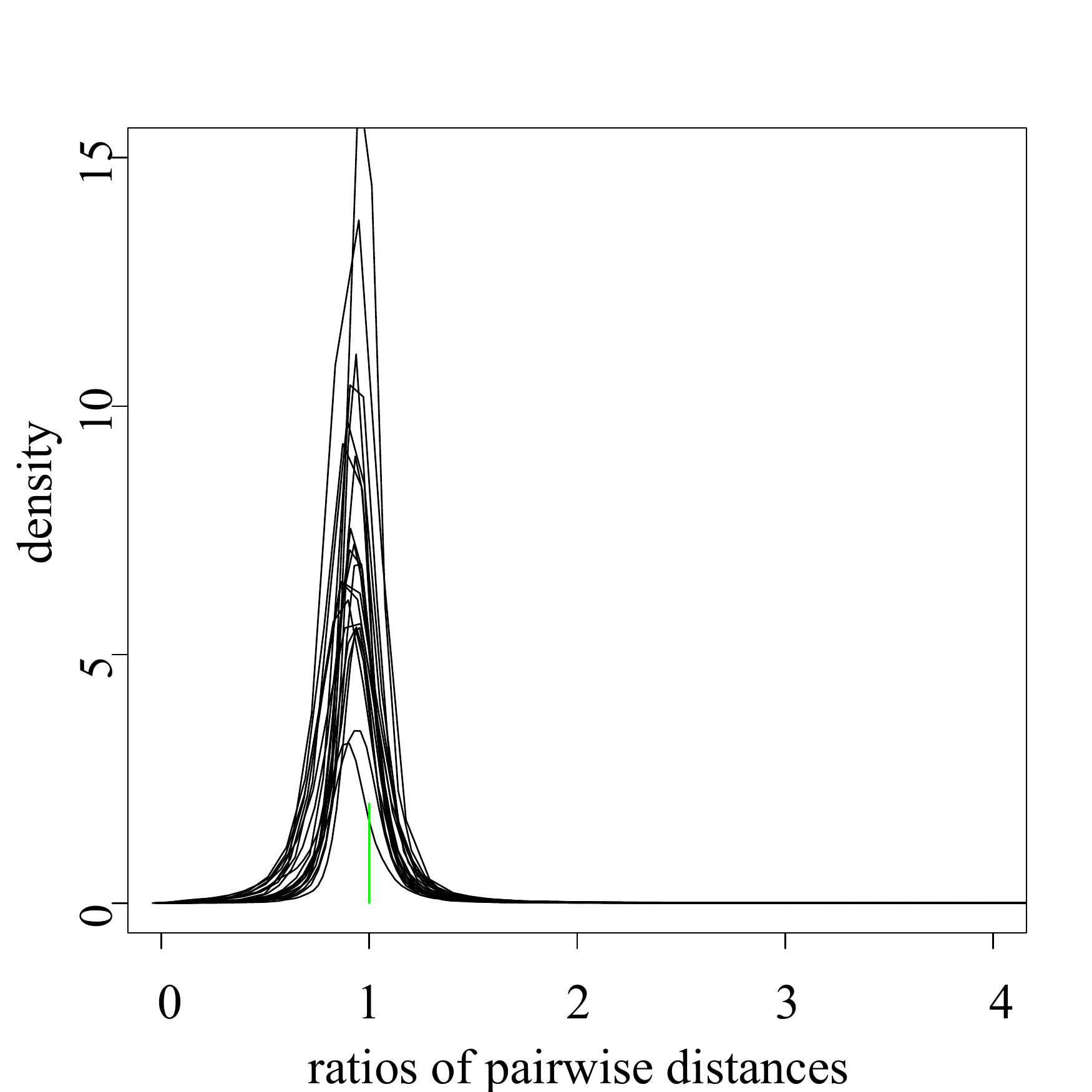}
		%\vspace{-1cm}
		\caption{Density plots of ratios of pairwise distances, comparing the estimated latent positions with true latent positions.}
		\label{PairwiseDist}
	\end{figure}

	We also plotted in Figure \ref{PairwiseDist} the distribution of pairwise distance ratios for each simulated network in the dense, small transition case with $n=100$.
	That is, we calculated the ratio $\frac{||\hat{\boldsymbol{\mu}}_{it}-\hat{\boldsymbol{\mu}}_{jt}||}{||\boldsymbol{X}_{it}-\boldsymbol{X}_{jt}||}$ for each $(i,j,t)$ of each simulated network, and plotted the density curve of these ratios. Most of these distributions are narrow and centered around 1, which indicates that the estimated latent positions	are close to the truth.
	
	%%%%%%%%%%%%%%%%%%%%%%%%%%%%%%%%%%%%%%%%%%%%%%%%%%%%%%%%%%%%%%%%%%%%%
	\section{Real Data Analysis}\label{real_data}
	\subsection{Dynamic Networks from ``Teenage Friends and Lifestyle Study"}
	\label{teenage}
	
	We analyzed a sequence of directed networks of friendship relations from the ``Teenage Friends and Lifestyle Study" dataset \citep{michell1997girls,michell1996peer,pearson2006homophily}. In this longitudinal study, a total of 160 pupils were studied over a three-year period since January 1995. At the measurement time point in each year, the pupils were asked to name up to twelve best friends, yielding three adjacency matrices. The $(i,j)$-th entry of the $t$-th adjacency matrix is 1 if pupil $i$ named pupil $j$ as one of his or her best friends at the $t$-th measurement time point, and 0 otherwise. The study also collected information on substance use and adolescent behavior, such as music preference, tobacco, alcohol and cannabis consumption.
	In the following study, we focused on the networks formed by the 129 pupils who were present at all three measurement time points (see the networks in the supplementary material). The average edge density of this network is 0.0274.

	We fitted the dynamic latent space model to the sequence of networks. To implement the proposed VB algorithm, we set a normal prior $\mathcal{N}(0,2)$ for the intercept $\beta$ and the hyperparameters $\sigma^2=0.5$ and $\tau^2=0.1$. The variational parameters $\left\{\tilde{\boldsymbol{\mu}}_{it}\right\}$'s were randomly initialized, and the initial values of other variational parameters were set to be $\tilde{\Sigma}_0=\mathbb{I}_{2}$, $\tilde{\xi}_0=0$, $\tilde{\psi}^2_{0}=2$. The algorithm converged in less than five seconds. The approximate posterior distribution of the intercept $\beta$ is $\mathcal{N}(-1.5092,0.0001)$. The AUC values of in-sample predictions for the three time points are 0.9364, 0.9497 and 0.9681, respectively.

\begin{figure}[h]
%\vspace{-0.4cm}
		\centering
		\includegraphics[width=0.8\textwidth]{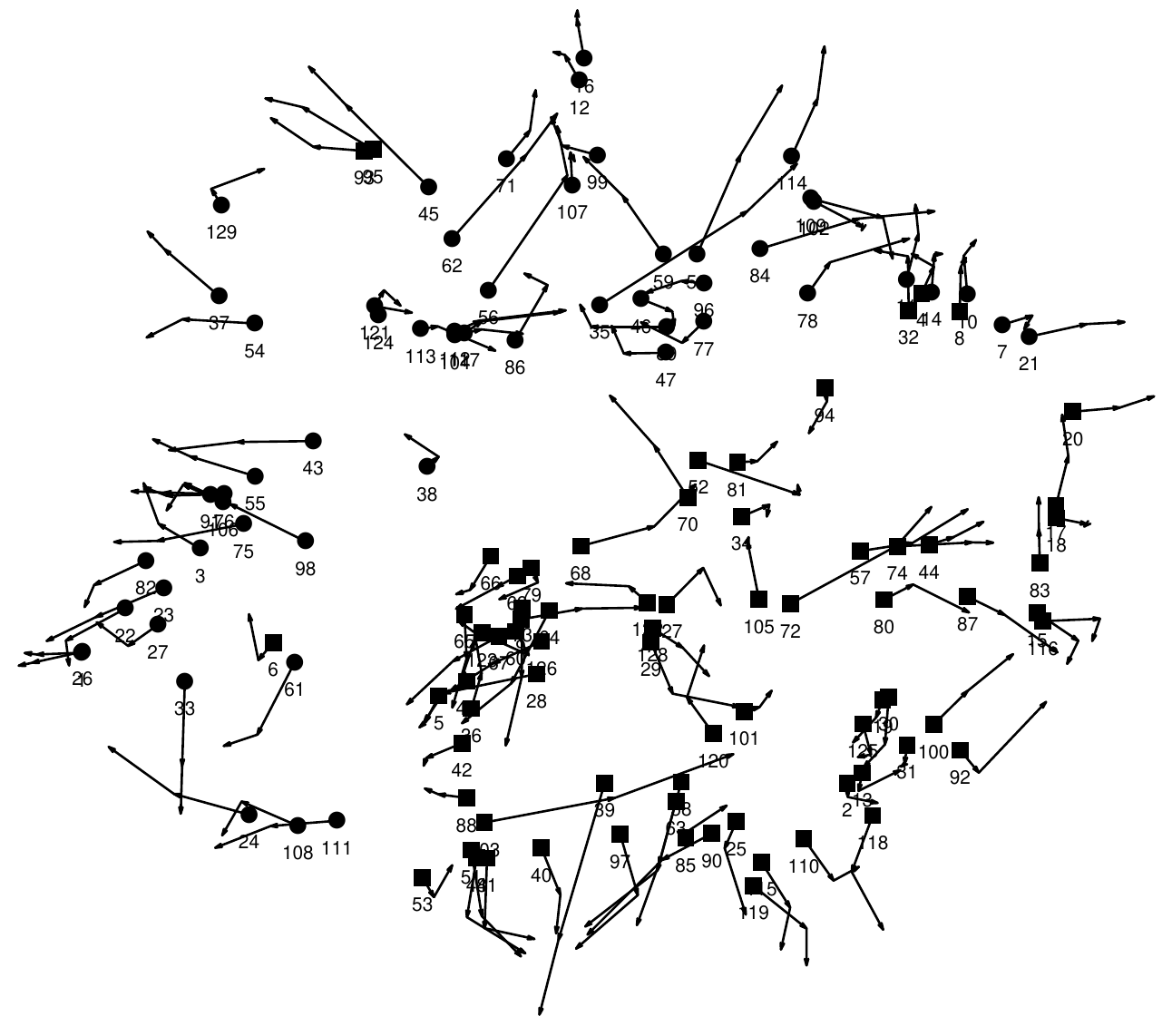}
		%\vspace{-1.1cm}
		\caption{Estimated latent positions of each node at the first time point. The arrows denote where the actors are moving to in the next two time points. Filled squares denote boys, and filled circles denote girls.}\label{traj129}
	\end{figure}

	Figure \ref{traj129} shows the estimated latent positions at the first time point based on the approximate posterior, as well as where the actors are moving to in the next two time points (indicated by the arrows). A longer arrow indicates a larger move in the latent space. Filled squares denote boys, and filled circles denote girls. We can see that girls lie on the top-left side of the latent space, while boys on the bottom-right side. Only a few actors (actors 6, 8, 32, 93, 95) are close to the opposite gender in the social space. This coincides with the claim in \cite{pearson2006homophily} that there is strong gender homophily in friendship selection. From the trajectories of the nodes, we can also see the formation of groups over time. For example, actors 15, 80, 87, 92, 100, 116 were forming a new group, while actors 35 and 50 were leaving their original social group.
	
\begin{figure}[h]
		\centering
		%\vspace{-0.9cm}
		\includegraphics[width=0.4\textwidth]{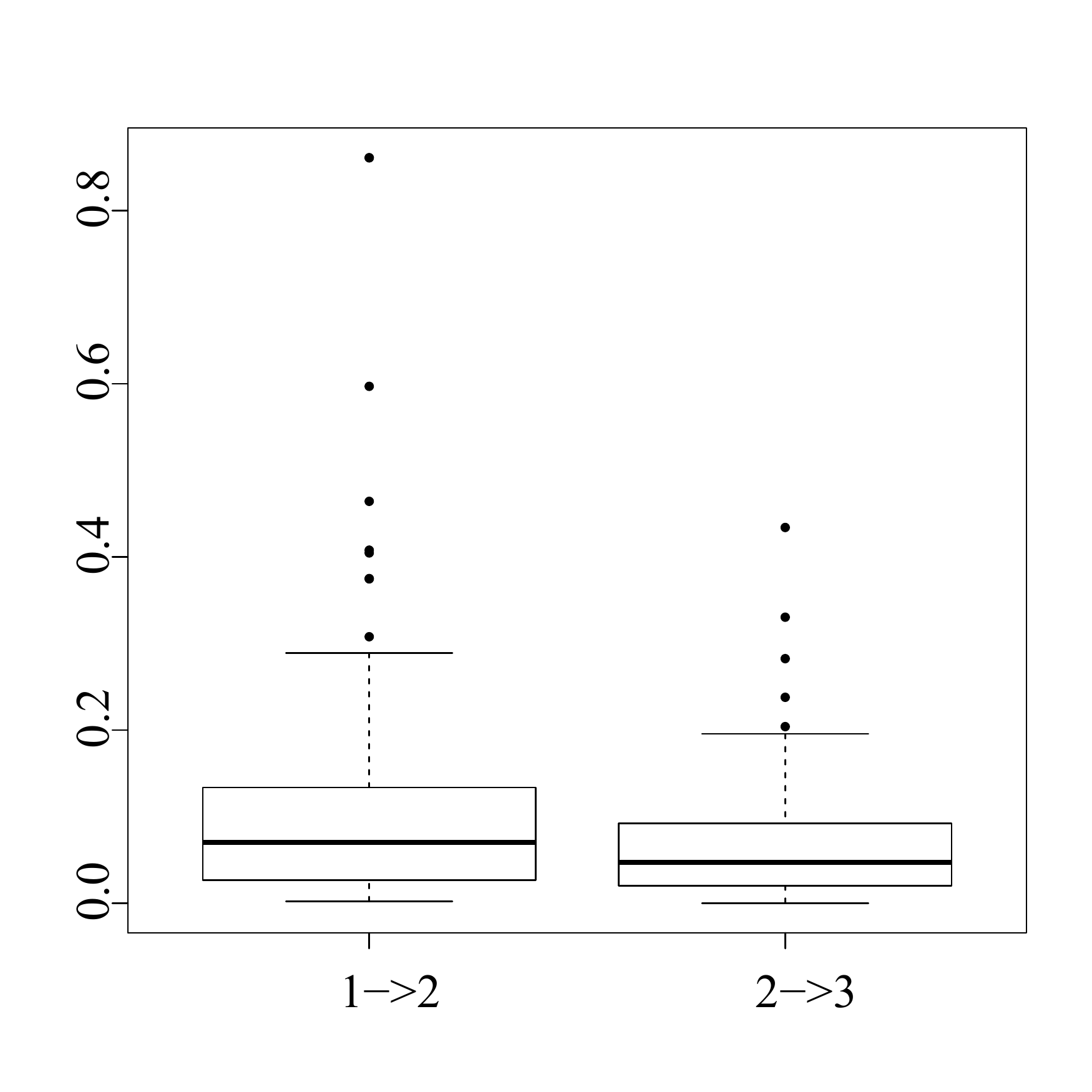}
		%\vspace{-1.4cm}
		\caption{Boxplots for the squared distances of movements for each node during two transitions.}\label{boxplot129}
	\end{figure}

	To study the dynamics of the network, we calculated the squared distances of movements for each node during the two transitions and created boxplots for them (Figure \ref{boxplot129}). From these plots, we can see that the median and variation of the moving distances of the first transition are larger than those of the second transition,	which indicates that the friendship network changed more during the first transition. \cite{pearson2006homophily} analyzed the same dataset and also claimed that the rate of network change is larger in the first transition than in the second transition.
	
\begin{figure}[h]
%\vspace{-0.6cm}
		\parbox[t]{3.17in}{
			\centerline{
				{
					\includegraphics[width=0.49\textwidth]{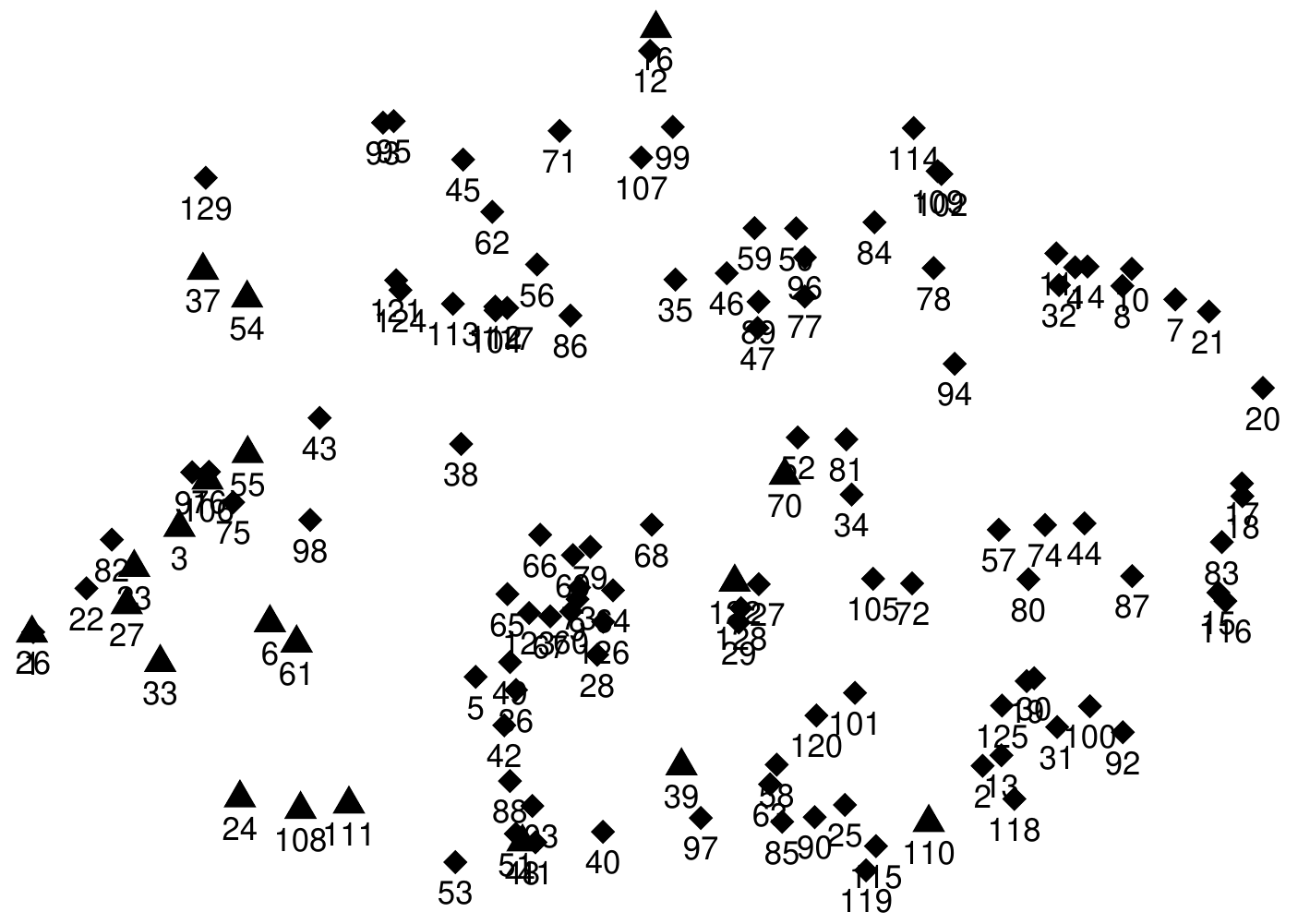}
				}
			}
		}
		\parbox[t]{3.17in}{
			\centerline{
				{
					\includegraphics[width=0.49\textwidth]{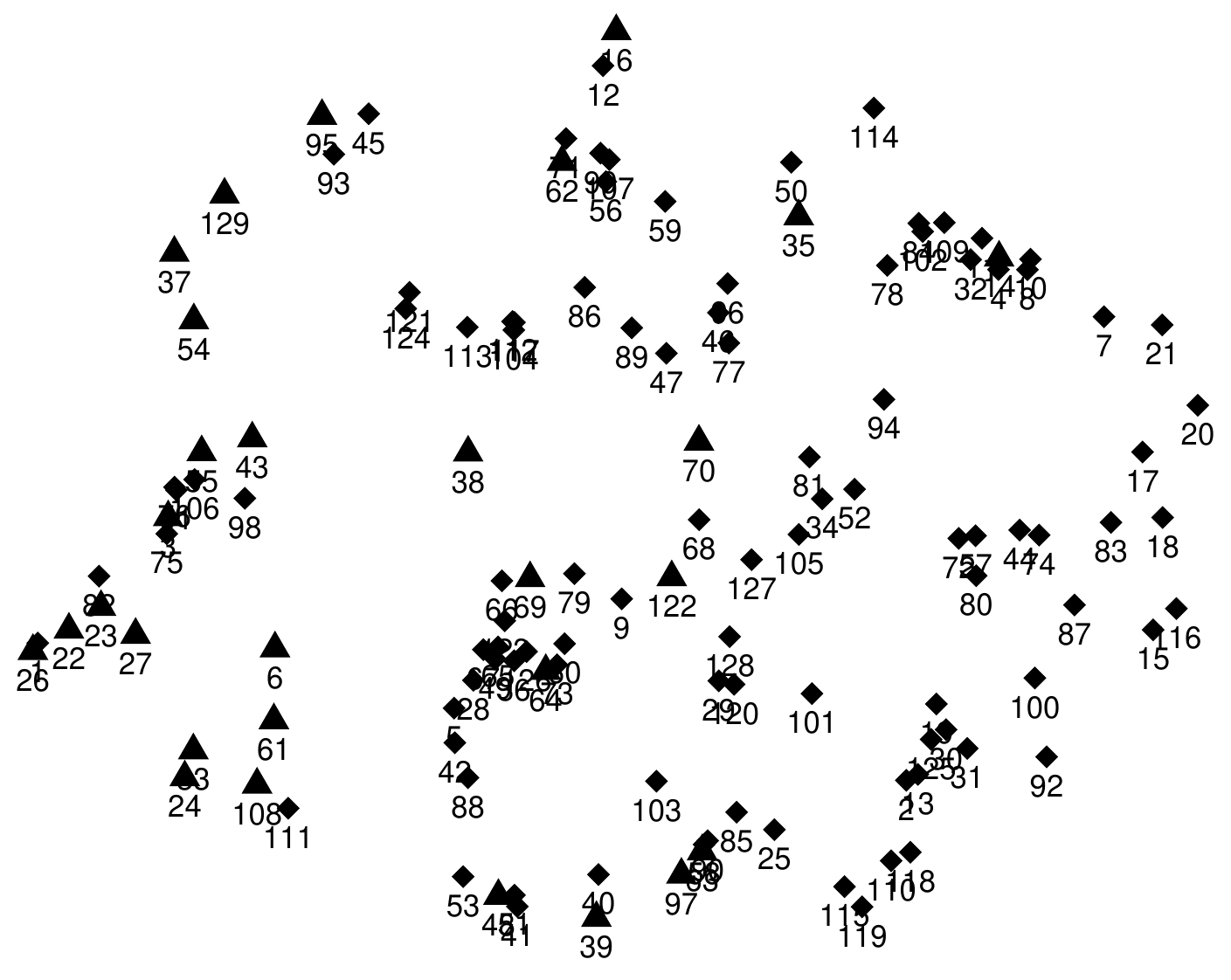}
				}
			}
		}
		\centerline{
			{
				\includegraphics[width=0.49\textwidth]{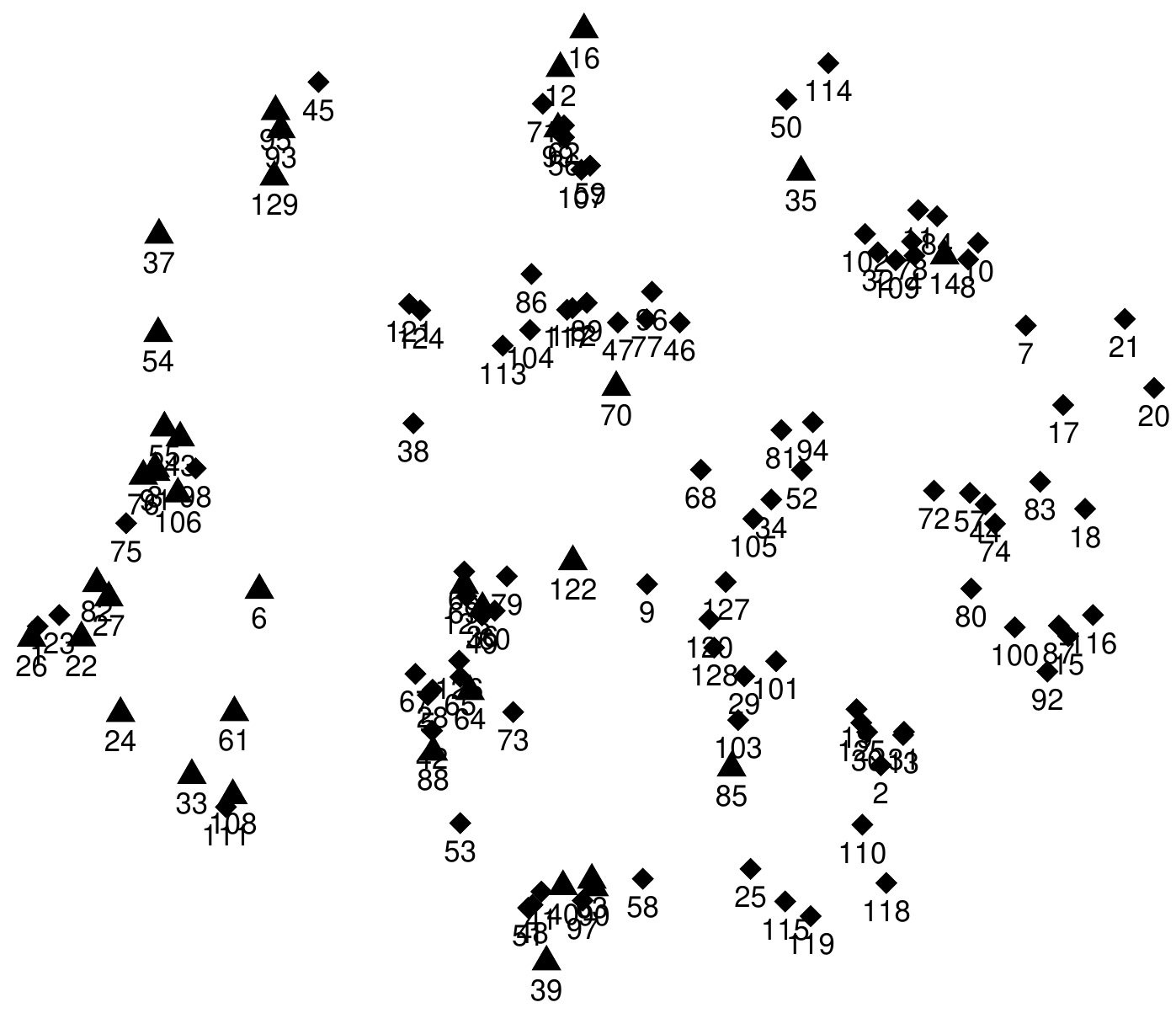}
			}
		}
		%\vspace{-2.2cm}
		\caption{The latent positions as well as tobacco consumption of the 129 pupils at times 1, 2 (top) and 3 (bottom). The filled triangles represent actors who used tobacco occasionally or regularly. The filled diamonds represent actors who never used tobacco or only tried once.}
		\label{tobacco129}
	\end{figure}

\begin{figure}[h!]
%\vspace{-0.5cm}
		\parbox[t]{3.17in}{
			\centerline{
				{
					\includegraphics[width=0.49\textwidth]{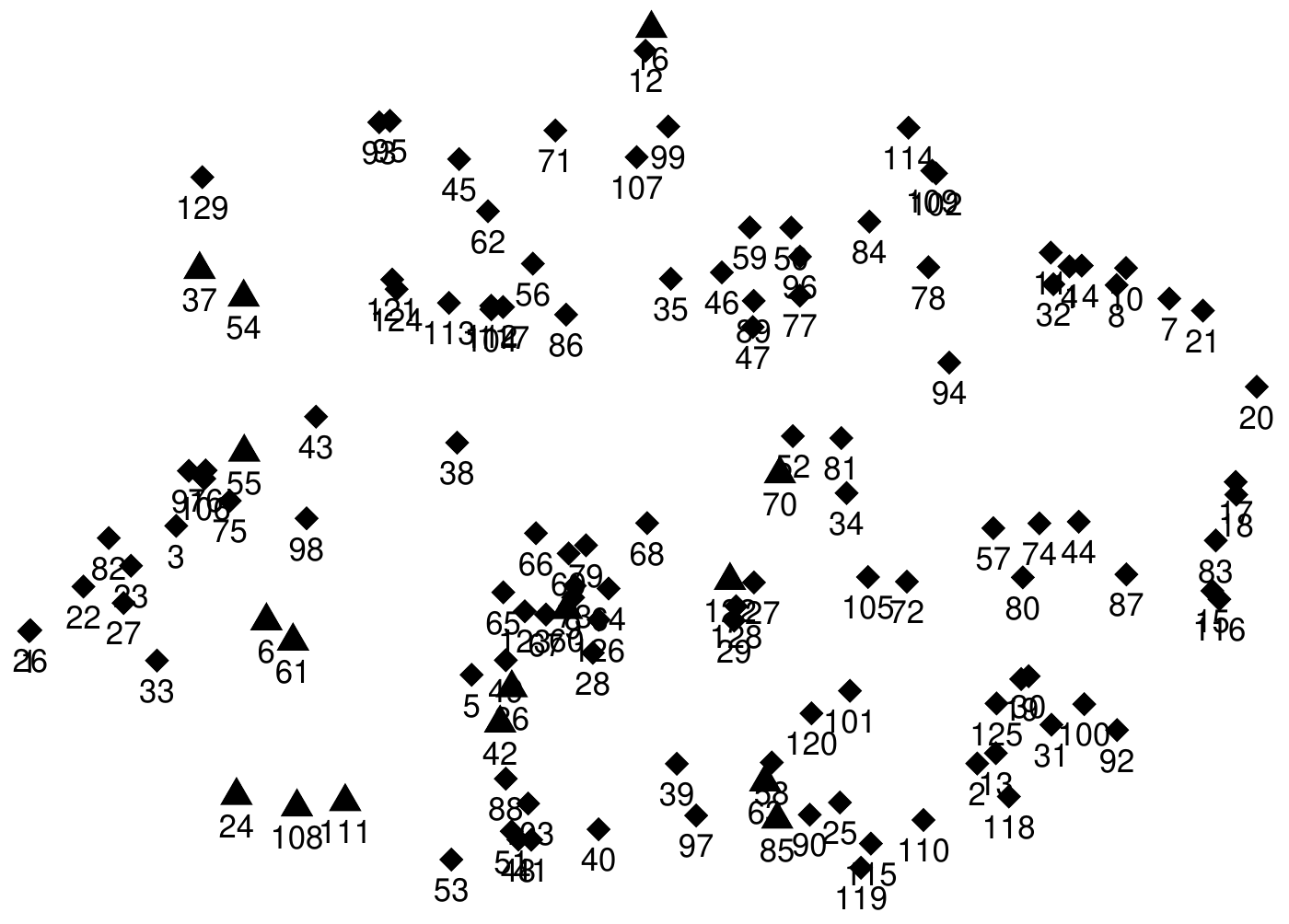}
				}
			}
		}
		\parbox[t]{3.17in}{
			\centerline{
				{
					\includegraphics[width=0.49\textwidth]{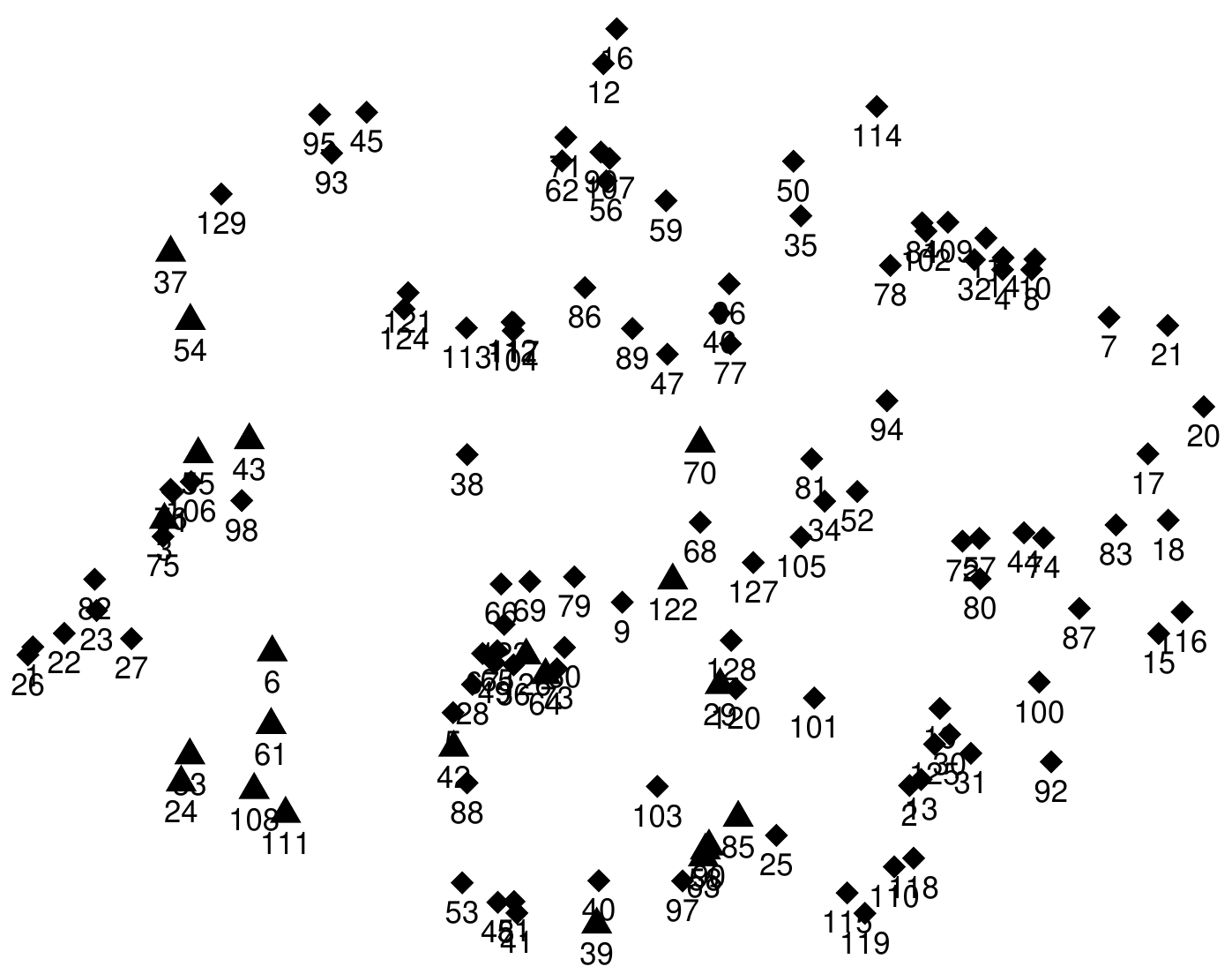}
				}
			}
		}
		\centerline{
			{
				\includegraphics[width=0.49\textwidth]{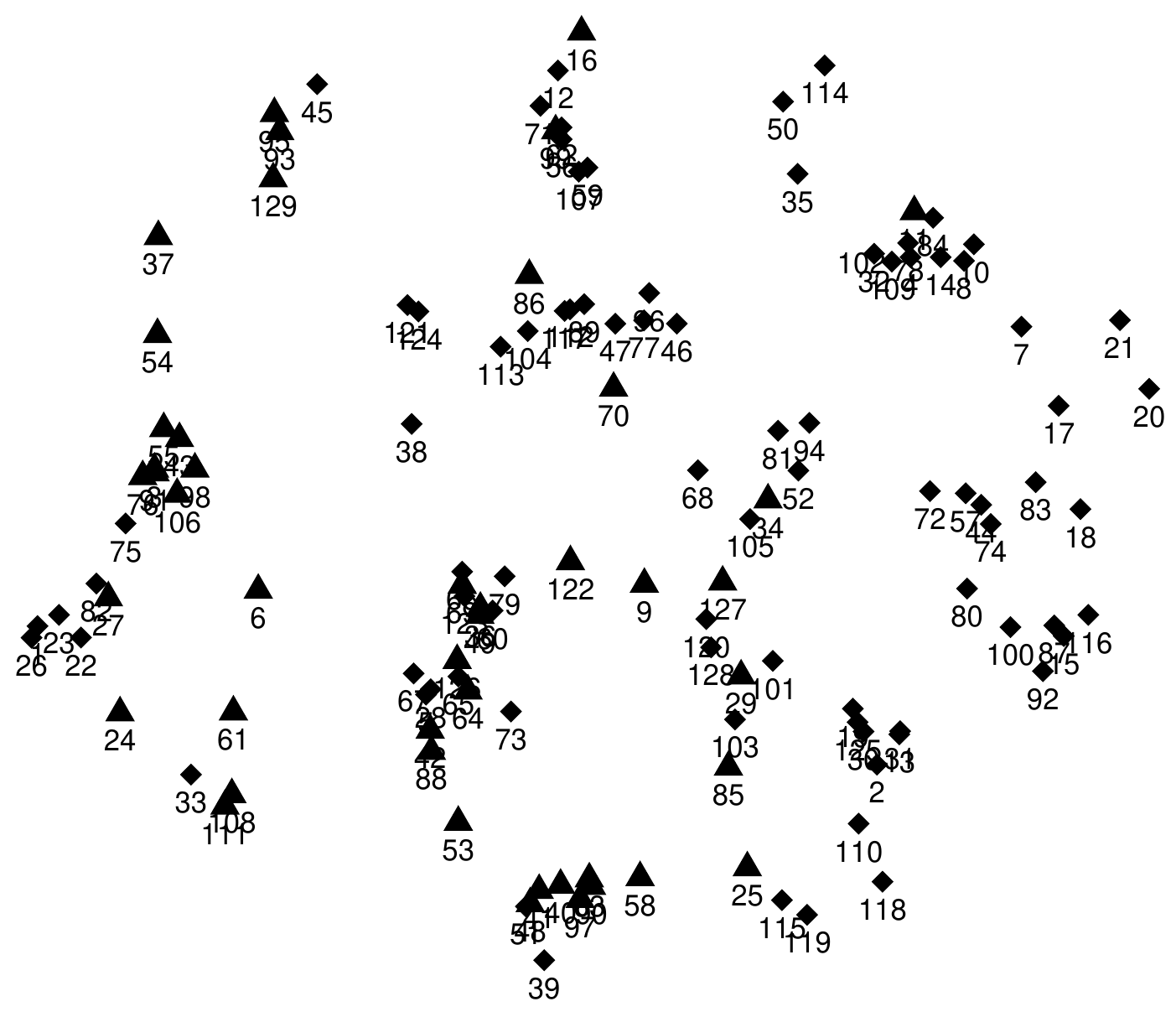}
			}
		}
		%\vspace{-2.2cm}
		\caption{The latent positions as well as cannabis consumption of the 129 pupils at times 1, 2 (top) and 3 (bottom). The filled triangles represent actors who used cannabis occasionally or regularly. The filled diamonds represent actors who never used cannabis or only tried once.}
		\label{cannabis129}
	\end{figure}

	We also examined tobacco and cannabis consumption of these 129 pupils. Figures \ref{tobacco129} and \ref{cannabis129} give the latent positions of the pupils, as well as their tobacco and cannabis consumption status at each time point. We notice that pupils with similar substance consumption behavior tended to move closer to each other. For example, actors 24, 61, 108 and 111 occasionally or regularly used tobacco and cannabis at the first time point, and they moved closer to each other in the latent space during the two transitions. Another interesting observation is that as actor 63, who occasionally or regularly used cannabis at the first time point, moved into the nearby social group, the whole group of pupils (actors 40, 41, 48, 58, 90, 97) became occasional or regular cannabis user at time point 3. This observation corroborates with the conclusion in \cite{pearson2006homophily} that there is a significant positive influence effect of friends on cannabis use.

	\subsection{The Wiki-Talk Temporal Network}\label{wikitalk}
	
	In this section, we analyze the Wiki-talk temporal network \citep{leskovec2010governance, paranjape2017motifs}. Wiki-talk pages are part of the Wikipedia administration system, where users are able to communicate with each other on possible improvements to Wikipedia pages. Each user page has an associated talk page. Wikipedia users can edit other user's talk page by leaving a message.
	
	The temporal network we analyzed represents the editing activities of Wikipedia users on each other's Wiki-talk pages. Nodes represent Wikipedia users, and an edge from node $i$ to node $j$ at time point $t$ means that user $i$ edited user $j$'s Wiki-talk page at time $t$. The original dataset contains 1,140,149 nodes and a time span of 2277 days. We took a subset of 5294 nodes and aggregated the temporal edges between these nodes in each year of the last three years to form three networks. All these 5294 nodes are present at all three time points. The edge density of the observed network at the three time points is 0.0022, 0.0030 and 0.0012, respectively.
	
	We implemented the proposed VB algorithm, with a normal prior $\mathcal{N}(0,0.01)$ on the intercept $\beta$ and the hyperparameters $\sigma^2=5$ and $\tau^2=0.01$. The variational parameters $\tilde{\boldsymbol{\mu}}_{it}$'s were initialized randomly, and the initial values of other variational parameters were set to be $\tilde{\Sigma}_0=\mathbb{I}_{2}$, $\tilde{\xi}_0=0$, $\tilde{\psi}^2_{0}=0.01$. The computation took about 20 minutes. The variational posterior of the intercept $\beta$ is $\mathcal{N}(-3.3054,\  5.5278\times 10^{-6})$. The AUC values of the in-sample predictions at each time step are 0.7260, 0.7486 and 0.6955, respectively.
	
\begin{figure}[h]
%\vspace{-0.1cm}
		\centering
		\begin{minipage}{\linewidth}
			\begin{minipage}{0.32\textwidth}
			%\vspace{-0.5cm}	
			\includegraphics[width=1.0\linewidth]{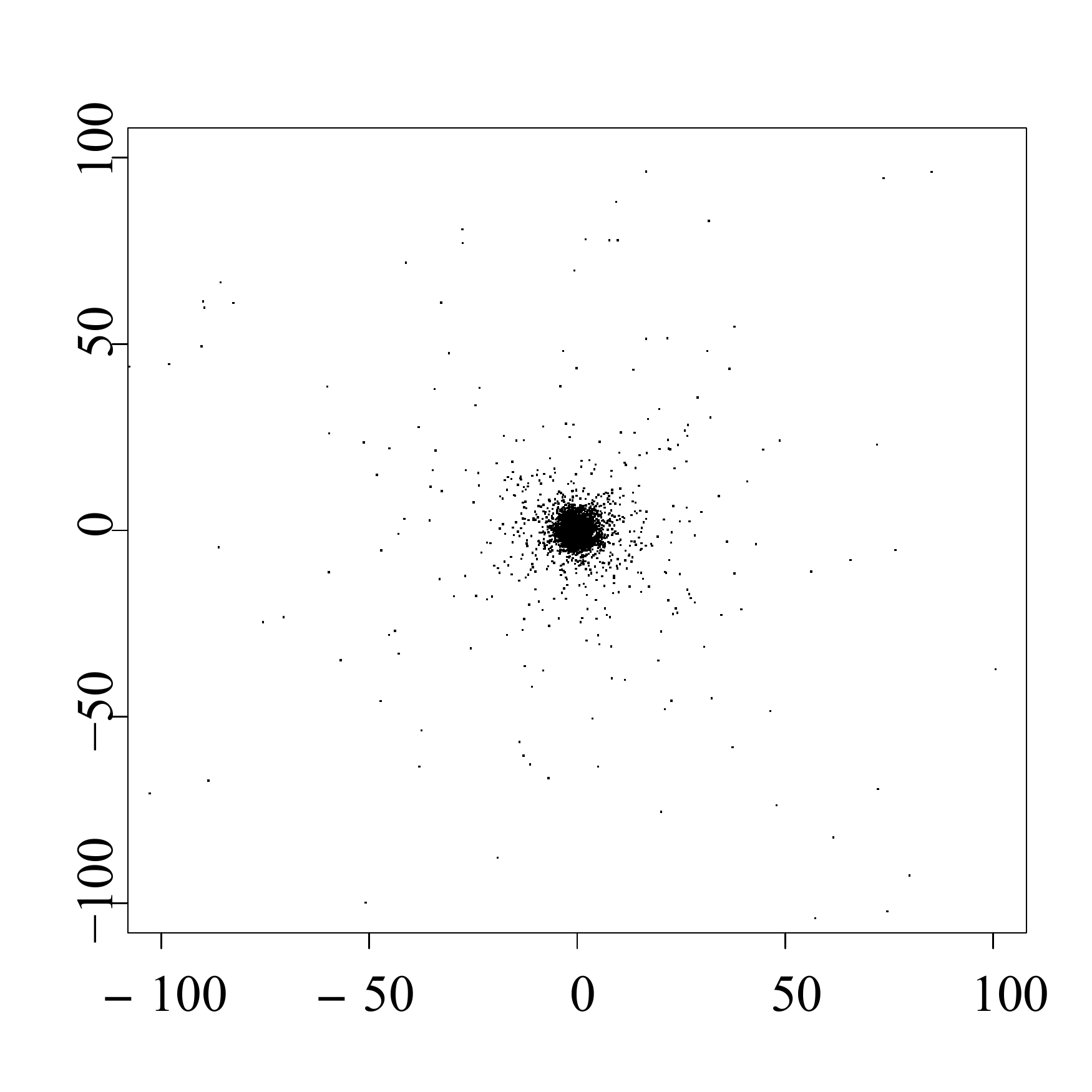}
			\end{minipage}
			\begin{minipage}{0.32\textwidth}
			%\vspace{-0.5cm}	
			\includegraphics[width=1.0\linewidth]{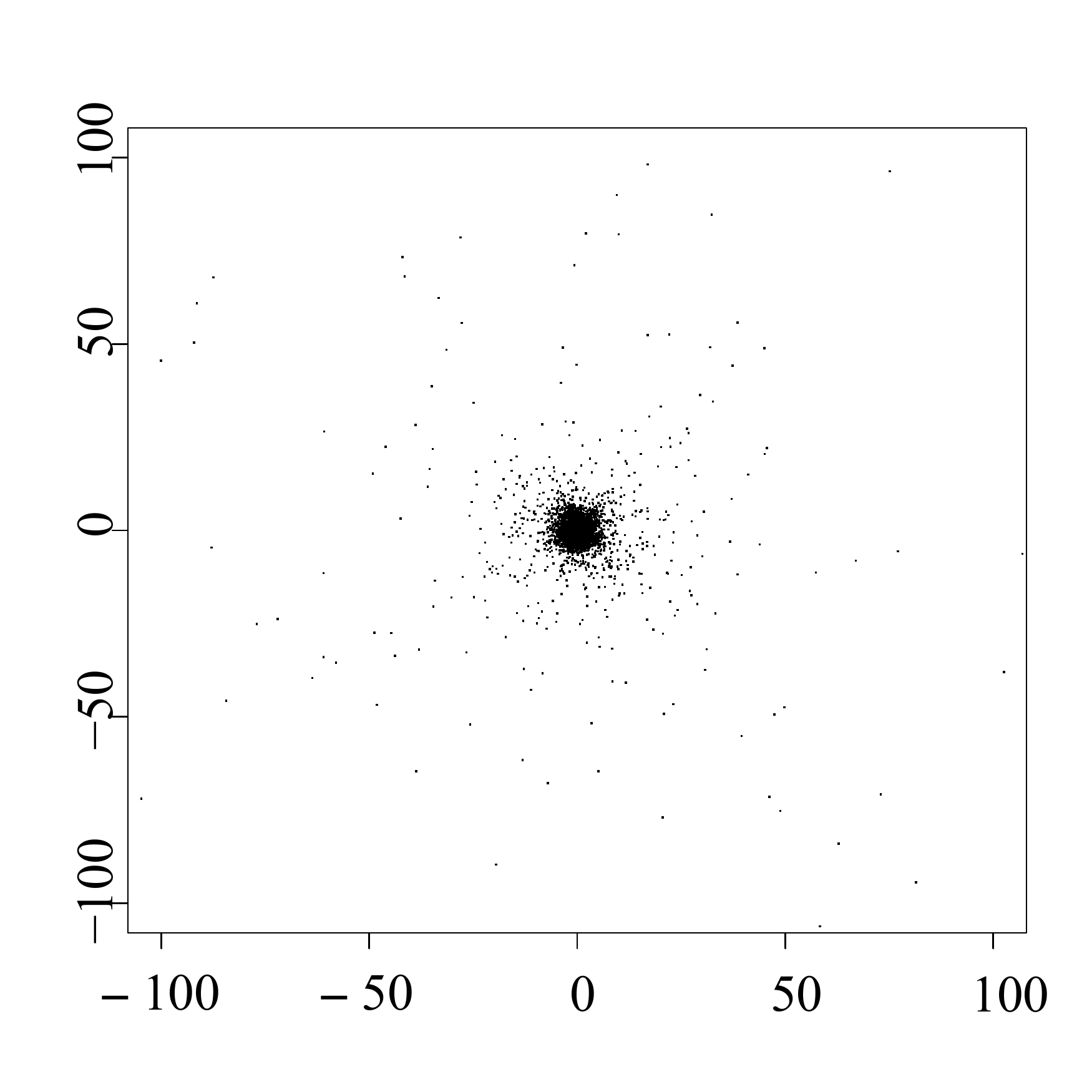}
			\end{minipage}
			%\begin{center}
			\begin{minipage}{0.32\textwidth}
			%\vspace{-0.5cm}	
			\includegraphics[width=1.0\linewidth]{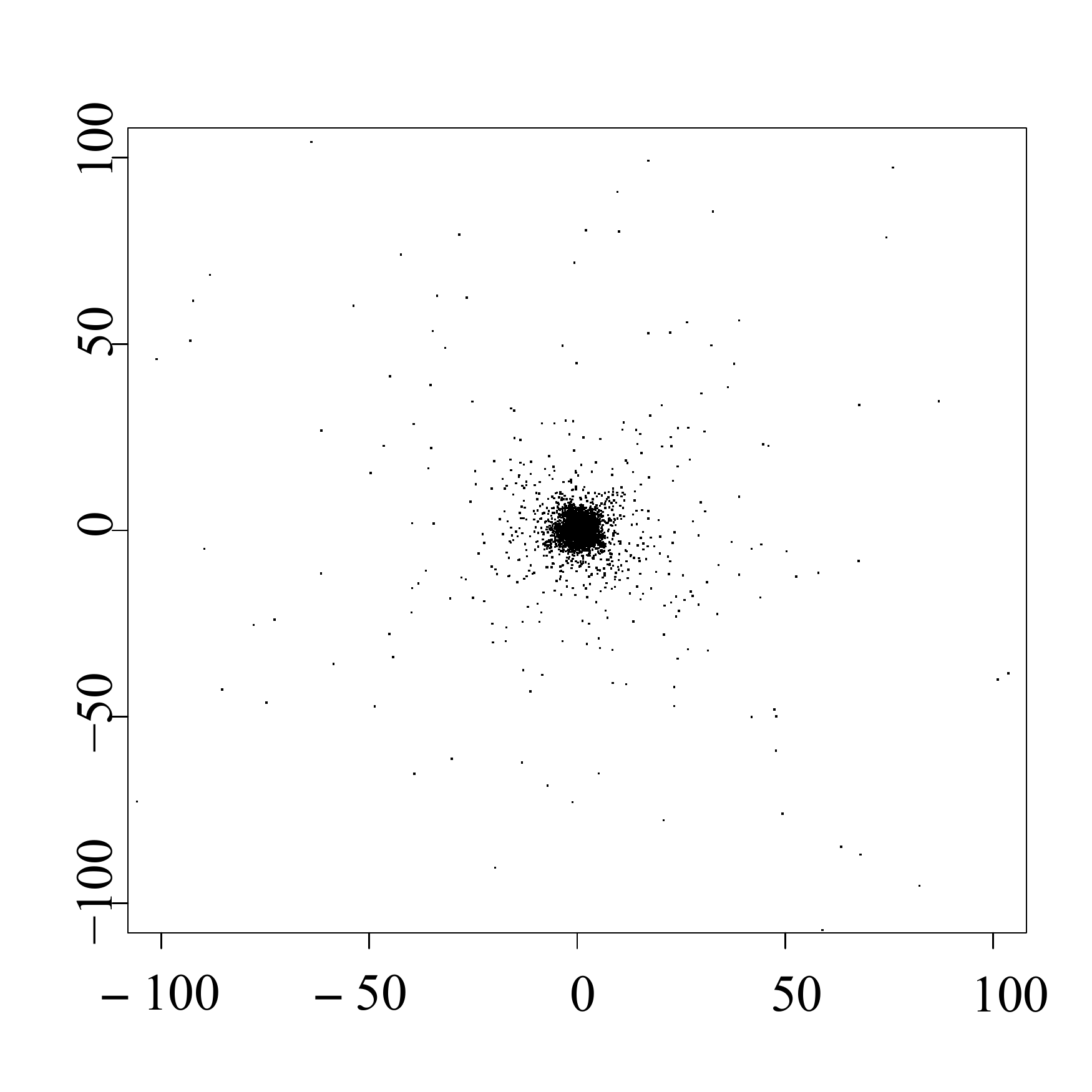}
			\end{minipage}
			%\end{center}
		\end{minipage}
%\vspace{-1.2cm}
		\caption{The estimated latent positions for the Wiki-talk dataset
			at times 1, 2 and 3.}
		\label{wikitalk_pos}
	\end{figure}

\begin{figure}[h]
%\vspace{-0.1cm}
		\centering
		\begin{minipage}{\linewidth}
			\begin{minipage}{0.5\textwidth}
			%\vspace{-1cm}	
			\includegraphics[width=0.7\linewidth]{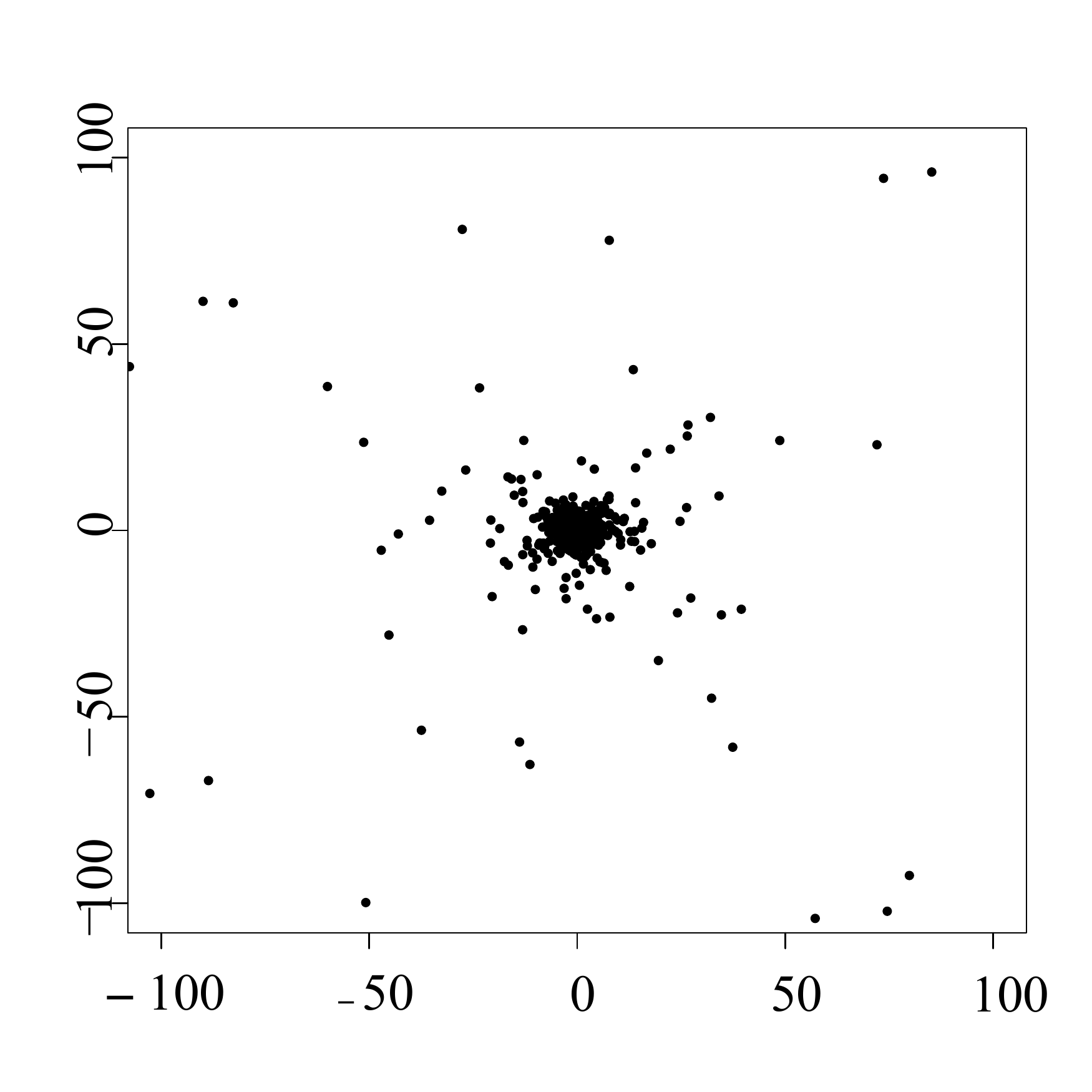}
			\end{minipage}
			\begin{minipage}{0.5\textwidth}
			%\vspace{-1cm}	
			\includegraphics[width=0.7\linewidth]{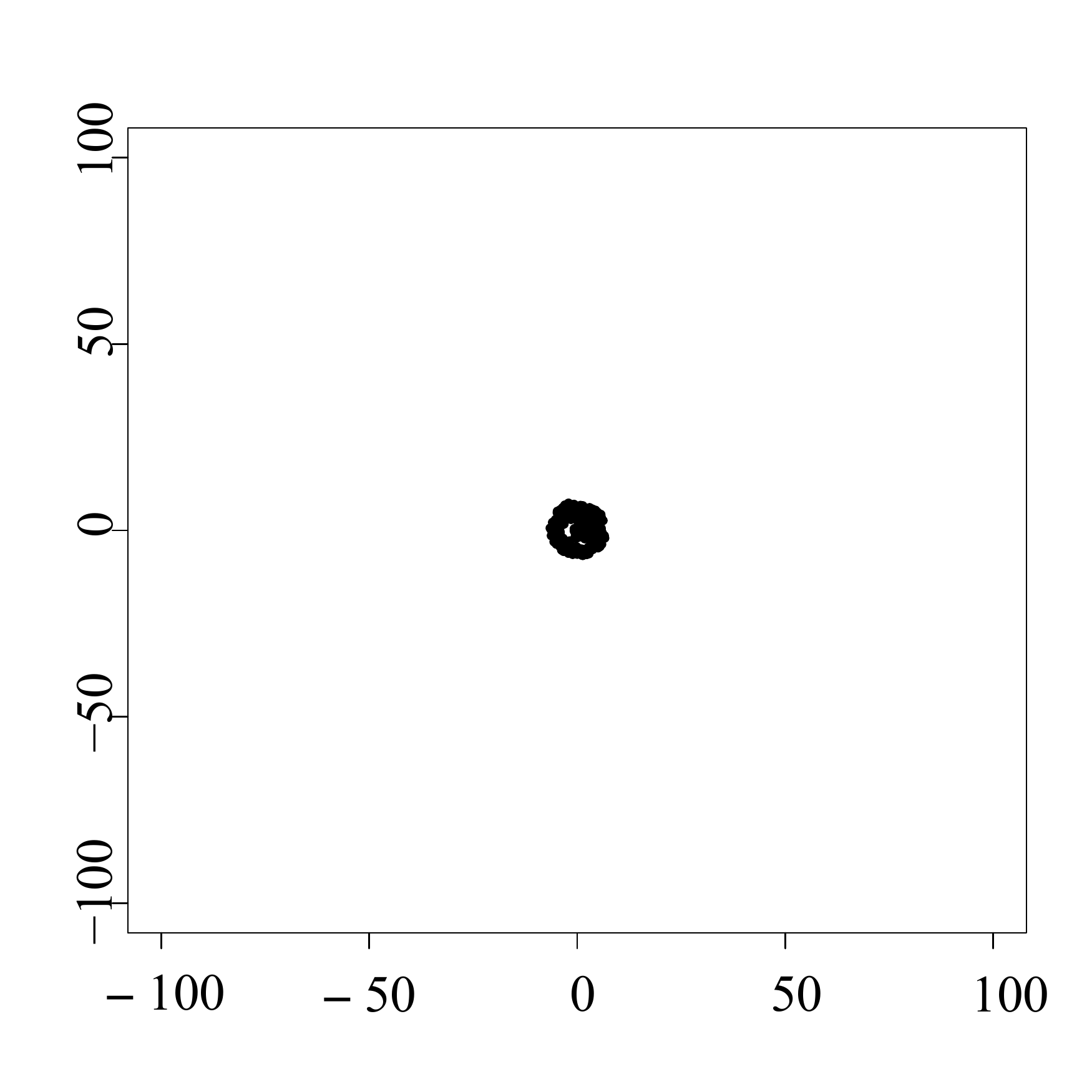}	
			\end{minipage}
%\vspace{-1.2cm}
			\caption{The initial latent positions of the nodes whose squared moving distances are greater than 5 (left) or smaller than 0.01 (right) during the first transition.
				\label{movements1}}
		\end{minipage}
	\end{figure}

\begin{figure}[h]
%\vspace{-0.3cm}
		\centering
		\begin{minipage}{\linewidth}
			\begin{minipage}{0.5\textwidth}
			%\vspace{-0.5cm}	
			\includegraphics[width=0.7\linewidth]{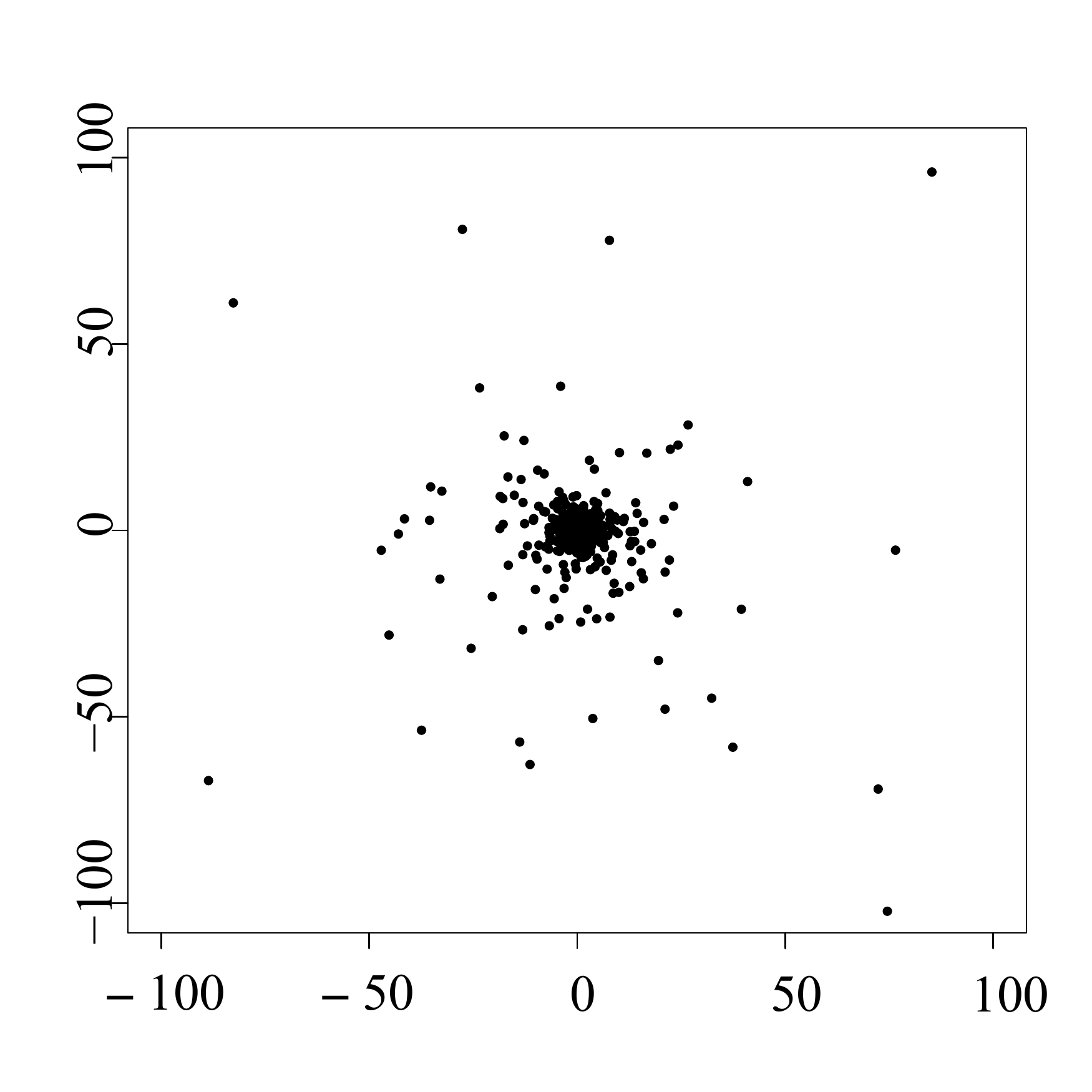}
			\end{minipage}
			\begin{minipage}{0.5\textwidth}
			%\vspace{-0.5cm}	
			\includegraphics[width=0.7\linewidth]{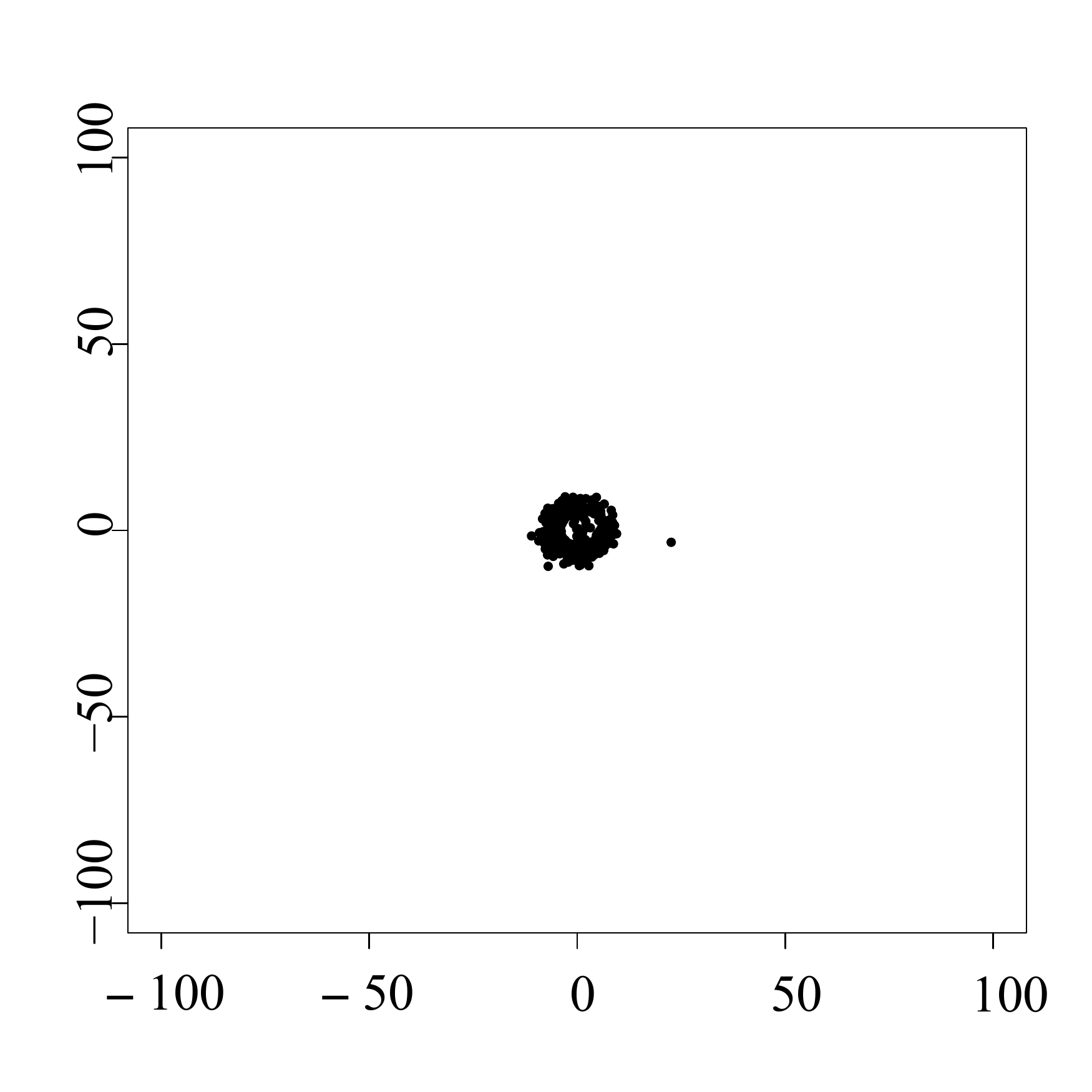}	
			\end{minipage}
			%\vspace{-1.2cm}
			\caption{The latent positions at the second time point of the nodes whose squared moving distances are greater than 5 (left) or smaller than 0.01 (right) during the second transition.
				\label{movements2}}
		\end{minipage}
	\end{figure}

	Figure \ref{wikitalk_pos} shows the estimated latent positions at each time step. Most nodes are concentrated around the center. With such a large number of nodes in the area close to the center, it is not easy to plot the movements of the nodes or their trajectories. Instead, we plotted the latent positions of the nodes whose squared moving distances are greater than 5 or smaller than 0.01 during the two transitions (Figures \ref{movements1} and \ref{movements2}). For both transitions, actors with large moving distances are more spread out in the latent space, while actors with small moving distances are all concentrated around the center of the latent space. They are very likely to be the most active users who have made great contribution to the website, or the administrative users who have been granted certain privilege.
	
\begin{figure}[h]
		\centering
		%\vspace{-0.7cm}
		\includegraphics[width=.4\linewidth]{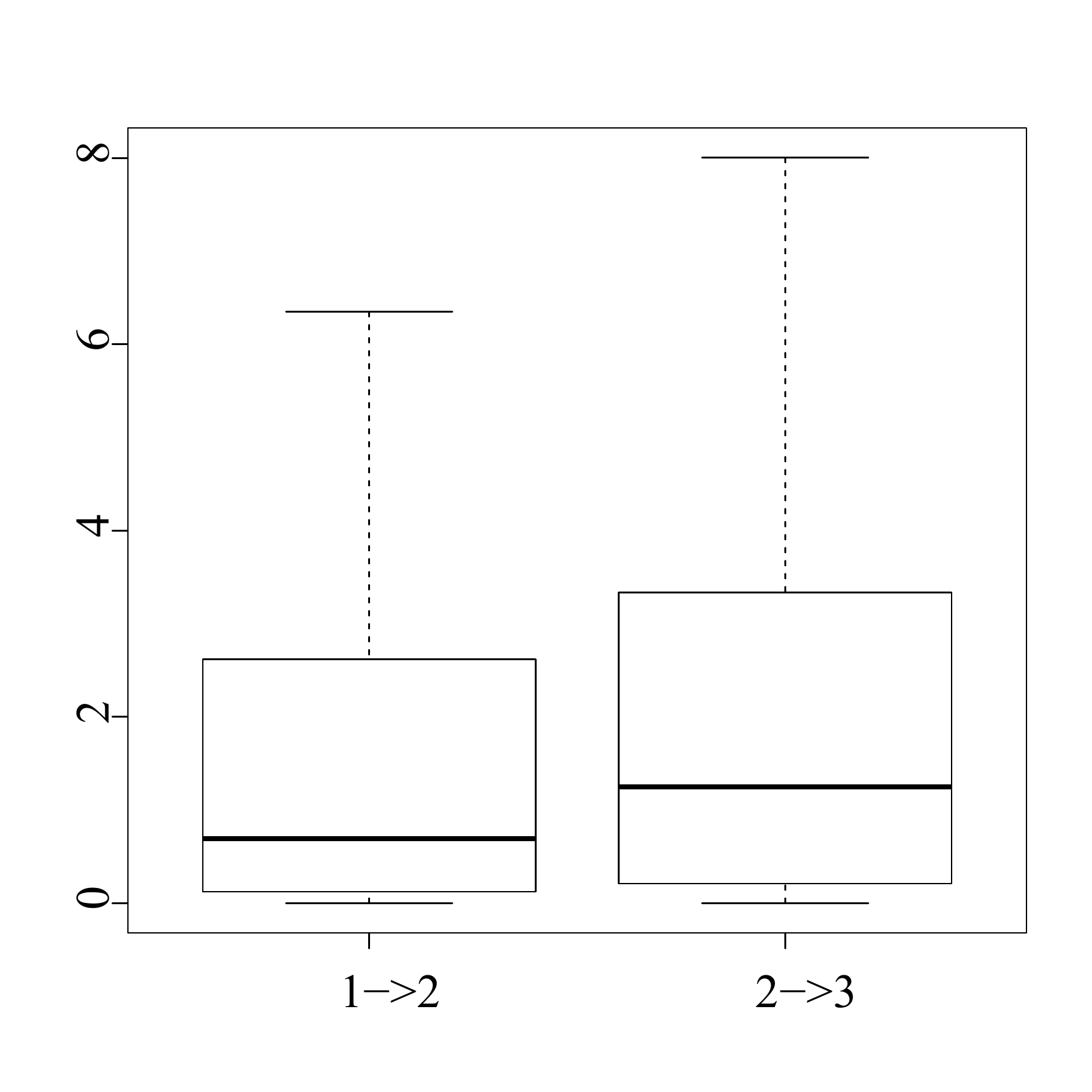}
		%\vspace{-1.2cm}
		\caption{Boxplots of the distances of movements of nodes in the latent space during the two transitions.\label{boxplot5294}}
	\end{figure}

	To study the dynamics of the network, we also calculated the squared distances of movements for each node during the two transitions and created boxplots for them (Figure \ref{boxplot5294}). From these plots, we can see that the ranges of these distances are larger in the second transition than in the first transition. The dynamics of the network did not seem to be stable during these transitions.

	%%%%%%%%%%%%%%%%%%%%%%%%%%%%%%%%%%%%%%%%%%%%%%%%%%%%%%%%%%%%%%%%%%%%%%%
	%\vspace{-0.5cm}
	\section{Discussion}\label{sec:discussion}
	%\vspace{-0.5cm}
	We proposed a variational Bayes algorithm for dynamic latent space models. The proposed algorithm is able to handle large-scale networks, and performs well for simulated and real-world networks. Furthermore, we prove that under certain conditions, the variational Bayes risk of the VB procedure goes to zero as the number of nodes goes to infinity. To the best of our knowledge, this is the first paper to propose a variational algorithm for dynamic distance model and to address its theoretical properties.
	
	We note that the lower bound of the expected log-likelihood is not unique (see \cite{jaakkola2000bayesian} for another possible lower bound). However, the one used in our derivation requires fewer approximation steps, and the resulting objective function is easier to optimize. It is also well-known that variational inference has a tendency to underestimate posterior variance due to approximating the posterior by a factorized distribution. Since we are focusing on point estimates, the underestimation of the variance will not affect our performance metric. It is of interest to investigate the	uncertainty estimation based on VB for latent space model.
	
	In the dynamic network context, the mean-field approximation we used may not be the most suited for the time series aspect of the model. For example, the generalized mean-field approach in \cite{xing2010state} factorizes the approximate posterior distribution into the product of several modules, and models each module by a state space model. Such variation is also possible for dynamic networks. However, while the state space models can better capture the dependence structure, it is harder to study theoretical properties of such algorithms.

	The proposed VB algorithm can be extended to more complicated models. For example, if we assume that the initial latent positions come from a mixture of Gaussian distributions and their cluster assignments change over time, our method can be modified to address the community detection problem for dynamic latent space models. More global parameters, such as the parameters characterizing popularity and social activity in \cite{sewell2015latent}, can also be incorporated into the model. Another possible extension is to incorporate dyadic-level covariate information into the dynamic network model, which extends the model proposed by \cite{krivitsky2009representing}. It is also possible to generalize the proposed method to dynamic multi-layer latent space models. It is of interest to develop a VB algorithm to accelerate the computation of such models.
	
	In latent space models, the choice of the likelihood function is flexible. The model we used in this paper can be categorized into a larger class of latent variable models (LVMs) (see \cite{rastelli2016properties}). \cite{rastelli2016properties} also introduced the Gaussian latent position model (GLPM) as a special class of the LVM, which replaces the logistic link function for the edges with a non-normalized multivariate Gaussian density. The proposed VB algorithm can be modified to apply to the GLPM and dynamic GLPM. The modified algorithm will involve a similar Jensen approximation as the proposed VB algorithm in this paper.

	%%%%%%%%%%%%%%%%%%%%%%%%%%%%%%%%%%%%%%%%%%%%%%%%%%%%%%%%%%%%%%%%%%%%%%%%%%%%%%%%%%%%%%%%%%%%%%%%%%%%%%%%%%%%%%%%%%%%%%%%%%%%
	
	\noindent {\large\bf Supplementary Materials}
	The supplementary material contains proofs of all theorems, details of the implementation for simulations, additional simulation studies to compare VB with MCMC, apply VB to networks with $n=5000$ nodes, investigate the effect of $\alpha$ in $\alpha$-VB, and verify the asymptotic behavior of the proposed algorithm.

	\par

%%%%%%%%%%%%%%%%%%%%%%%%%%%%%%%%%%%%%%%%%%%%%%%%%%%%%%%%%%%%%%%%%%%%%%%%%%%%%%%%%%%%%%%%%%%%%%%%%%%%%%%%%%%%%%%%%%%%%%%%%%%%
	%\vskip 14pt
	\noindent {\large\bf Acknowledgements}
	This work was supported in part by National Science Foundation grant DMS-2015561. The authors would like to thank Professor Yun Yang for helpful discussion of theoretical results.
	\par
	
	%%%%%%%%%%%%%%%%%%%%%%%%%%%%%%%%%%%%%%%%%%%%%%%%%%%%%%%%%%%%%%%%%%%%%%%%%%%%%%%%%%%%%%%%%%%%%%%%%%%%%%%%%%%%%%%%%%%%%%%%%%%%
	\markboth{\hfill{\footnotesize\rm Yan Liu AND Yuguo Chen} \hfill}
	{\hfill {\footnotesize\rm VB for Dynamic Latent Space Models} \hfill}
	
	%\iffalse
	\bibhang=1.7pc
	\bibsep=2pt
	\fontsize{9}{14pt plus.8pt minus .6pt}\selectfont
	\renewcommand\bibname{\large \bf References}
	%\begin{thebibliography}{11}
	\expandafter\ifx\csname
	natexlab\endcsname\relax\def\natexlab#1{#1}\fi
	\expandafter\ifx\csname url\endcsname\relax
	\def\url#1{\texttt{#1}}\fi
	\expandafter\ifx\csname urlprefix\endcsname\relax\def\urlprefix{URL}\fi
	%\fi
	
	\lhead[\footnotesize\thepage\fancyplain{}\leftmark]{}\rhead[]{\fancyplain{}\rightmark\footnotesize{} }%Put this line in Page 2
	%%%%%%%%%%%%%%%%%%%%%%%%%%%%%%%%%%%%%%%%%%%%%%%%%%%%%%%%%%

%%%%%%%%%%%%%%%%%%%%%%%%%%%%%%%%%%%%%%%%%%%%%%%%%%%%%%%%%%
\clearpage

\section*{\large Supplementary Materials}
\subsection*{Proof of Theorem 1}

First, by Theorem 3.1 of \cite{yang2017alpha}, we have the following variational risk bound for the $\alpha<1$ case:
\begin{eqnarray*}
	&& \int \frac{1}{n(n-1)T} D_{\alpha}^{(n)}(\theta,\theta^*) \hat{q}_{\theta,\alpha}(d\theta)\\
	&\leq& \frac{\alpha}{n(n-1)T(1-\alpha)}\Psi_{n,\alpha}(q_{\theta},q_{\boldsymbol{\mathcal{X}}})
	+ \frac{1}{n(n-1)T(1-\alpha)} \log\left(\frac{1}{\zeta}\right),
\end{eqnarray*}
with probability at least $(1-\zeta)$ for any $\zeta\in(0,1)$.
This result establishes a connection between the variational Bayes risk and the $\alpha$-VB objective function, which implies that minimizing the $\alpha$-VB objective function $\Psi_{n,\alpha}$ will also minimize the variational Bayes risk.

The next step of the proof is to further simplify the above upper bound based on a certain choice of the variational family of the latent variable $\boldsymbol{\mathcal{X}}$ and the model parameter $\theta$. Recall that $\theta=(\beta,\sigma^2,\tau^2)$ denotes all the model parameters, and $\theta^*=(\beta^*,\sigma^{*2},\tau^{*2})$ is their true values. From now on, we use $\pi=(\sigma^2,\tau^2)$ to denote the parameters that characterize the distribution of latent variables, and use $\pi^*=(\sigma^{*2},\tau^{*2})$ to denote their true values.

We still use the mean-field decomposition $q(\boldsymbol{\mathcal{X}},\theta) = q(\boldsymbol{\mathcal{X}})q(\theta)$. However, the dependence structure between the observations and the latent variables in our model is different from the
simplifying assumptions in \cite{yang2017alpha}, since we no longer have i.i.d. observations or observation-specific latent variables. In our case, for any fixed $q_{\theta}$, we choose the variational distribution $q(\boldsymbol{\mathcal{X}})$ in the following way:
$$
q(\boldsymbol{\mathcal{X}}) \propto
\left(
\prod_{t=1}^{T}\prod_{i\neq j} p(Y_{ijt}|\beta^*,\boldsymbol{X}_{it},\boldsymbol{X}_{jt})
\right)
\times \left(
\prod_{i=1}^{n} \left(
p(\boldsymbol{X}_{i1}|\pi^*) \prod_{t=1}^{T} p(\boldsymbol{X}_{it}|\pi^*,\boldsymbol{X}_{i(t-1)})
\right)
\right),
$$
where the normalizing constant is $p(\boldsymbol{\mathcal{Y}}|\theta^*)$.

With this choice of variational family, the $\alpha$-VB objective function becomes
\begin{align*}
\Psi_{n,\alpha}(q_{\theta},q_{\boldsymbol{\mathcal{X}}})
&= -\int_{\Theta} \left(l_n(\theta)-l_n(\theta^*)\right) q(d\theta)
+ \Delta_{J}(q_{\theta},q_{\boldsymbol{\mathcal{X}}})
+ \frac{1}{\alpha}D(q_{\theta}||p_{\theta})\\
&= -\int_{\Theta} \left(l_n(\theta)-l_n(\theta^*)+\hat{l}_n(\theta)-l_n(\theta)\right) q(d\theta)
+ \frac{1}{\alpha}D(q_{\theta}||p_{\theta})\\
&= -\int_{\Theta} \left(\hat{l}_n(\theta)-l_n(\theta^*)\right) q(d\theta)
+ \frac{1}{\alpha}D(q_{\theta}||p_{\theta}),
\end{align*}
where the first term on the right hand side is
\begin{align*}
& -\int_{\Theta} \left(\hat{l}_n(\theta)-l_n(\theta^*)\right) q(d\theta)\\
= &-\int_{\Theta} \left(
\int_{\boldsymbol{\mathcal{X}}} \log
\frac{p(\boldsymbol{\mathcal{Y}}|\boldsymbol{\mathcal{X}},\beta) p(\boldsymbol{\mathcal{X}}|\pi)}
{q(\boldsymbol{\mathcal{X}})} q(d\boldsymbol{\mathcal{X}})
- \log\left(
\int_{\boldsymbol{\mathcal{X}}}
\frac{p(\boldsymbol{\mathcal{Y}}|\boldsymbol{\mathcal{X}},\beta^*) p(\boldsymbol{\mathcal{X}}|\pi^*)}
{q(\boldsymbol{\mathcal{X}})} q(d\boldsymbol{\mathcal{X}})
\right)
\right) q(d\theta)\\
= &-\int_{\Theta} \left(
\int_{\boldsymbol{\mathcal{X}}}
\sum_{t=1}^{T}\sum_{i\neq j} \log
\frac{p(Y_{ijt}|\beta,\boldsymbol{X}_{it},\boldsymbol{X}_{jt})}
{p(Y_{ijt}|\beta^*,\boldsymbol{X}_{it},\boldsymbol{X}_{jt})}
q(d\boldsymbol{\mathcal{X}})
- D(p(\boldsymbol{\mathcal{X}}|\pi^*)||p(\boldsymbol{\mathcal{X}}|\pi))
\right)
q(d\theta)\\
&+ \left(
\int_{\Theta}\int_{\boldsymbol{\mathcal{X}}}
p(\boldsymbol{\mathcal{Y}}|\theta^*) q(d\boldsymbol{\mathcal{X}})q(d\theta)
-\int_{\Theta} \log\left(
\int_{\boldsymbol{\mathcal{X}}}
\frac{p(\boldsymbol{\mathcal{Y}}|\boldsymbol{\mathcal{X}},\beta^*) p(\boldsymbol{\mathcal{X}}|\pi^*)}
{q(\boldsymbol{\mathcal{X}})} q(d\boldsymbol{\mathcal{X}})
\right) q(d\theta)
\right).
\end{align*}
Note that the last expression in the equation above contains two terms and the second term is 0. Therefore, the variational risk bound becomes
\begin{align}\label{riskbound1}
& \int \frac{1}{n(n-1)T} D_{\alpha}^{(n)}(\theta,\theta^*) \hat{q}_{\theta,\alpha}(d\theta)\nonumber\\
\leq & \frac{\alpha}{n(n-1)T(1-\alpha)}\Psi_{n,\alpha}(q_{\theta},q_{\boldsymbol{\mathcal{X}}})
+ \frac{1}{n(n-1)T(1-\alpha)} \log\left(\frac{1}{\zeta}\right) \nonumber\\
=& -\frac{\alpha}{n(n-1)T(1-\alpha)}
\int_{\Theta}\int_{\boldsymbol{\mathcal{X}}} \biggr(
\sum_{t=1}^{T} \sum_{i\neq j}
\log\frac{p(Y_{ijt}|\boldsymbol{X}_{it},\boldsymbol{X}_{jt},\beta)}{p(Y_{ijt}|\boldsymbol{X}_{it},\boldsymbol{X}_{jt},\beta^*)} \nonumber\\
&\qquad\qquad\qquad\qquad\qquad\qquad\qquad -\sum_{i=1}^{n} D(p(\boldsymbol{X}_{it}|\pi^*)||p(\boldsymbol{X}_{it}|\pi))
\biggr)q(d\boldsymbol{\mathcal{X}})q(d\theta)\nonumber\\
&+ \frac{1}{n(n-1)T(1-\alpha)} \left(
D(q(\theta)||p(\theta)) + \log (1/\zeta)
\right) \nonumber\\
:=& -\frac{\alpha}{n(n-1)T(1-\alpha)} W_1
+ \frac{1}{n(n-1)T(1-\alpha)} \left(
D(q(\theta)||p(\theta)) + \log (1/\zeta)
\right).
\end{align}

The variational risk bound (\ref{riskbound1}) can be viewed as an analogue of Corollary 3.2 in \cite{yang2017alpha}. Now in order to apply Chebyshev's inequality to obtain the desired result, we need to bound the first and second moments of the first term of the right-hand side of (\ref{riskbound1}).

Recall the following definition of the KL-neighborhoods of the true model parameters given in the statement of Theorem \ref{theorem1}:
$$
\mathcal{B}_{n}(\pi^*,\epsilon_{\pi})
:= \left\{ \pi:
D\left(p(\boldsymbol{\mathcal{X}}_{1}|\pi^*)||p(\boldsymbol{\mathcal{X}}_{1}|\pi)\right)\leq \epsilon_{\pi}^2, \quad
V\left(p(\boldsymbol{\mathcal{X}}_{1}|\pi^*)||p(\boldsymbol{\mathcal{X}}_{1}|\pi)\right)\leq \epsilon_{\pi}^2
\right\},
$$
\begin{align*}
\mathcal{B}_{n}(\beta^*,\epsilon_{\beta})
:= \biggr\{ \beta: &\sup_{\boldsymbol{X}_{11},\boldsymbol{X}_{21}}D(p(Y_{121}|\beta^*,\boldsymbol{X}_{11},\boldsymbol{X}_{21})||p(Y_{121}|\beta,\boldsymbol{X}_{11},\boldsymbol{X}_{21}))\leq \epsilon_{\beta}^2,\\ &\sup_{\boldsymbol{X}_{11},\boldsymbol{X}_{21}}V(p(Y_{121}|\beta^*,\boldsymbol{X}_{11},\boldsymbol{X}_{21})||p(Y_{121}|\beta,\boldsymbol{X}_{11},\boldsymbol{X}_{21}))\leq \epsilon_{\beta}^2 \biggr\},
\end{align*}
where
$
V(p||q) := \int p \log^2(\frac{p}{q}) d\mu.
$
Then we choose $q_{\theta}(\theta)$ as the probability density function (pdf) $q_{\theta}^*$, which is the pdf of the product measure of restrictions of the priors of the model parameters to KL-neighborhoods $\mathcal{B}_{n}(\pi^*,\epsilon_{\pi})$ and $\mathcal{B}_{n}(\beta^*,\epsilon_{\beta})$.

By Fubini's Theorem,
\begin{align}
&\mathbb{E}_{\theta^*}[W_1] \nonumber\\
= &\mathbb{E}_{\theta^*}
\left[
\int_{\Theta}q_{\theta}^*(\theta)
\int_{\boldsymbol{\mathcal{X}}}\left(
\sum_{t=1}^{T} \sum_{i\neq j} \log\frac{P(Y_{ijt}|\boldsymbol{X}_{it},\boldsymbol{X}_{jt},\beta)}{P(Y_{ijt}|\boldsymbol{X}_{it},\boldsymbol{X}_{jt},\beta^*)}
-\sum_{i=1}^{n}D(p(\boldsymbol{\mathcal{X}}_i|\pi^*)||p(\boldsymbol{\mathcal{X}}_i|\pi))
\right)
q(d\boldsymbol{\mathcal{X}}) d\theta
\right]\nonumber\\
= &\int_{\Theta}\left\{
\mathbb{E}_{\theta^*}\left[
\int_{\boldsymbol{\mathcal{X}}}\left(
\sum_{t=1}^{T} \sum_{i\neq j} \log\frac{P(Y_{ijt}|\boldsymbol{X}_{it},\boldsymbol{X}_{jt},\beta)}{P(Y_{ijt}|\boldsymbol{X}_{it},\boldsymbol{X}_{jt},\beta^*)}
-\sum_{i=1}^{n}D(p(\boldsymbol{\mathcal{X}}_i|\pi^*)||p(\boldsymbol{\mathcal{X}}_i|\pi))
\right)
q(d\boldsymbol{\mathcal{X}})
\right]
\right\}
q_{\theta}^*(\theta) d\theta \nonumber\\
= &\int_{\Theta}\biggr\{
- \sum_{i=1}^{n}D(p(\boldsymbol{\mathcal{X}}_i|\pi^*)||p(\boldsymbol{\mathcal{X}}_i|\pi)) \nonumber\\
&\qquad - n(n-1)T \int_{\boldsymbol{\mathcal{X}}}D(p(\cdot|\boldsymbol{X}_{11},\boldsymbol{X}_{21},\beta^*)||p(\cdot|\boldsymbol{X}_{11},\boldsymbol{X}_{21},\beta)) q(d\boldsymbol{\mathcal{X}})
\biggr\}
q_{\theta}^*(\theta) d\theta. \nonumber
\end{align}
Since $q_{\theta}^*$ is the restriction of $p(\theta)$ into the KL-neighborhoods defined above, we have
$$
-\frac{\alpha}{n(n-1)T(1-\alpha)} \mathbb{E}_{\theta^*}[W_1]
\leq \frac{\alpha\left(n(n-1)T \epsilon_{\beta}^2 + n\epsilon_{\pi}^2\right)}{n(n-1)T(1-\alpha)}.
$$

Furthermore, the variance
\begin{align}
&\text{Var}_{\theta^*}[W_1] \nonumber\\
=& \text{Var}_{\theta^*}
\left[
\int_{\Theta}q_{\theta}^*(\theta)
\int_{\boldsymbol{\mathcal{X}}}\left(
\sum_{t=1}^{T} \sum_{i\neq j} \log\frac{P(Y_{ijt}|\boldsymbol{X}_{it},\boldsymbol{X}_{jt},\beta)}{P(Y_{ijt}|\boldsymbol{X}_{it},\boldsymbol{X}_{jt},\beta^*)}
-\sum_{i=1}^{n}D(p(\boldsymbol{\mathcal{X}}_i|\pi^*)||p(\boldsymbol{\mathcal{X}}_i|\pi))
\right)
q(d\boldsymbol{\mathcal{X}}) d\theta
\right]\nonumber\\
=& \text{Var}_{\theta^*}
\left[
\mathbb{E}_{q(\boldsymbol{\mathcal{X}}),q_{\theta}^*(\theta)}\left[
\sum_{t=1}^{T} \sum_{i\neq j} \log\frac{P(Y_{ijt}|\boldsymbol{X}_{it},\boldsymbol{X}_{jt},\beta)}{P(Y_{ijt}|\boldsymbol{X}_{it},\boldsymbol{X}_{jt},\beta^*)}
-\sum_{i=1}^{n}D(p(\boldsymbol{\mathcal{X}}_i|\pi^*)||p(\boldsymbol{\mathcal{X}}_i|\pi))
\right]
\right]\nonumber\\
=& \text{Var}_{\theta^*}
\left[
\sum_{t=1}^{T} \sum_{i\neq j} \mathbb{E}_{q(\boldsymbol{\mathcal{X}}),q_{\theta}^*(\theta)}\left[
\log\frac{P(Y_{ijt}|\boldsymbol{X}_{it},\boldsymbol{X}_{jt},\beta)}{P(Y_{ijt}|\boldsymbol{X}_{it},\boldsymbol{X}_{jt},\beta^*)}
\right]
-\sum_{i=1}^{n} \mathbb{E}_{q_{\theta}^*(\theta)}\left[
D(p(\boldsymbol{\mathcal{X}}_i|\pi^*)||p(\boldsymbol{\mathcal{X}}_i|\pi))
\right]
\right]
\nonumber\\
=& \sum_{t,s=1}^{T} \sum_{i\neq j}\sum_{k\neq l}
\text{Cov}\left(
\mathbb{E}_{q(\boldsymbol{\mathcal{X}}),q_{\theta}^*(\theta)}\left[
\log\frac{P(Y_{ijt}|\boldsymbol{X}_{it},\boldsymbol{X}_{jt},\beta)}{P(Y_{ijt}|\boldsymbol{X}_{it},\boldsymbol{X}_{jt},\beta^*)}
\right],\
\mathbb{E}_{q(\boldsymbol{\mathcal{X}}),q_{\theta}^*(\theta)}\left[
\log\frac{P(Y_{kls}|\boldsymbol{X}_{ks},\boldsymbol{X}_{ls},\beta)}{P(Y_{kls}|\boldsymbol{X}_{ks},\boldsymbol{X}_{ls},\beta^*)}
\right]
\right)\label{1}\\
&+ \sum_{i,j=1}^{n} \text{Cov}\left(
\mathbb{E}_{q_{\theta}^*(\theta)}\left[
D(p(\boldsymbol{\mathcal{X}}_i|\pi^*)||p(\boldsymbol{\mathcal{X}}_i|\pi))
\right], \
\mathbb{E}_{q_{\theta}^*(\theta)}\left[
D(p(\boldsymbol{\mathcal{X}}_j|\pi^*)||p(\boldsymbol{\mathcal{X}}_j|\pi))
\right]
\right)\label{2}\\
&-\sum_{t=1}^{T} \sum_{i\neq j}\sum_{k=1}^{n} \text{Cov}\left(
\mathbb{E}_{q(\boldsymbol{\mathcal{X}}),q_{\theta}^*(\theta)}\left[
\log\frac{P(Y_{ijt}|\boldsymbol{X}_{it},\boldsymbol{X}_{jt},\beta)}{P(Y_{ijt}|\boldsymbol{X}_{it},\boldsymbol{X}_{jt},\beta^*)}
\right],\
\mathbb{E}_{q_{\theta}^*(\theta)}\left[
D(p(\boldsymbol{\mathcal{X}}_k|\pi^*)||p(\boldsymbol{\mathcal{X}}_k|\pi))
\right]
\right)\label{3}.
\end{align}

First, note that for any $1\leq i\neq j \leq n$, $t=1,\dots,T$,
\begin{align*}
\log\frac{P(Y_{ijt}|\boldsymbol{X}_{it},\boldsymbol{X}_{jt},\beta)}{P(Y_{ijt}|\boldsymbol{X}_{it},\boldsymbol{X}_{jt},\beta^*)}
=& Y_{ijt}(\beta-||\boldsymbol{X}_{it}-\boldsymbol{X}_{jt}||^2)
- Y_{ijt}(\beta^*-||\boldsymbol{X}_{it}-\boldsymbol{X}_{jt}||^2)\\
&-\log \left(
1+e^{\beta-||\boldsymbol{X}_{it}-\boldsymbol{X}_{jt}||^2}
\right)
+ \log \left(
1+e^{\beta^*-||\boldsymbol{X}_{it}-\boldsymbol{X}_{jt}||^2}
\right)\\
=& Y_{ijt}(\beta-\beta^*)
-\log\left(
\frac{1+e^{\beta-||\boldsymbol{X}_{it}-\boldsymbol{X}_{jt}||^2}}{1+e^{\beta^*-||\boldsymbol{X}_{it}-\boldsymbol{X}_{jt}||^2}}
\right).
\end{align*}
Thus, after taking expectation with respect to $\boldsymbol{\mathcal{X}}$ and $\beta$, only the first term is random. Let
$$
\mathbb{E}_{q(\boldsymbol{\mathcal{X}}),q_{\theta}^*(\theta)}\left[
\log\frac{P(Y_{ijt}|\boldsymbol{X}_{it},\boldsymbol{X}_{jt},\beta)}{P(Y_{ijt}|\boldsymbol{X}_{it},\boldsymbol{X}_{jt},\beta^*)}
\right]
:= c_1 Y_{ijt} + c_2(i,j,t),
$$
where $c_1$ and $c_2$ are constants that depend on the variational distribution $q$.

Then the term (\ref{1}) becomes
$$
\sum_{t,s=1}^{T} \sum_{i\neq j}\sum_{k\neq l}\text{Cov}\left[
c_1 Y_{ijt} + c_2(i,j,t),\ c_1 Y_{kls} + c_2(k,l,s)
\right]
= c_1^2 \sum_{t,s=1}^{T}\sum_{i\neq j}\sum_{k\neq l} \text{Cov}\left[Y_{ijt},Y_{kls}\right].
$$

The number of terms in the summation is $n^2 (n-1)^2 T^2$, but for any $t$ and $s$, $\text{Cov}\left[Y_{ijt},Y_{kls}\right]=0$ when $i\neq k$ and $j\neq l$. Also, for any $i,j=1,\dots,n$ and $t=1,\dots,T$, the variance term can be bounded in the following way:
\begin{align*}
& \text{Var}_{\theta^*}\left[
\mathbb{E}_{q(\boldsymbol{\mathcal{X}}),q_{\theta}^*(\theta)}\left[
\log \frac{P(Y_{ijt}|\boldsymbol{X}_{it},\boldsymbol{X}_{jt},\beta)}{P(Y_{ijt}|\boldsymbol{X}_{it},\boldsymbol{X}_{jt},\beta^*)}
\right]
\right]\\
=& \text{Var}_{\theta^*}\left[
\int_{\Theta} q_{\theta}^*(\theta)
\int_{\boldsymbol{\mathcal{X}}} q_{\boldsymbol{\mathcal{X}}}(\boldsymbol{\mathcal{X}})
\log \frac{P(Y_{ijt}|\boldsymbol{X}_{it},\boldsymbol{X}_{jt},\beta)}{P(Y_{ijt}|\boldsymbol{X}_{it},\boldsymbol{X}_{jt},\beta^*)}
d\boldsymbol{\mathcal{X}} d\theta
\right]\\
\leq & \mathbb{E}_{\theta^*}\left[
\int_{\Theta} q_{\theta}^*(\theta)
\int_{\boldsymbol{\mathcal{X}}} q_{\boldsymbol{\mathcal{X}}}(\boldsymbol{\mathcal{X}})
\log \frac{P(Y_{ijt}|\boldsymbol{X}_{it},\boldsymbol{X}_{jt},\beta)}{P(Y_{ijt}|\boldsymbol{X}_{it},\boldsymbol{X}_{jt},\beta^*)}
d\boldsymbol{\mathcal{X}} d\theta
\right]^2\\
\leq & \mathbb{E}_{\theta^*}\left[
\int_{\Theta} q^*_{\theta}(\theta)
\left[
\int_{\boldsymbol{\mathcal{X}}} q_{\boldsymbol{\mathcal{X}}}(\boldsymbol{\mathcal{X}})
\log \frac{P(Y_{ijt}|\boldsymbol{X}_{it},\boldsymbol{X}_{jt},\beta)}{P(Y_{ijt}|\boldsymbol{X}_{it},\boldsymbol{X}_{jt},\beta^*)} d\boldsymbol{\mathcal{X}}
\right]^2
d\theta
\right]\\
= & \int_{\Theta} \mathbb{E}_{\theta^*}\left[
\int_{\boldsymbol{\mathcal{X}}} q_{\boldsymbol{\mathcal{X}}}(\boldsymbol{\mathcal{X}})
\log \frac{P(Y_{ijt}|\boldsymbol{X}_{it},\boldsymbol{X}_{jt},\beta)}{P(Y_{ijt}|\boldsymbol{X}_{it},\boldsymbol{X}_{jt},\beta^*)} d\boldsymbol{\mathcal{X}}
\right]^2 q^*_{\theta}(\theta) d\theta\\
\leq & \int_{\Theta} \mathbb{E}_{\theta^*}\left[
\int_{\boldsymbol{\mathcal{X}}} q_{\boldsymbol{\mathcal{X}}}(\boldsymbol{\mathcal{X}})
\log^2 \frac{P(Y_{ijt}|\boldsymbol{X}_{it},\boldsymbol{X}_{jt},\beta)}{P(Y_{ijt}|\boldsymbol{X}_{it},\boldsymbol{X}_{jt},\beta^*)} d\boldsymbol{\mathcal{X}}
\right] q^*_{\theta}(\theta) d\theta\\
= & \int_{\Theta} \left[
\int_{\boldsymbol{\mathcal{X}}} V\left[
p(\cdot|\beta^*,\boldsymbol{X}_{11},\boldsymbol{X}_{21})||p(\cdot|\beta,\boldsymbol{X}_{11},\boldsymbol{X}_{21})
\right] q_{\boldsymbol{\mathcal{X}}}(\boldsymbol{\mathcal{X}}) d\boldsymbol{\mathcal{X}}
\right]
q^*_{\theta}(\theta) d\theta\\
\leq & \epsilon_{\beta}^2,
\end{align*}
where the second and third inequalities are due to Jensen's inequality, and the second equality is due to Fubini's theorem. Thus, by Cauchy-Schwarz inequality,
\begin{eqnarray*}
	&& \mbox{term}\ (\ref{1}) \\
	&\leq& \sum_{t,s=1}^{T}\sum_{i\neq j}\sum_{k\neq l} \sqrt{
		\text{Var}_{\theta^*}\left[
		\mathbb{E}\left[
		\log \frac{P(Y_{ijt}|\boldsymbol{X}_{it},\boldsymbol{X}_{jt},\beta)}{P(Y_{ijt}|\boldsymbol{X}_{it},\boldsymbol{X}_{jt},\beta^*)}
		\right]
		\right]\cdot
		\text{Var}_{\theta^*}\left[
		\mathbb{E}\left[
		\log \frac{P(Y_{kls}|\boldsymbol{X}_{ks},\boldsymbol{X}_{ls},\beta)}{P(Y_{kls}|\boldsymbol{X}_{ks},\boldsymbol{X}_{ls},\beta^*)}
		\right]
		\right]
	}\\
	&\leq& (n(n-1)T^2 + n(n-1)(n-2)T^2)\epsilon_{\beta}^2\\
	&\leq& n(n-1)^2 T^2 (\epsilon_{\beta}^2 + \epsilon_{\pi}^2).
\end{eqnarray*}

The other two terms: $\mbox{term}\ (\ref{2})=\mbox{term}\ (\ref{3})=0$. By Chebyshev's inequality, for any fixed $(\epsilon_{\beta},\epsilon_{\pi})\in (0,1)$ and any $D>1$,
\begin{align}\label{chebyshev}
\mathbb{P}_{\theta^*}\left(
W_1
\leq -Dn(n-1)T(\epsilon_{\beta}^2+\epsilon_{\pi}^2)
\right)
&\leq
\mathbb{P}_{\theta^*}\left(
W_1
-\mathbb{E}[W_1]
\leq -(D-1)n(n-1)T(\epsilon_{\beta}^2+\epsilon_{\pi}^2)
\right)\nonumber\\
&\leq
\frac{\text{Var}\left[W_1\right]}{(D-1)^2 n^2 (n-1)^2 T^2 (\epsilon_{\beta}^2+\epsilon_{\pi}^2)^2}\nonumber\\
&\leq
\frac{1}{(D-1)^2 n(\epsilon_{\beta}^2+\epsilon_{\pi}^2)}.
\end{align}

Since the variational family of $\theta$ is the restriction of the prior on the KL-neighborhoods $\mathcal{B}_n(\pi^*,\epsilon_{\pi})$ and $\mathcal{B}_n(\beta^*,\epsilon_{\beta})$, we have
$$
D(q_{\theta}^*(\theta)||p_{\theta}(\theta))
= -\log P_{\pi}\left[\mathcal{B}_n(\pi^*,\epsilon_{\pi})\right]
-\log P_{\beta}\left[\mathcal{B}_n(\beta^*,\epsilon_{\beta})\right],
$$
where $P_{\pi}$ and $P_{\beta}$ denote the probability measures corresponding to the priors of $\pi$ and $\beta$, respectively.
This together with inequality (\ref{chebyshev}) implies that for any fixed $(\epsilon_{\pi}^2,\epsilon_{\beta}^2) \in (0,1)^2$ and $D>1$, it holds with probability at least $1-\frac{2}{(D-1)^2 n(\epsilon_{\beta}^2+\epsilon_{\pi}^2)}$ that
\begin{align*}
\int \frac{1}{n(n-1)T} D_{\alpha}^{(n)}(\theta,\theta^*) \hat{q}_{\theta,\alpha}(d\theta)
\leq & \frac{D\alpha}{1-\alpha}(\epsilon_{\pi}^2 + \epsilon_{\beta}^2)
- \frac{1}{n(n-1)T(1-\alpha)}\log P_{\pi}(\mathcal{B}_{n}(\pi^*,\epsilon_{\pi})) \nonumber\\
& - \frac{1}{n(n-1)T(1-\alpha)}\log P_{\beta}(\mathcal{B}_{n}(\beta^*,\epsilon_{\beta})).
\end{align*}

\subsection*{Proof of Theorem 2}

For any $i,j=1,\dots,n$ and $t=1,\dots,T$, the KL divergence
\begin{align*}
D\left(
p(\cdot|\beta^*,\boldsymbol{X}_{it},\boldsymbol{X}_{jt})
||p(\cdot|\beta,\boldsymbol{X}_{it},\boldsymbol{X}_{jt})
\right)
&= \mathbb{E}\left[Y_{ijt}\right] (\beta^*-\beta)
- \log \left( \frac{1+e^{\beta^*-||\boldsymbol{X}_{it}-\boldsymbol{X}_{jt}||^2}}{1+e^{\beta-||\boldsymbol{X}_{it}-\boldsymbol{X}_{jt}||^2}}
\right)\\
&\leq |\beta-\beta^*|
-\log e^{-|\beta-\beta^*|} \leq 2|\beta-\beta^*|,
\end{align*}
and the $V$-divergence
\begin{align*}
&V\left(
p(\cdot|\beta^*,\boldsymbol{X}_{it},\boldsymbol{X}_{jt})
||p(\cdot|\beta,\boldsymbol{X}_{it},\boldsymbol{X}_{jt})
\right) \\
= &\mathbb{E}_{\beta^*} \left[\left(
Y_{ijt}(\beta^*-\beta) - \log\left(
\frac{1+e^{\beta^*-||\boldsymbol{X}_{it}-\boldsymbol{X}_{jt}||^2}}{1+e^{\beta-||\boldsymbol{X}_{it}-\boldsymbol{X}_{jt}||^2}}
\right)\right)^2
\right]\\
= &\mathbb{E}\left[Y_{ijt}^2\right]|\beta^*-\beta|^2
+ \log^2 \left(
\frac{1+e^{\beta^*-||\boldsymbol{X}_{it}-\boldsymbol{X}_{jt}||^2}}{1+e^{\beta-||\boldsymbol{X}_{it}-\boldsymbol{X}_{jt}||^2}}
\right)
- (\beta^*-\beta) \log \left(
\frac{1+e^{\beta^*-||\boldsymbol{X}_{it}-\boldsymbol{X}_{jt}||^2}}{1+e^{\beta-||\boldsymbol{X}_{it}-\boldsymbol{X}_{jt}||^2}}
\right)\mathbb{E}[Y_{ijt}]\\
\leq & |\beta^*-\beta|^2
+ |\beta^*-\beta|^2 = 2|\beta^*-\beta|^2.
\end{align*}
Note that $(\beta^*-\beta) \log \left(
\frac{1+e^{\beta^*-||\boldsymbol{X}_{it}-\boldsymbol{X}_{jt}||^2}}{1+e^{\beta-||\boldsymbol{X}_{it}-\boldsymbol{X}_{jt}||^2}}
\right)\mathbb{E}[Y_{ijt}] \geq 0$, and the last inequality is based on the following fact
$$
e^{-|x-y|} \leq \frac{1+e^{x}}{1+e^{y}} \leq e^{|x-y|}.
$$

This implies that the KL neighborhood $\mathcal{B}_n (\beta^*,\epsilon_{\beta})$ contains the set $\left\{\beta: |\beta-\beta^*|\leq \frac{c}{2}\epsilon_{\beta}^2 \right\}$ for some constant $c$, and the volume of this set is at most of the order $\mathcal{O}(\epsilon_{\beta}^2)$. Consequently, by the thick prior assumption, the prior mass of this set $P_{\beta}(\left\{\beta: |\beta-\beta^*|\leq \frac{c}{2}\epsilon_{\beta}^2 \right\})$ is at least of the order $\mathcal{O}(1/\epsilon_{\beta}^2)$.

Similarly, for any $i=1,\ldots,n$,
\begin{align*}
D(p(\boldsymbol{\mathcal{X}}_i|\sigma^{*2},\tau^{*2})||p(\boldsymbol{\mathcal{X}}_i|\sigma^2,\tau^2))
=& \left[-\frac{d}{2}\log\left(\frac{\sigma^{*2}}{\sigma^2}\right)
- d\left(\frac{\sigma^{*2}}{2\sigma^{*2}}-\frac{\sigma^{*2}}{2\sigma^2}\right) \right]\\
&+ \left[-\frac{d}{2}\log\left(\frac{\tau^{*2}}{\tau^2}\right)
- d\left(\frac{\tau^{*2}}{2\tau^{*2}}-\frac{\tau^{*2}}{2\tau^2}\right) \right]
\cdot (T-1).
\end{align*}

Since we restrict model parameters in a compact set, by Lipschitz continuity, there exists some constant $C$, such that
$$
\mathcal{B}_n (\pi^*,\epsilon_{\pi})
\supset \left\{
(\sigma^2,\tau^2): |\sigma^2-\sigma^{*2}|\leq C\epsilon_{\pi}^2, \quad
|\tau^2-\tau^{*2}|\leq C\epsilon_{\pi}^2
\right\}.
$$
The volume of the set on the right-hand side is $(2C\epsilon_{\pi}^2)^2$. Thus, for any $D>1$, we have the following risk bound with probability tending to 1 as $n\rightarrow\infty$,
\begin{eqnarray*}
	&& \int \frac{1}{n(n-1)T} D_{\alpha}^{(n)}(\theta,\theta^*) \hat{q}_{\theta,\alpha}(d\theta)\\
	&\leq& \frac{D\alpha}{1-\alpha}(\epsilon_{\pi}^2 + \epsilon_{\beta}^2)
	- \frac{2\log \epsilon_{\beta}}{n(n-1)T(1-\alpha)}
	- \frac{2\log(2C\epsilon_{\pi}^2)}{n(n-1)T(1-\alpha)}.
\end{eqnarray*}
Choosing $\epsilon_{\beta}=\epsilon_{\pi}=\frac{1}{\sqrt{n}}$, then the risk bound is
of the order
$\mathcal{O}(\frac{1}{n})$.

\subsection*{Proof of Theorem 3}

As stated in the main text, showing theoretical properties of the regular VB requires extra conditions. We first restate the two assumptions proposed by \cite{yang2017alpha}.

\begin{Assumption}\label{identifiability}
	For some $\epsilon_{n}>0$ and any $\epsilon>\epsilon_{n}$, there exist a subset of the parameter space $\mathcal{F}_{n,\epsilon} \subset \Theta$ and a test function $\phi_{n,\epsilon}$ such that
	\begin{align*}
	P_{\theta}(\mathcal{F}_{n,\epsilon}^c)
	&\leq e^{-cn(n-1) \epsilon^2},\\
	\mathbb{E}_{\theta^*}\left[\phi_{n,\epsilon}\right]
	&\leq e^{-cn(n-1) \epsilon_{n}^2},\\
	\mathbb{E}_{\theta}\left[1-\phi_{n,\epsilon}\right]
	&\leq e^{-cn(n-1) h^2(\theta||\theta^*)}, \quad
	\forall\ \theta\in\mathcal{F}_{n,\epsilon}
	\quad \text{such that} \quad h^2(\theta||\theta^*)\geq \epsilon^2.
	\end{align*}
\end{Assumption}

\begin{Assumption}\label{concentration}
	There exists a constant $C>0$ such that
	\begin{align*}
	P_{\theta}\left[ \mathcal{B}_{n}(\pi^*,\epsilon_{n}) \right]
	&\geq e^{-Cn\epsilon_{n}^2},\\
	P_{\theta}\left[ \mathcal{B}_{n}(\beta^*,\epsilon_{n}) \right]
	&\geq e^{-Cn\epsilon_{n}^2}.
	\end{align*}
\end{Assumption}

Assumption \ref{identifiability} is the statistical identifiability condition characterized by the test function condition (see \cite{ghosal2007convergence}). Since we restrict the parameter space to a compact set, such a test function exists and Assumption \ref{identifiability} is automatically satisfied.

Assumption \ref{concentration} is the prior concentration condition. With Assumptions \ref{thickness} and \ref{P} in the main text, it can be shown that Assumption \ref{concentration} here is satisfied.

Under Assumption \ref{P}, with a similar proof as the proof of (\ref{chebyshev}) in Theorem \ref{theorem1}, we can show that for any $D>1$, there exists an event $\mathcal{A}_n$ such that
$$
P_{\theta^*}(\mathcal{A}_n)
\geq 1- \frac{1}{2(D-1)^2 n\epsilon_n^2},
$$
and there exist variational distributions $(q_{\theta}^*,q_{\boldsymbol{\mathcal{X}}}^*)$, such that under event $\mathcal{A}_n$,
\begin{align*}
\Psi_n(q_{\theta}^*,q_{\boldsymbol{\mathcal{X}}}^*)
&\leq 2Dn(n-1)T\epsilon_{n}^2
-\log P_{\pi}\left(\mathcal{B}_n(\pi^*,\epsilon_{\pi})\right)
-\log P_{\beta}\left(\mathcal{B}_n(\beta^*,\epsilon_{\beta})\right)\\
&\leq 2Dn(n-1)T\epsilon_{n}^2 + 2C\epsilon_{n}^2,
\end{align*}
where $\Psi_n$ denotes the regular VB objective function.

Under Assumption \ref{identifiability}, by Theorem 3.5 of \cite{yang2017alpha}, for any $\epsilon \geq \epsilon_{n}$, there exists an event $\mathcal{B}_\epsilon$, such that
$$
P_{\theta^*}(\mathcal{B}_\epsilon)
\geq 1- 2e^{-cn(n-1)T\epsilon_{n}^2},
$$
and under event $\mathcal{B}_\epsilon$, we have the following upper bound to the variational Bayes risk for any $(q_{\theta},q(\boldsymbol{\mathcal{X}}))$ in the variational family
\begin{align}
& \hat{Q}_{\theta}(\mathcal{F}_{n,\epsilon}^c)
\log\frac{\hat{Q}_{\theta}(\mathcal{F}_{n,\epsilon}^c)}{P_{\theta}(\mathcal{F}_{n,\epsilon}^c)}
+ (1-\hat{Q}_{\theta}(\mathcal{F}_{n,\epsilon}^c))
\log\frac{1-\hat{Q}_{\theta}(\mathcal{F}_{n,\epsilon}^c)}{1-P_{\theta}(\mathcal{F}_{n,\epsilon}^c)}\nonumber\\
&+ cn(n-1)T \int_{\theta\in\mathcal{F}_{n,\epsilon},h^2(\theta||\theta^*)\geq\epsilon^2}
h^2(\theta||\theta^*)\hat{Q}_{\theta}(d\theta) \nonumber\\
\leq & \Psi_n(q_{\theta},q_{\boldsymbol{\mathcal{X}}})
+ \frac{cn(n-1)T\epsilon_{n}^2}{2}
+ \log 2, \label{regularVBbound}
\end{align}
where $\hat{Q}_{\theta}(\cdot)$ is the probability measure corresponding to the VB solution $\hat{q}_{\theta}$.

Thus, under the event $\mathcal{A}_n\cap\mathcal{B}_{\epsilon}$, we have
\begin{align}\label{alpha=1_bound}
& \hat{Q}_{\theta}(\mathcal{F}_{n,\epsilon}^c)
\log\frac{\hat{Q}_{\theta}(\mathcal{F}_{n,\epsilon}^c)}{P_{\theta}(\mathcal{F}_{n,\epsilon}^c)}
+ (1-\hat{Q}_{\theta}(\mathcal{F}_{n,\epsilon}^c))
\log\frac{1-\hat{Q}_{\theta}(\mathcal{F}_{n,\epsilon}^c)}{1-P_{\theta}(\mathcal{F}_{n,\epsilon}^c)}\nonumber\\
& + cn(n-1)T \int_{\theta\in\mathcal{F}_{n,\epsilon},h^2(\theta||\theta^*)\geq\epsilon^2}
h^2(\theta||\theta^*)\hat{Q}_{\theta}(d\theta) \nonumber\\
\leq & Cn(n-1)T\epsilon_{n}^2,
\end{align}
where $C>0$ is a constant.

Note that both the sum of the first terms and the third term in the left hand side of (\ref{alpha=1_bound}) are nonnegative, so there exist constants $C'$, $C''$, such that
\begin{align*}
\hat{Q}_{\theta}(\theta\in\mathcal{F}_{n,\epsilon},h^2(\theta||\theta^*)\geq\epsilon^2)
\leq & \frac{1}{\epsilon^2} \int_{\theta\in\mathcal{F}_{n,\epsilon},h^2(\theta||\theta^*)\geq\epsilon^2}
h^2(\theta||\theta^*)\hat{Q}_{\theta}(d\theta)
\leq  C' \frac{\epsilon_{n}^2}{\epsilon^2}, \nonumber \\
\hat{Q}_{\theta}(\mathcal{F}_{n,\epsilon}^c)
\leq & C'' \frac{\epsilon_{n}^2}{\epsilon^2}.
\end{align*}
The inequality in the second expression above is due to the following facts:
\begin{align*}
\hat{Q}_{\theta}(\mathcal{F}_{n,\epsilon}^c) \log \hat{Q}_{\theta}(\mathcal{F}_{n,\epsilon}^c)
+ (1-\hat{Q}_{\theta}(\mathcal{F}_{n,\epsilon}^c)) \log(1-\hat{Q}_{\theta}(\mathcal{F}_{n,\epsilon}^c))
&\geq -\log 2, \\
-\hat{Q}_{\theta}(\mathcal{F}_{n,\epsilon}^c) \log \hat{P}_{\theta}(\mathcal{F}_{n,\epsilon}^c)
-(1-\hat{Q}_{\theta}(\mathcal{F}_{n,\epsilon}^c)) \log(1-\hat{P}_{\theta}(\mathcal{F}_{n,\epsilon}^c))
&\geq -\hat{Q}_{\theta}(\mathcal{F}_{n,\epsilon}^c) \log \hat{P}_{\theta}(\mathcal{F}_{n,\epsilon}^c) \\
&\geq cn(n-1)T\epsilon^2.
\end{align*}

Let $\epsilon=k\epsilon_{n}$ for $k=1,2,\dots,\left\lfloor{e^{cn(n-1)T\epsilon_{n}^2/4}} \right\rfloor$, we can show that the following inequality holds with probability at least $1-\frac{1}{2(D-1)^2n \epsilon_n^2}-2e^{-cn(n-1)T\epsilon_{n}^2/4}\geq 1-\frac{1}{(D-1)^2n \epsilon_n^2}$,
$$
\hat{Q}_{\theta}(h^2(\theta||\theta^*)\geq\epsilon^2)
\leq \hat{Q}_{\theta}(\theta\in\mathcal{F}_{n,\epsilon},h^2(\theta||\theta^*)\geq\epsilon^2)
+ \hat{Q}_{\theta}(\mathcal{F}_{n,\epsilon}^c)
\leq (C'+C'')\frac{\epsilon_{n}^2}{\epsilon^2}.
$$

Let $\epsilon=n^{1/4} \epsilon_{n}$, then
$$
\hat{Q}_{\theta}(h^2(\theta||\theta^*)\geq\epsilon^2)
\leq (C'+C'')\frac{\epsilon_{n}^2}{\epsilon^2}
\rightarrow 0.
$$

Therefore, for any $R<e^{2cn(n-1)T\epsilon_n^2}$, the variational Bayes risk
\begin{align*}
\int_{\left\{
	h^2(\theta||\theta^*)\leq R^2	
	\right\}}
h^2(\theta||\theta^*) \hat{q}_{\theta}(d\theta)
&= \int_{0}^{R^2} \hat{Q}_{\theta}(h^2(\theta||\theta^*)\geq t) dt\\
&= \int_{0}^{\epsilon_n^2} \hat{Q}_{\theta}(h^2(\theta||\theta^*)\geq t) dt
+ \int_{\epsilon_n^2}^{R^2} \hat{Q}_{\theta}(h^2(\theta||\theta^*)\geq t) dt\\
&\leq \epsilon_n^2 + \int_{\epsilon_n^2}^{R^2} \hat{Q}_{\theta}(h^2(\theta||\theta^*)\geq t) dt \\
&= \epsilon_n^2 + 2\int_{\epsilon_n}^{R} s \hat{Q}_{\theta}(h^2(\theta||\theta^*)\geq s^2) ds\\
&\leq \epsilon_{n}^2 + 2\int_{\epsilon_n}^{R} s \cdot (C'+C'')\frac{\epsilon_n^2}{s^2} ds\\
&\leq C\epsilon_{n}^2 (1+\log\frac{R}{\epsilon_n}).
\end{align*}

\subsection*{Additional Simulation: VB vs MCMC}\label{sim1}

In this simulation we compare the performance of the proposed VB algorithm with a Metropolis within Gibbs MCMC sampling algorithm. 
We used parallel tempering to facilitate the mixing of MCMC. Details of this MCMC algorithm are given in the next section.
To make the computation feasible for MCMC, we simulated 20 dynamic networks, each with only $n=50$ nodes and $T=10$ time points.
We considered two cases for the variance of the transition distribution:
$\tau^2=0.0004$ for the small transition case and
$\tau^2=0.01$ for the large transition case.
We considered networks with different edge density by setting
$\beta=0.5$, $-0.5$, $-1.5$ for the dense, moderate and sparse cases, respectively.

\begin{figure}[h!]
	\vspace{-1cm}
	\centering
	\begin{minipage}{\linewidth}
		\begin{minipage}{0.5\textwidth}
			\includegraphics[width=0.86\linewidth]{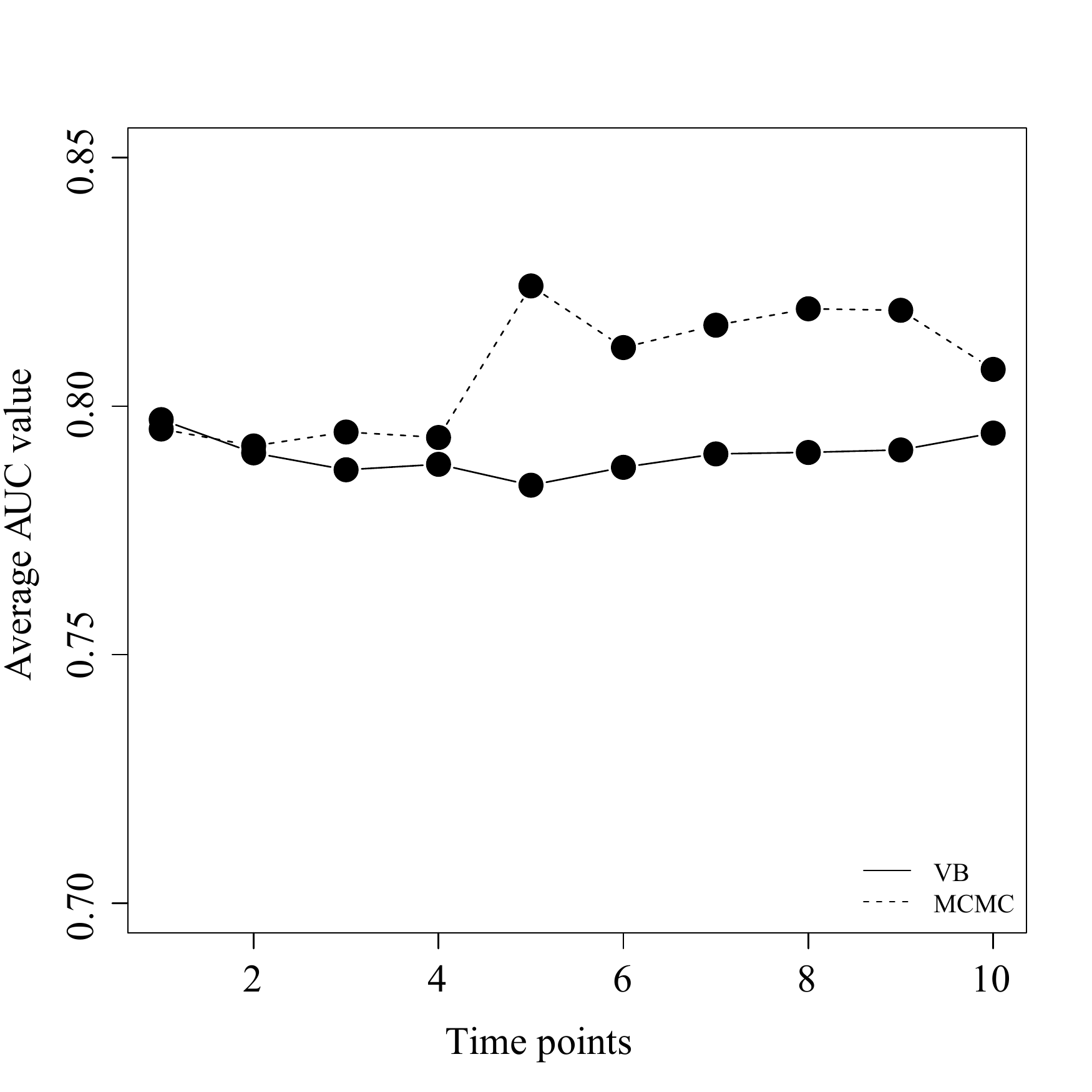}
		\end{minipage}
		\begin{minipage}{0.5\textwidth}
			\includegraphics[width=0.86\linewidth]{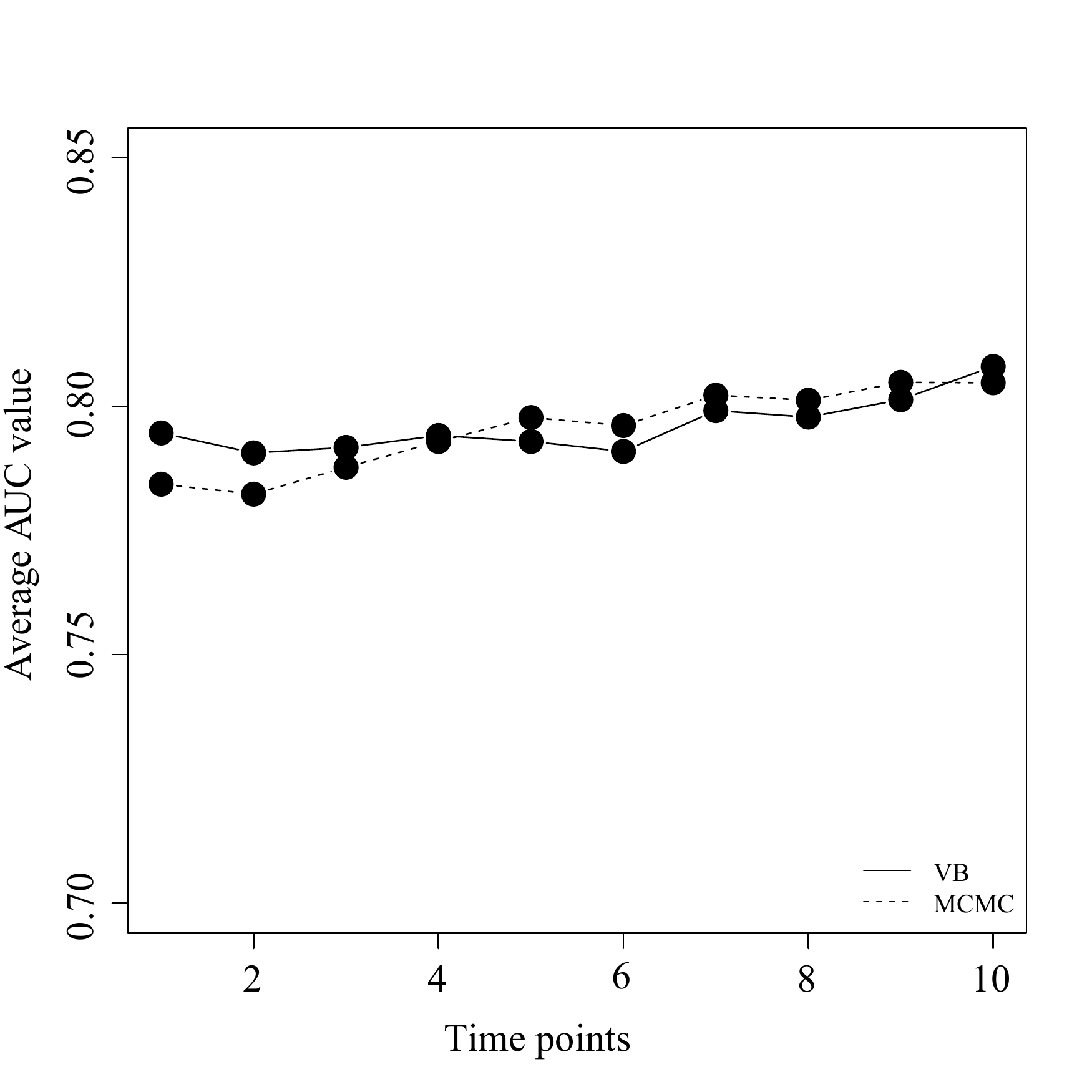}
		\end{minipage}
		\begin{minipage}{0.5\textwidth}
			\includegraphics[width=0.86\linewidth]{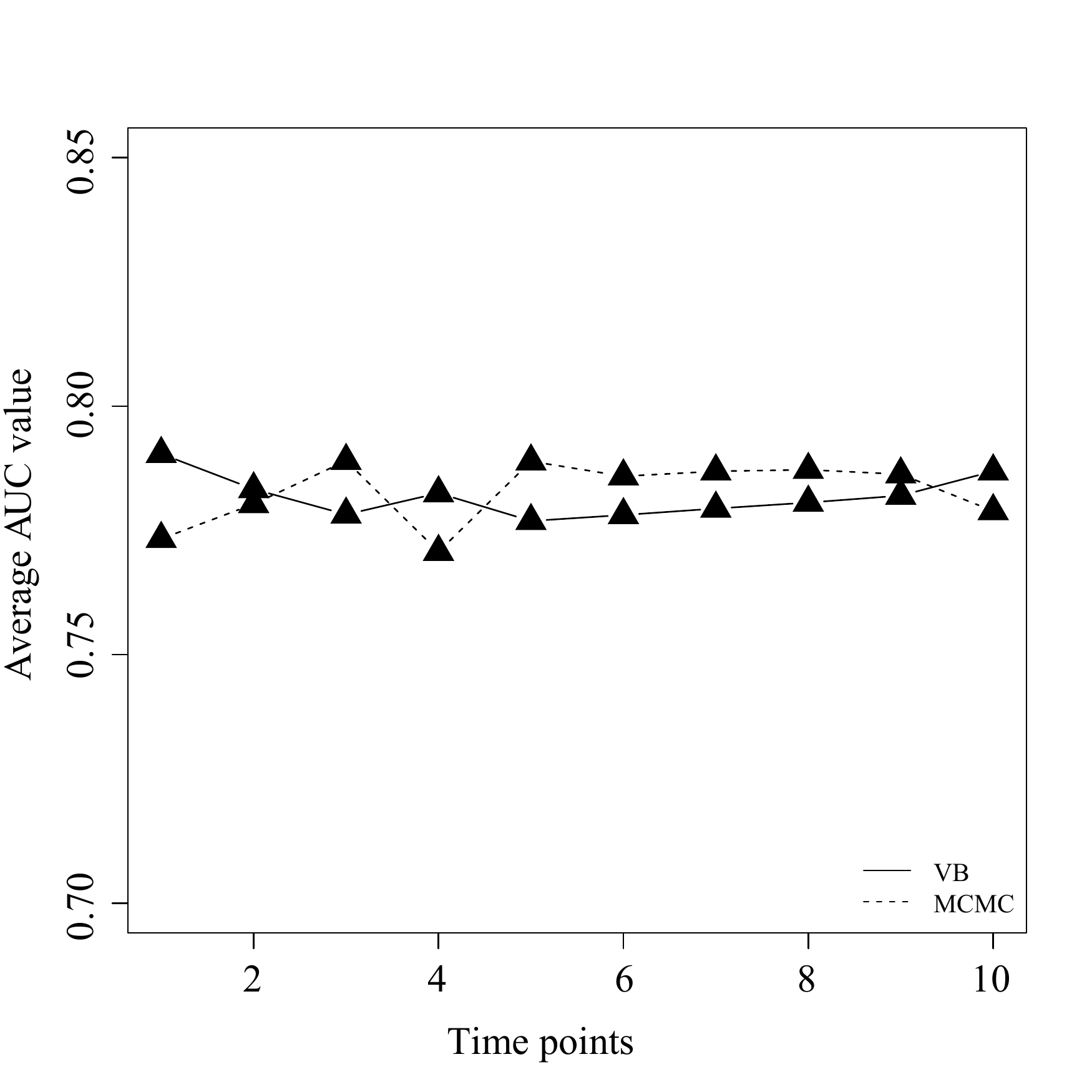}
		\end{minipage}
		\begin{minipage}{0.5\textwidth}
			\includegraphics[width=0.86\linewidth]{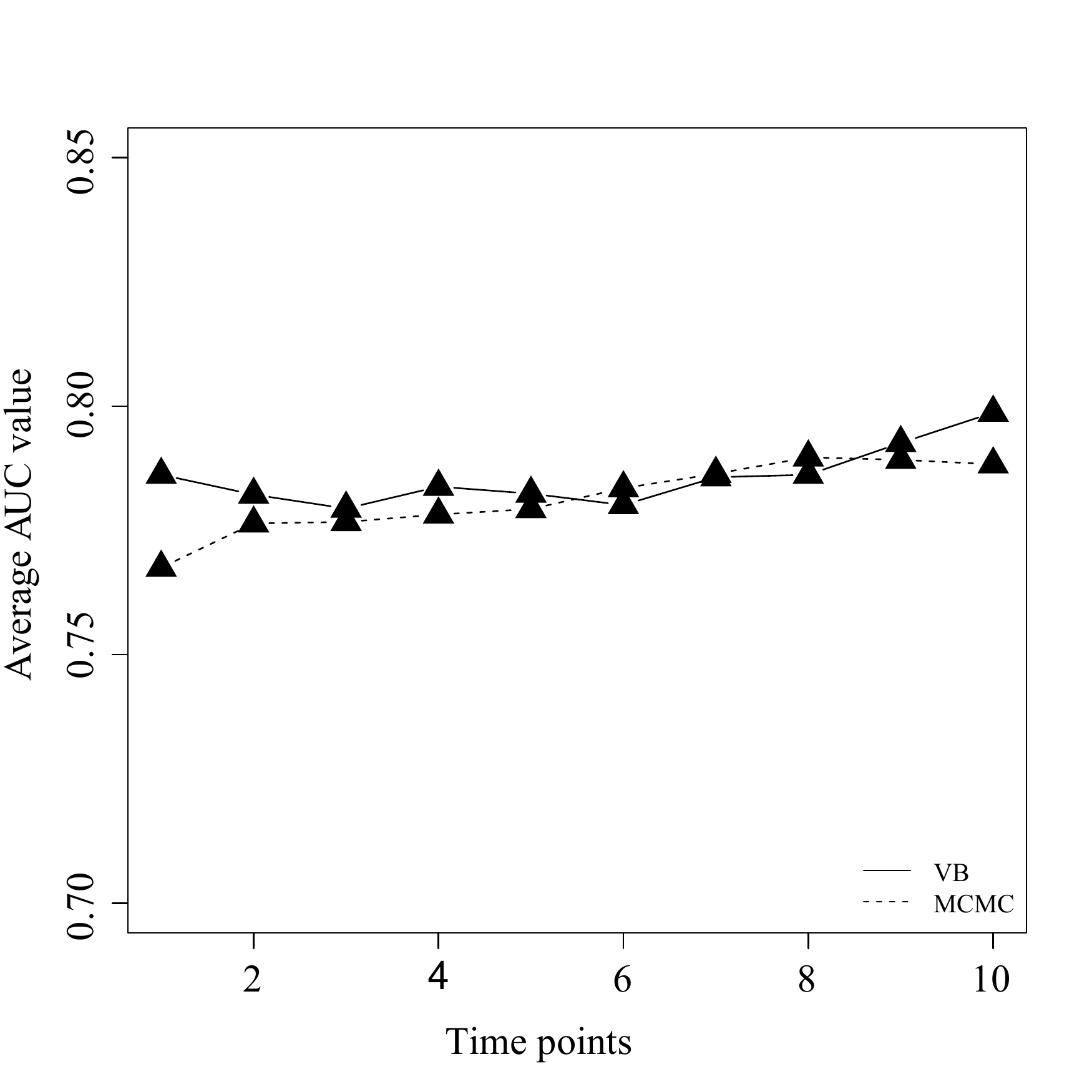}
		\end{minipage}
		\begin{minipage}{0.5\textwidth}
			\includegraphics[width=0.86\linewidth]{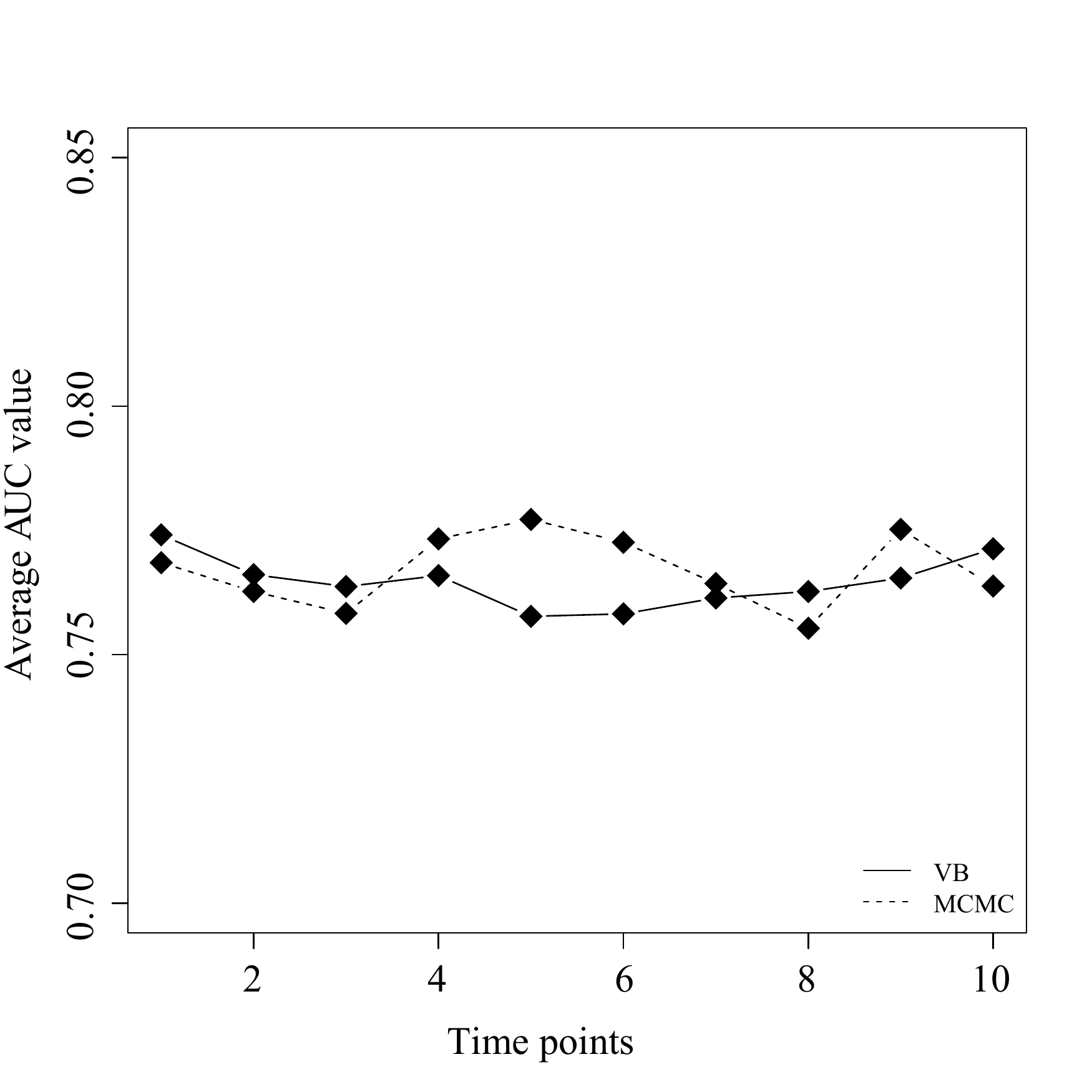}
		\end{minipage}
		\begin{minipage}{0.5\textwidth}
			\includegraphics[width=0.86\linewidth]{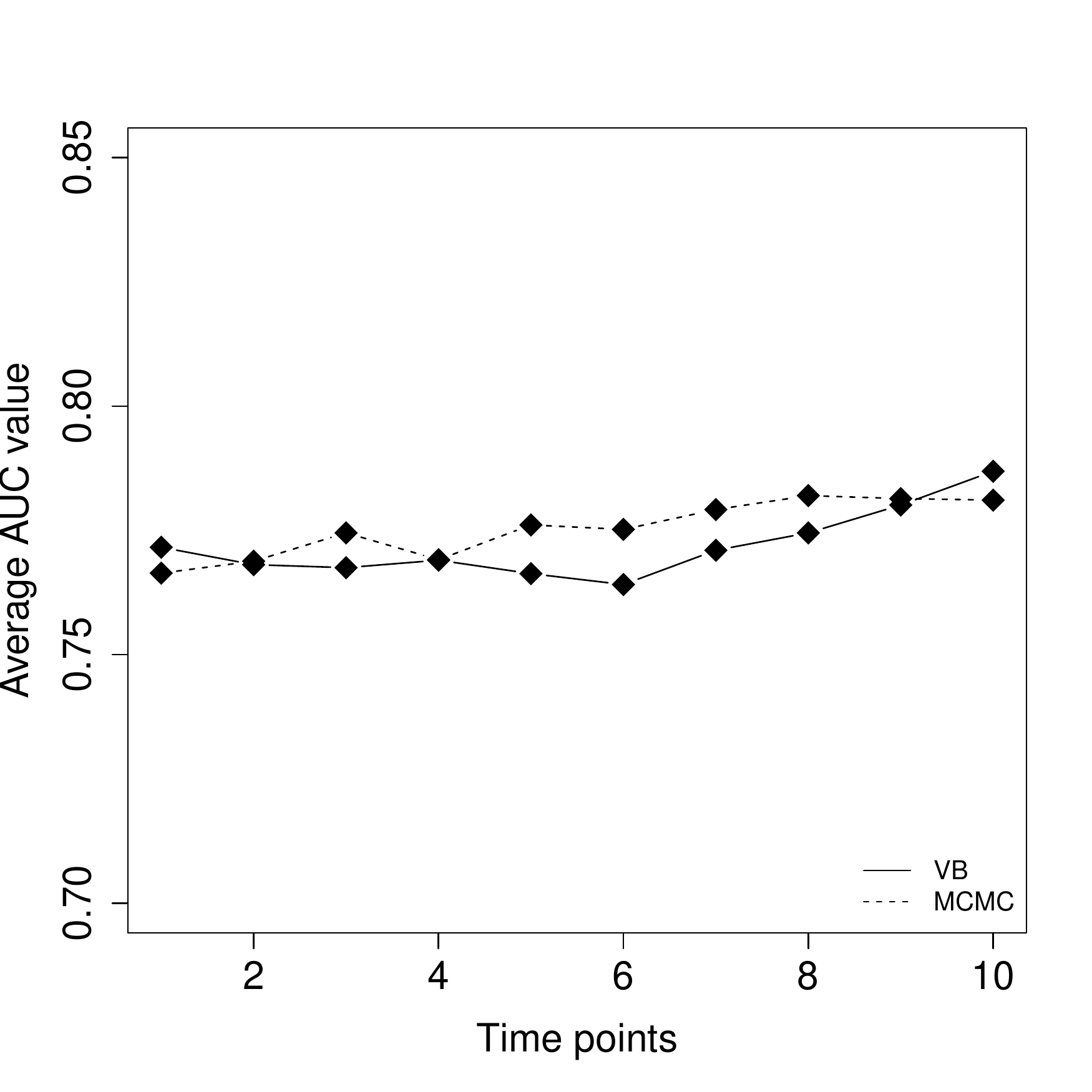}
		\end{minipage}
	\end{minipage}
	\caption{The average in-sample AUC values for VB and MCMC on simulated networks.
		(Left column: small transition; right column: large transition. First row: dense networks; second row: moderate networks; third row: sparse networks.)}\label{sim50}
\end{figure}

The parallel tempering algorithm took about half an hour on average to obtain 100,000 samples (10,000 for burn-in and 90,000 for inference), while the proposed VB algorithm only took several seconds.
The performance was evaluated by the AUC values of in-sample predictions.
The results of the two algorithms are
summarized in Figure \ref{sim50}.
Their performances in terms of AUC values are close. Compared with MCMC, the variational algorithm achieves similar performance with much less computation time.

\subsection*{Details of the Parallel Tempering Algorithm}

We used the following parallel tempering algorithm in the comparison between VB and MCMC in Section \ref{sim1}.

\begin{itemize}
	\item For $k=1,\dots,K$, construct target distributions
	$\pi_k \propto \exp\left\{ \frac{\log(p(\boldsymbol{\mathcal{Y}}|\beta,\boldsymbol{\mathcal{X}})p(\boldsymbol{\mathcal{X}}|\sigma^2,\tau^2)p(\beta))}{T_k} \right\}$, where $T_K>\dots>T_1$ are the temperatures. The lowest temperature $T_1=1$ corresponds to the target distribution we are interested in.
	\item Initialize $\boldsymbol{\mathcal{X}}^{(0)}_{k}$ and $\beta^{(0)}_{k}$
	randomly for $k=1,\dots,K$. Here the subscript $k$ indicates the sample in the $k$-th temperature level.
	\item Suppose the sample at step $t$ is $(\boldsymbol{\mathcal{X}}^{(t)}_{k},\beta^{(t)}_{k})$, $k=1,\dots,K$.  
	\begin{itemize}
		\item Draw $u\sim \text{Uniform}(0,1)$.
		\item If $u\leq a_0$, for $k=1,\dots,K$, draw $(\boldsymbol{\mathcal{X}}^{(t+1)}_{k},\beta^{(t+1)}_{k})$ using the 
		Metropolis-Hastings within Gibbs algorithm (given below) with target distribution $\pi_k$.
		\item Otherwise, randomly choose a neighboring pair of temperatures $T_i$ and $T_{i+1}$, and swap $(\boldsymbol{\mathcal{X}}^{(t)}_{i},\beta^{(t)}_{i})$ and $(\boldsymbol{\mathcal{X}}^{(t)}_{i+1},\beta^{(t)}_{i+1})$ with probabilty $\min\left\{
		1, \frac{\pi_{i}\left(\boldsymbol{\mathcal{X}}^{(t)}_{i+1},\beta^{(t)}_{i+1}\right) \pi_{i+1}\left(\boldsymbol{\mathcal{X}}^{(t)}_{i},\beta^{(t)}_{i}\right)}{\pi_{i}\left(\boldsymbol{\mathcal{X}}^{(t)}_{i},\beta^{(t)}_{i}\right) \pi_{i+1}\left(\boldsymbol{\mathcal{X}}^{(t)}_{i+1},\beta^{(t)}_{i+1}\right)}
		\right\}$.
	\end{itemize}
\end{itemize}
In our simulations, we set the number of temperatures $K=3$, and the three temperatures are $T_1=1$, $T_2=10$ and $T_3=20$. The tuning parameter $a_0=0.9$.

Now we give the detail of the Metropolis-Hastings within Gibbs algorithm for temperature $T_1=1$. The algorithms for other temperatures are similar and
will be omitted. Recall the likelihood function of the model and the priors are given by
$$
p(\boldsymbol{\mathcal{Y}}|\beta,\boldsymbol{\mathcal{X}})
= \prod_{t=1}^{T}\prod_{i\neq j}
\frac{e^{Y_{ijt}(\beta-||\boldsymbol{X}_{it}-\boldsymbol{X}_{jt}||^2)}}{1+e^{\beta-||\boldsymbol{X}_{it}-\boldsymbol{X}_{jt}||^2}},
$$

$$
p(\boldsymbol{\mathcal{X}}|\sigma^2,\tau^2)
\propto
\prod_{i=1}^{n} \left(
e^{-\frac{||X_{i1}||^2}{2\sigma^2}}
\cdot \prod_{t=2}^{T} e^{-\frac{||X_{it}-X_{i(t-1)}||^2}{2\tau^2}}
\right),
\quad
p(\beta) \propto
e^{-\frac{(\beta-\xi)^2}{2\psi^2}}.
$$

The full conditional distributions of the latent variables and the intercept are given by
\begin{itemize}
	\item For $t=1$, $i=1,\dots,n$:
	$$
	p(\boldsymbol{X}_{i1}|\cdot)
	\propto
	\left(
	\prod_{i=1}^{n} e^{-\frac{||\boldsymbol{X}_{i1}||^2}{2\sigma^2}-\frac{||\boldsymbol{X}_{i2}-\boldsymbol{X}_{i1}||^2}{2\tau^2}}
	\right) \cdot
	\left(
	\prod_{i\neq j}
	\frac{e^{Y_{ij1}(\beta-||\boldsymbol{X}_{i1}-\boldsymbol{X}_{j1}||^2)}}{1+e^{\beta-||\boldsymbol{X}_{i1}-\boldsymbol{X}_{j1}||^2}}
	\right)
	$$
	
	\item For $t=2,\dots,T$:
	$$
	p(\boldsymbol{X}_{it}) \propto
	\left(
	\prod_{i=1}^{n} e^{-\frac{||\boldsymbol{X}_{it}-\boldsymbol{X}_{i(t-1)}||^2 + ||\boldsymbol{X}_{i(t+1)}-\boldsymbol{X}_{it}||^2}{2\tau^2}}
	\right) \cdot
	\left(
	\prod_{i\neq j}
	\frac{e^{Y_{ijt}(\beta-||\boldsymbol{X}_{it}-\boldsymbol{X}_{jt}||^2)}}{1+e^{\beta-||\boldsymbol{X}_{it}-\boldsymbol{X}_{jt}||^2}}
	\right)
	$$
	
	\item
	$$
	p(\beta|\cdot) \propto
	e^{-\frac{(\beta-\xi)^2}{2\psi^2}} \cdot
	\left(
	\prod_{t=1}^{T}\prod_{i\neq j}
	\frac{e^{Y_{ijt}(\beta-||\boldsymbol{X}_{it}-\boldsymbol{X}_{jt}||^2)}}{1+e^{\beta-||\boldsymbol{X}_{it}-\boldsymbol{X}_{jt}||^2}}
	\right)
	$$
\end{itemize}
All of these full conditional distributions are drawn via Metropolis-Hastings with normal random walk proposals. In order to resolve the non-identifiability issue associated with latent space models, we perform a Procrustes transformation after we draw a new set of $\left\{\boldsymbol{X}_{it}\right\}$'s.

\subsection*{Details on the Implementation of the Algorithm in Simulation Studies}
In the simulation of Section \ref{sim1}, the initial latent positions were drawn from a mixture Gaussian distribution with equal probability on two components
centered at $(-0.5,0)$ and $(0.5,0)$, respectively, and the variance of 
both components was set to be $\sigma^2=0.5$.
In the simulation of Section \ref{sim}, the initial latent positions were drawn from a mixture Gaussian distribution with equal probability on two components
centered at $(-1.5,0)$ and $(1.5,0)$, respectively, and the variance of
both components was set to be $\sigma^2=0.5$.
The variance of the transition distribution was set to be $\tau^2=0.01$ for the small transition case and $\tau^2=0.16$ for the large transition case.

The edge density of the network can be controlled by the intercept $\beta$. In the simulation of Section \ref{sim1}, we set $\beta=0.5$, $-0.5$, $-1.5$ for the dense, moderate and sparse cases, respectively. The corresponding edge density is around 0.24, 0.10 and 0.06, respectively. The prior distribution for $\beta$
was set to be $\mathcal{N}(0,2)$. In the simulation of Section \ref{sim}, the average degree of dense, moderate and sparse networks are approximately 7.5, 4, and 1.8, respectively. The prior for $\beta$ was set to be $\mathcal{N}(0,2)$.

The VB algorithm requires initial values for the variational parameters.
In the simulation of Section \ref{sim1}, the initial values of $\left\{\tilde{\boldsymbol{\mu}}_{it}\right\}$ ($i=1,\dots,50$, $t=1,\dots,10$) were obtained through multi-dimensional scaling (MDS). The initial value for the covariance matrix $\tilde{\Sigma}$ was set to be the identity matrix $\mathbb{I}_{2}$. The initial values for $\tilde{\xi}$ and $\tilde{\psi}$ were 0 and 2, respectively.
In the simulation of Section \ref{sim}, the variational parameters $\left\{\tilde{\boldsymbol{\mu}}_{it}\right\}$ ($i=1,\dots,n$, $t=1,\dots,T$) were randomly initialized. The initial values of other variational parameters were set to be the same as the simulation of Section \ref{sim1}.

\subsection*{Simulation for Networks with 5000 Nodes}

We carried out simulation studies for networks with $n=5000$ nodes under two
settings.
To control the density of the networks,
we set the intercept $\beta=-2.5$ for the dense case, and $\beta=-4.5$ for the sparse case. All other settings were the same as the simulation in Section \ref{sim}.
The prior for $\beta$ was set to be $\mathcal{N}(0,0.01)$. The variational parameters
$\left\{\tilde{\boldsymbol{\mu}}_{it}\right\}$ ($i=1,\dots,n$, $t=1,\dots,T$)
were still randomly initialized. The average AUC values and their standard errors are given in Table \ref{sim5000}.
We can see that the variational method still performed well with these large networks, and the performance on dense networks is better than sparse ones.

\begin{table}[h]
	\begin{scriptsize}
		\begin{tabular}{l|cccccccccc}
			\hline
			Time                                                                                   & 1      & 2      & 3      & 4      & 5      & 6      & 7      & 8      & 9      & 10     \\ \hline
			\multirow{2}{*}{\begin{tabular}[c]{@{}l@{}}Dense, \\ small $\tau^2$\end{tabular}}    & 0.8496 & 0.8488 & 0.8488 & 0.8495 & 0.8497 & 0.8489 & 0.8507 & 0.8536 & 0.8545  & 0.8582 \\
			& (0.0021) & (0.0034) & (0.0028) & (0.0025) & (0.0031) & (0.0020) & (0.0023) & (0.0026) & (0.0014) & (0.0015) \\ \hline
			\multirow{2}{*}{\begin{tabular}[c]{@{}l@{}}Sparse, \\ large $\tau^2$\end{tabular}}   & 0.7561 & 0.7564 & 0.7580 & 0.7608 & 0.7635 & 0.7663 & 0.7676 & 0.7699 & 0.7725 & 0.7762 \\
			& (0.0015)  & (0.0025) & (0.0024) & (0.0020) & (0.0018) & (0.0014) & (0.0013)  & (0.0011) & (0.0024) & (0.0014) \\ \hline
		\end{tabular}
		\caption{The average AUC values and standard errors (in parentheses) for VB on simulated networks with $n=5000$ nodes.}
		\label{sim5000}
	\end{scriptsize}
\end{table}

\subsection*{Additional Simulation: the Effect of $\alpha$ in $\alpha$-VB}
In this simulation, we studied the effect of $\alpha$ in the $\alpha$-VB algorithm. While the authors provided some simulation results for the $\alpha$-VB algorithm in \cite{yang2017alpha}, we focus on the effect of the choice of $\alpha$ in the dynamic latent space model.

In the $\alpha$-VB framework, the upper bound of the KL-divergence is given by
\begin{align*}
D_{\alpha}
\leq &-\frac{nT}{2}\log(\det(\tilde{\Sigma}))
+ \left(\frac{n}{2\sigma^2}+\frac{n(T-1)}{\tau^2}\right)\tr(\tilde{\Sigma})\\ \nonumber
& + \frac{1}{2\sigma^2}\sum_{i=1}^{n}\tilde{\boldsymbol{\mu}}_{i1}^T \tilde{\boldsymbol{\mu}}_{i1}
+ \frac{1}{2\tau^2}\sum_{t=2}^{T}\sum_{i=1}^{n}(\tilde{\boldsymbol{\mu}}_{it}-\tilde{\boldsymbol{\mu}}_{i(t-1)})^T (\tilde{\boldsymbol{\mu}}_{it}-\tilde{\boldsymbol{\mu}}_{i(t-1)})\\ \nonumber
& + \frac{1}{2\alpha}\left(\frac{\tilde{\psi}^2}{\psi^2}-\log\frac{\tilde{\psi}^2}{\psi^2} + \frac{(\tilde{\xi}-\xi)^2}{\psi^2} \right)
- \sum_{t=1}^{T}\sum_{i\neq j}\biggr\{ Y_{ijt}
\left(
\tilde{\xi}-2\tr(\tilde{\Sigma})-||\tilde{\boldsymbol{\mu}}_{it}-\tilde{\boldsymbol{\mu}}_{jt}||^2
\right) \\
& -\log\left(
1+\frac{\exp\{\tilde{\xi}+\frac{1}{2}\tilde{\psi}^2\}}{\det(\mathbb{I}+4\tilde{\Sigma})^{1/2}}
\cdot
\exp\{
-(\tilde{\boldsymbol{\mu}}_{it}-\tilde{\boldsymbol{\mu}}_{jt})^T
(\mathbb{I}+4\tilde{\Sigma})^{-1}
(\tilde{\boldsymbol{\mu}}_{it}-\tilde{\boldsymbol{\mu}}_{jt})
\}
\right)
\biggr\} + constant.
\end{align*}

Note that the update equations for variational parameter $\tilde{\boldsymbol{\mu}}$ and $\tilde{\Sigma}$ stay the same. The update equations for $\tilde{\xi}$ and $\tilde{\psi}^2$ are listed as follows.
\begin{itemize}
	\item Update of $\tilde{\xi}$:
	$$
	\tilde{\xi}^{(s+1)} \leftarrow
	\left(1+\alpha\psi^2 f''(\tilde{\xi}^{(s)})\right)^{-1}
	\left[
	\xi+\alpha\psi^2
	\left(
	\sum_{t=1}^{T}\sum_{i\neq j}Y_{ijt} + f''(\tilde{\xi}^{(s)})\tilde{\xi}^{(s)}-f'(\tilde{\xi}^{(s)})
	\right)
	\right].
	$$
	
	\item Update of $\tilde{\psi}^2$:
	$
	\tilde{\psi}^{2~(s+1)} \leftarrow
	\left(\frac{1}{\psi^2}+2\alpha f'(\tilde{\psi}^{2~(s)})\right)^{-1}.
	$
\end{itemize}

In this simulation study, we tried four different $\alpha$ values: $0.2$, $0.5$, $0.9$, $1.0$. The proposed VB algorithm corresponds to the $\alpha=1.0$ case. For each case, twenty datasets were simulated, each with $n=100$ and $T=10$. The data generating process was the same as the one for dense networks with small transition in the simultion in Section \ref{sim}. The average AUC values for each choice of $\alpha$ are given in Table \ref{sim-alphaVB}.

\begin{table}[h!]
	\begin{tabular}{ccccccccccc}
		\hline
		$T$     & 1      & 2      & 3      & 4      & 5      & 6      & 7      & 8      & 9      & 10     \\ \hline
		$\alpha=0.2$ & 0.8961 & 0.8952 & 0.8960 & 0.8968 & 0.8971 & 0.8965 & 0.8970 & 0.8982 & 0.8984 & 0.9016 \\ \hline
		$\alpha=0.5$ & 0.8961 & 0.8952 & 0.8960 & 0.8968 & 0.8971 & 0.8965 & 0.8970 & 0.8982 & 0.8984 & 0.9016 \\ \hline
		$\alpha=0.9$ & 0.8961 & 0.8952 & 0.8960 & 0.8968 & 0.8971 & 0.8965 & 0.8970 & 0.8982 & 0.8984 & 0.9017 \\ \hline
		$\alpha=1.0$ & 0.8961 & 0.8952 & 0.8960 & 0.8968 & 0.8971 & 0.8965 & 0.8970 & 0.8982 & 0.8984 & 0.9016 \\ \hline
	\end{tabular}
	\caption{The average AUC values given by the $\alpha$-VB algorithm with four different choices of $\alpha$.}\label{sim-alphaVB}
\end{table}

As shown in Table \ref{sim-alphaVB}, the performance of the $\alpha$-VB algorithm for the dynamic latent space model is not very sensitive to the choice of $\alpha$. This is due to the fact that the $\alpha$-VB penalization is only used on the intercept $\beta$ here, and the majority of variational parameters related to the latent positions are not affected.

As mentioned in \cite{yang2017alpha}, in general one may want to choose an $\alpha$ value that is close to 1 in practice (e.g., $\alpha=0.9$). In this case, the algorithm will enjoy theoretical guarantees without requiring extra assumptions, and at the same time the parameter estimation will be close to the $\alpha=1$ case.

\subsection*{Addition Simulation: Asymptotic Behaviors}
In this simulation study, we first verified the consistency of parameter estimation of the proposed VB algorithm as the number of nodes in the network goes to infinity. Note that both VB and MCMC are schemes for approximating the posterior distribution, but the true posterior distribution is unknown. Our theoretical results indicate that we can compare the VB estimate with the true parameter value.

In all simulations, the true value of the intercept $\beta=-2$. Other than this, the data generating process was the same as the one for dense networks with small transitions in Section \ref{sim}. The average AUC values and mean squared errors (MSEs) ($\mathbb{E}[(\hat{\beta}-\beta)^2]$) are given in Tables \ref{asym_table1} and \ref{asym_table2}.
As the sample size increases, the estimation accuracy of the model parameter $\beta$ improves.

\begin{table}[h!]
	\centering
	\begin{tabular}{c|cccccccccc}
		\hline
		$T$     & 1      & 2      & 3      & 4      & 5      & 6      & 7      & 8      & 9      & 10     \\ \hline
		$n=100$ & 0.6657 & 0.6297 & 0.6623 & 0.6407 & 0.6282 & 0.6297 & 0.6364 & 0.6724 & 0.6614 & 0.6632 \\ \hline
		$n=200$ & 0.7343 & 0.7299 & 0.7389 & 0.7341 & 0.7346 & 0.7381 & 0.7322 & 0.7478 & 0.7482 & 0.7560 \\ \hline
		$n=400$ & 0.7612 & 0.7572 & 0.7582 & 0.7553 & 0.7576 & 0.7679 & 0.7651 & 0.7663 & 0.7683 & 0.7743 \\ \hline
		$n=800$ & 0.7562 & 0.7583 & 0.7586 & 0.7625  & 0.7648 &     0.7635 & 0.7667 & 0.7682 & 0.7704 & 0.7727 \\ \hline
	\end{tabular}
	\caption{The average AUC values given by the VB algorithm with increasing number of nodes.}
	\label{asym_table1}
\end{table}

\begin{table}[h!]
	\centering
	\begin{tabular}{c|cccc}
		\hline
		$n$ & 100 & 200 & 400 & 800 \\ \hline
		MSE & 0.4567  & 0.0876  & 0.0228  & 0.0052  \\ \hline
	\end{tabular}
	\caption{The MSEs of the intercept $\beta$ with increasing number of nodes.}
	\label{asym_table2}
\end{table}

We also ran another simulation study to explore the behavior of the proposed algorithm when the number of time steps $T\rightarrow\infty$. In this simulation, we fixed the number of nodes $n=50$, and tried several different $T$ values. For each case, we calculated the average AUC value for all snapshots. Results in Table \ref{asym_table3} are based on 20 simulations.
The performance of the proposed algorithm does not change much
as the number of time steps increases.

\begin{table}[h!]
	\centering
	\begin{tabular}{c|cccccc}
		\hline
		$T$         & 10     & 20     & 40 & 80 & 160 & 320 \\ \hline
		Average AUC & 0.8987 & 0.8979 &  0.9024  &  0.9021  & 0.9078    & 0.9136    \\ \hline
	\end{tabular}
	\caption{The average AUC values given by the VB algorithm with increasing number of time steps.}
	\label{asym_table3}
\end{table}

\subsection*{Teenage Friendship Network Data}

\begin{figure}[h!]
	\vspace{-1.2cm}
	\parbox[t]{3.15in}{
		\centerline{
			{
				\includegraphics[width=0.7\textwidth]{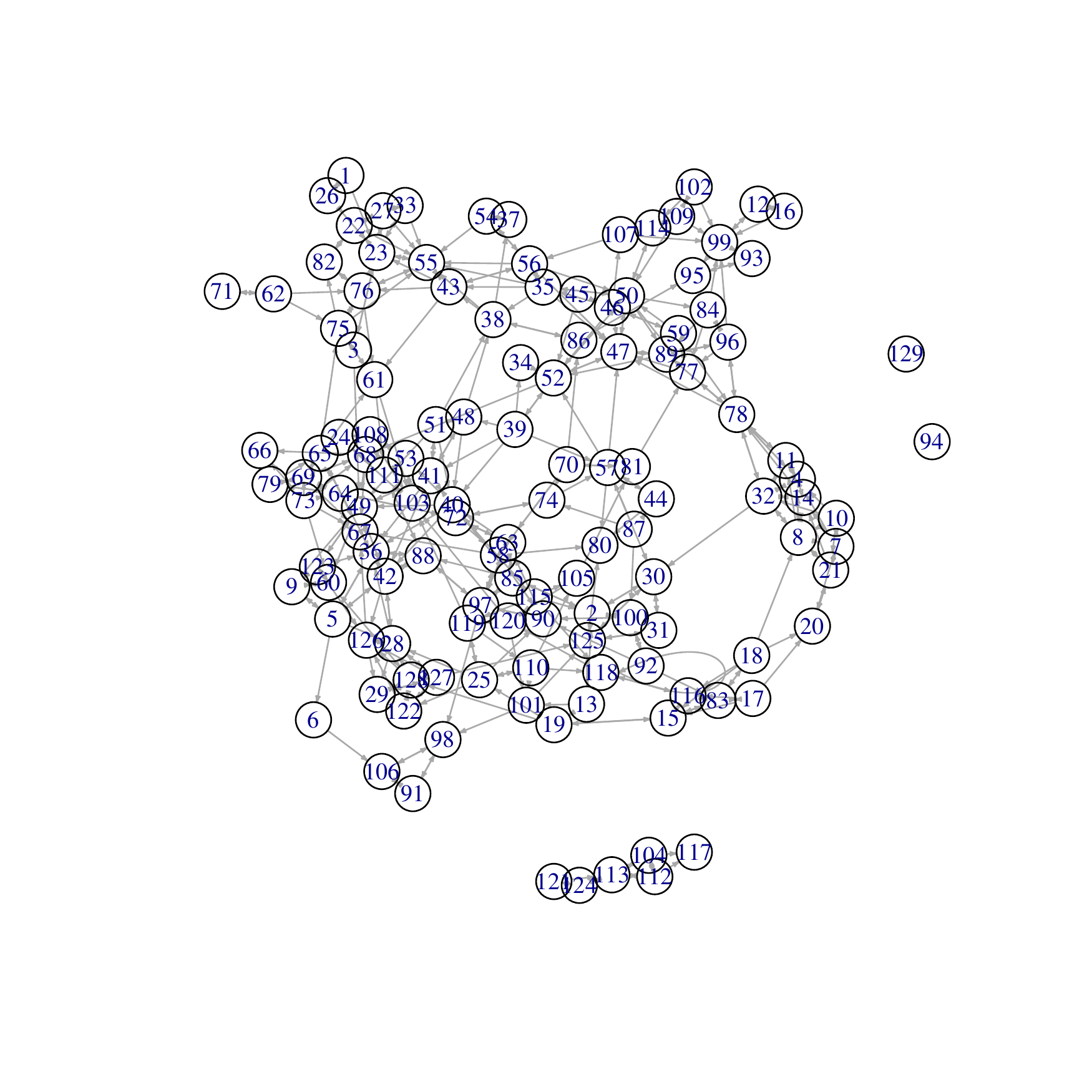}
			}
		}
	}
	\parbox[t]{3.15in}{
		\centerline{
			{
				\includegraphics[width=0.7\textwidth]{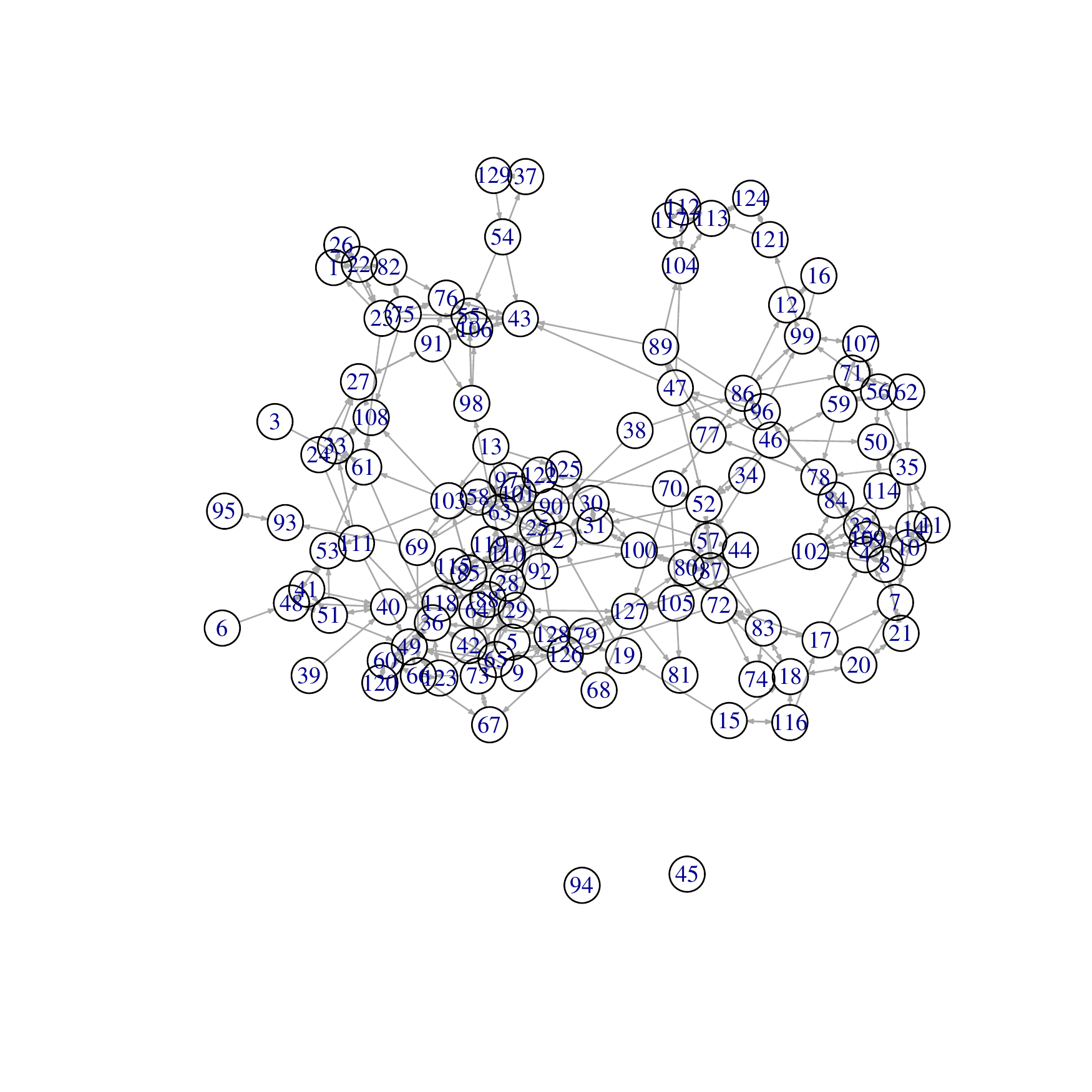}
			}
		}
	}
	\centerline{
		\vspace{-5cm}
		{
			\includegraphics[width=0.7\textwidth]{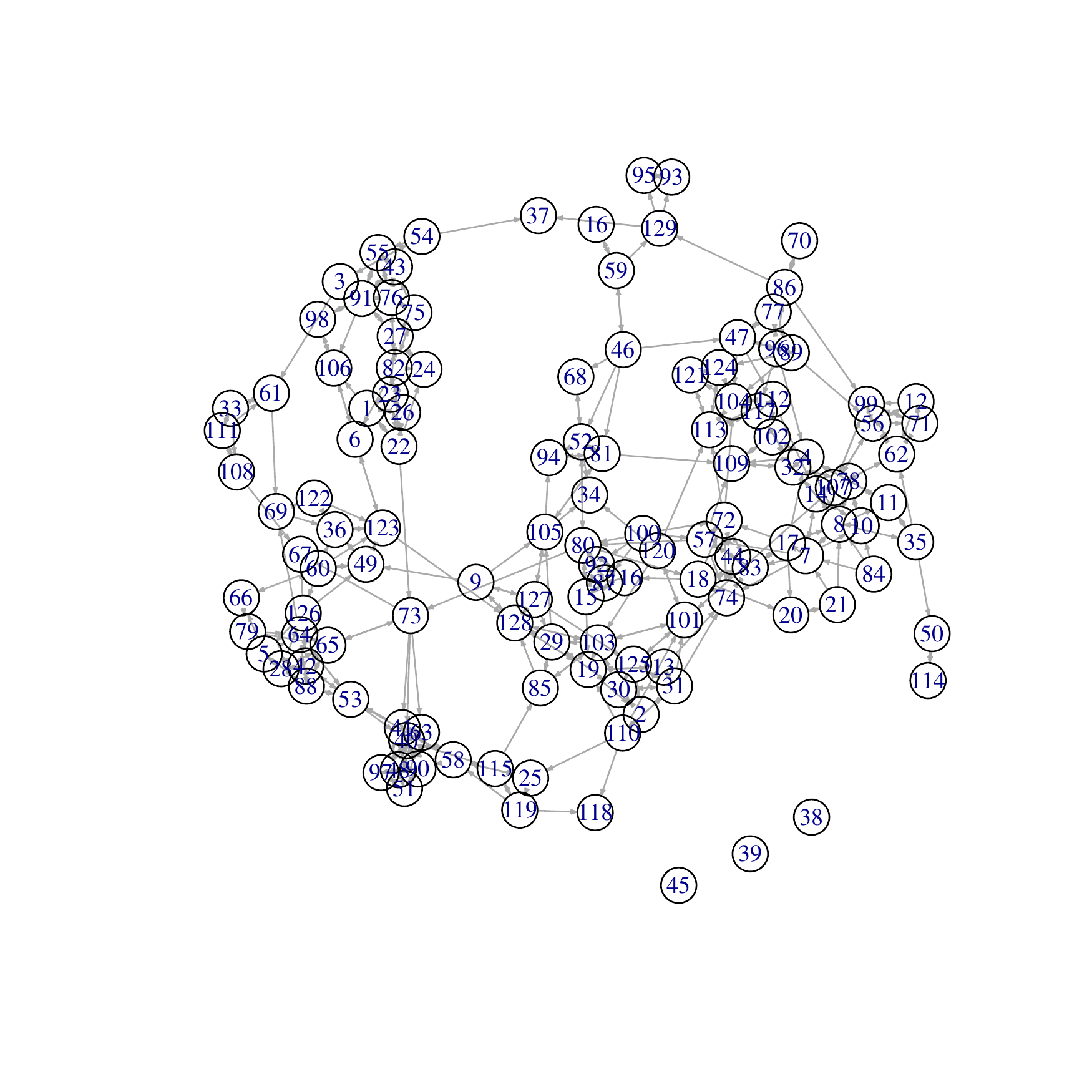}
		}
	}
	\vspace{-2.5cm}
	\caption{Friendship networks from ``Teenage Friends and Lifestyle Study" at
		times 1, 2 (top) and 3 (bottom).}
	\label{s129netplot}
\end{figure}

Figure \ref{s129netplot} shows the networks formed by the 129 pupils who were present at all three measurement time points from the ``Teenage Friends and Lifestyle Study" dataset. This network data is analyzed in Section \ref{teenage}.

\clearpage
\bibliographystyle{asa}
	\bibliography{DynamicNetwork}

	%\begin{thebibliography}{}
	%
	%\bibitem[\protect\citeauthoryear{Curtis}{Curtis}{1943}]{1943}
	%Curtis, M. (1943).
	%\newblock {\em Documents on International Affairs, 1938}, Volume~II.
	%\newblock London: Oxford University Press.
	%
	%\bibitem[\protect\citeauthoryear{Eubank}{Eubank}{2004}]{Eubank}
	%Eubank, K. (2004).
	%\newblock {\em The origins of World War II\/} (3 ed.).
	%\newblock Wheeling, Ill.: Harlan Davidson.
	%
	%\bibitem[\protect\citeauthoryear{Gellately}{Gellately}{1988}]{1988}
	%Gellately, R. (1988).
	%\newblock The gestapo and german society: Political denunciation in the gestapo
	%  case files.
	%\newblock {\em The Journal of Modern History\/}~{\em 60}, pp. 654--694.
	%
	%\bibitem[\protect\citeauthoryear{Noakes and Pridham}{Noakes and
	%  Pridham}{2001}]{Noakes}
	%Noakes, J. and G.~Pridham (2001).
	%\newblock {\em Nazism, 1919-1945. Vol. 3: Foreign Policy, War and Racial
	%  Extermination}.
	%\newblock Exeter: University of Exeter Press.
	%
	%\end{thebibliography}

	%%%%%%%%%%%%%%%%%%%%%%%%%%%%%%%%%%%%%%%%%%%%%%%%%%%%%%%%%%%%%%%%%%%%%%%%%%%%%%%%%%%%%%%%%%%%%%%%%%%%%%%%%%%%%%%%%%%%%%%%%%%%
	\vskip .65cm
	\noindent
	Department of Statistics, University of Illinois at Urbana-Champaign, Champaign, IL 61820
	\vskip 2pt
	\noindent
	E-mail: yanl5@illinois.edu
	\vskip 2pt
	
	\noindent
	Department of Statistics, University of Illinois at Urbana-Champaign, Champaign, IL 61820
	\vskip 2pt
	\noindent
	E-mail: yuguo@illinois.edu
	% \vskip .3cm
	%\centerline{(Received ???? 20??; accepted ???? 20??)}\par
\end{document}